\documentclass[aps,pre,twocolumn,floatfix,groupedaddress,superscriptaddress, showpacs]{revtex4-1} 

\usepackage{amsmath}
\usepackage{amssymb}
\usepackage{lipsum}
\usepackage{mathrsfs}

\usepackage{graphicx}
\usepackage{bm}
\usepackage[T1]{fontenc}
\usepackage{url}
\usepackage{rotating}
\usepackage{soul}
\usepackage{scalefnt}
\usepackage[hidelinks]{hyperref}
\usepackage[utf8]{inputenc}
\usepackage{enumerate}
\usepackage{float}
\usepackage{epigraph}

\usepackage{subcaption}
\usepackage{booktabs}

\usepackage{amsthm}

\theoremstyle{plain}

\usepackage{color}
\definecolor{redcolor}{rgb}{1.0,0.,0.}

\begin{document}
\title{Numerical Frequency Transfer Function Analysis of a Leaky Integrate-and-Fire Neuron}
\author{Felipe Gewers}
\email{felipe.gewers@usp.br}
\author{Luciano da F. Costa}
\affiliation{Instituto de F\'{\i}sica de S\~{a}o Carlos, Universidade de S\~{a}o Paulo, S\~{a}o Carlos, S\~ao Paulo, Brazil\\}
\graphicspath{{Figures/}}

\begin{abstract}
This work reports a transfer function-based approach to characterizing the operation of single neuronal cells in terms of the instantaneous frequency of the input and output signals. We adopt the leaky integrate-and-fire model. The transfer relationship is obtained by performing successive numeric-computational simulations and statistical regressions. Several interesting results are reported, including the identification and characterization of linearity in the transfer relationship, as well as the identification of regions in the parameter space characterized by sharper transfer functions. These properties can be used to facilitate simulations under certain circumstances and to validate the model by comparison with biological results.  
\end{abstract}

\maketitle

\section{Introduction}

Neuronal systems have been continuous and intensely investigated along the last decades mostly because of their essential role in underlying human intelligence. The several approaches attempted so far vary in many aspects, focusing on smaller or larger scales of time, space and detail \cite{dayan2003theoretical,kandel2000principles,paun2010oxford,ganong1995review,hobson2002cognitive}. Because a single neuronal cell constitutes the basic processing element of the nervous system, it has received particular attention from both theoretical and experimental points of view. One approach that has been often adopted consists in modeling the electrochemical activity of a single neuronal cell in terms of a set of mathematical dynamical equations \cite{hodgkin1952quantitative,pospischil2008minimal, izhikevich2003simple}. Because these approaches demand substantial computational cost, simplified investigations have been proposed involving only the transmembrane potential value, not taking into account the ionic currents, as in the case of integrate-and-fire models \cite{burkitt2006review}. Neuronal simulations performed in this way typically involve the dynamic unfolding of the neuronal state along time, under control of some mathematical rules such as a differential equation \cite{izhikevich2004model,brette2007simulation}. The analysis of the simulated results obtained by using this type of approaches is often performed taking into account the frequency of neuronal spikes, as this is known to correspond to an important aspect underlying information transfer and processing in neuronal systems \cite{baddeley1997responses,salinas2000periodicity,ferster1995cracking,lu2001temporal,colgin2009frequency}.

Several devices in electronics, such as transistors, are characterized in terms of an input/output formulation in terms of \emph{transfer function} \cite{mitra2006digital,keesman2011system}. More specifically, if the input signal is represented as the time function $x(t)$, and the output as $y(t)$, the transfer function $y(t) = F (x(t))$ defines the mapping from the input into the output signals. This approach is particularly useful because it summarize the operation of the device in terms of a simple functional relationship. Consequently, it becomes particularly straightforward to simulate the device operation in terms of this methodology. In addition, the shape of the transfer function itself can provide several insights about the processing being performed by the device, including its linearity.

Despite the interesting advantages provided by the transfer function approach, it has rarely been applied to characterizing and modeling the nervous system. In the present work, we develop a transfer function approach to characterize the operation of single neuronal cells, with respect to the instantaneous frequency of the action potentials. More specifically, the input signal $x(t)$ is taken as the instantaneous frequency of the presynaptic spikes, while the output signal $y(t)$ corresponds to the instantaneous frequency of the postsynaptic action potentials. We adopt the leaky integrate-and-fire model, and several numeric-computational simulations are performed for several model parameters and inputs. This allows us to estimate the transfer function of the instantaneous frequencies. Because of the memory property inherent to these systems, the obtained relationship is not exact in the sense that more than one distinct output can be obtained for the same input signal. In spite of this limitation, it has been possible to infer a linear pattern of relationship for several parametric configurations, which allowed us to estimate the neuronal operation with a good level of accuracy. 

This approach also has benefits for the description and validation of the neuronal model by comparing the results obtained with
biological equivalents; for the formulation of a methodology based solely on coding by instantaneous frequencies of the spikes, as the results obtained allow the input and output frequencies to directly related.  It is also interesting for the simulations of neuronal experiments that employ the concept of instantaneous frequency, because it provides a way more efficient for obtaining the output instantaneous frequency of a neuron than computing all the neuronal dynamics.

This article starts by presenting the adopted leaky integrate-and-fire model, as well as the adopted considerations and how the synapses have been approached and modeled. Next, we introduce the concept of dynamic frequency transfer function, and discuss why it is not possible to have a proper function in our case. Then, we present the parameters adopted in the simulations, including the model parameters, time resolution and frequency time window. The obtained results are then reported and discussed, with emphasis on the linearity of the transfer function, and perspectives for future developments are then proposed.

\section{The Leaky Integrate-and-Fire Model}

The leaky integrate-and-fire model (LIF) is probably one of the most well-known spiking neuron model, being widely used in computational neuroscience \cite{brunel2007lapicque,quiroga2013principles}. The equations describing the model are simple, which allows simulations of networks with a large numbers of neurons, as well as analytical treatments \cite{vreeswijk1998chaotic,brunel2000dynamics}. Despite their simplicity, integrate-and-fire networks have been able to describe a broad spectrum of neuronal dynamics and cortical functions, from the emergence of up-down states \cite{compte2003cellular,parga2007network,holcman2006emergence}, self-sustaining synchronous and asynchronous states \cite{kriener2014dynamics,hopfield1995rapid,griffith1963stability,el2009master}, working memory \cite{compte2000synaptic,wang1999synaptic,mongillo2008synaptic}, attention \cite{buehlmann2010optimal,deco2011cholinergic}, decision making \cite{wang2002probabilistic}, rhythmogenesis \cite{brunel2003determines}, and coding of sensory information \cite{buehlmann2010optimal,mazzoni2008encoding,mazzoni2011cortical}.

In the integrate-and-fire model, the neuron state is determined only by its membrane potential, that is altered by the synaptic inputs and/or injected current, each of these synaptic inputs are weighted by their respective synaptic efficacy \cite{hansel1998numerical}. When depolarized, the membrane potential decays exponentially with a characteristic time constant, the membrane time constant, and the neuron is said to be \emph{leaky}. An integrate-and-fire neuron is understood as a 'point neuron' since its spatial structure is not considered \cite{burkitt2006review}.

Integrating everything that was so far presented, the membrane potential of the leaky integrate-and-fire neuron model is defined by the following differential equation:

\begin{equation}\label{eq:lif}
	\tau_m\frac{dv(t)}{dt}=v_r+R_mI(t)-v(t),
\end{equation}

\noindent where $v(t)$ is the membrane potential, $v_r$ is the resting potential and $I(t)$ is the total current through the membrane. The parameter $\tau_m$ is the membrane time constant, $R_m$ is the membrane resistance and $C_m$ is the membrane capacitance, with the relation between them given as:

\begin{equation}
	\tau_m=R_mC_m.
\end{equation}

\begin{figure}[!htb]
	\centering
	\includegraphics[width=0.95\columnwidth]{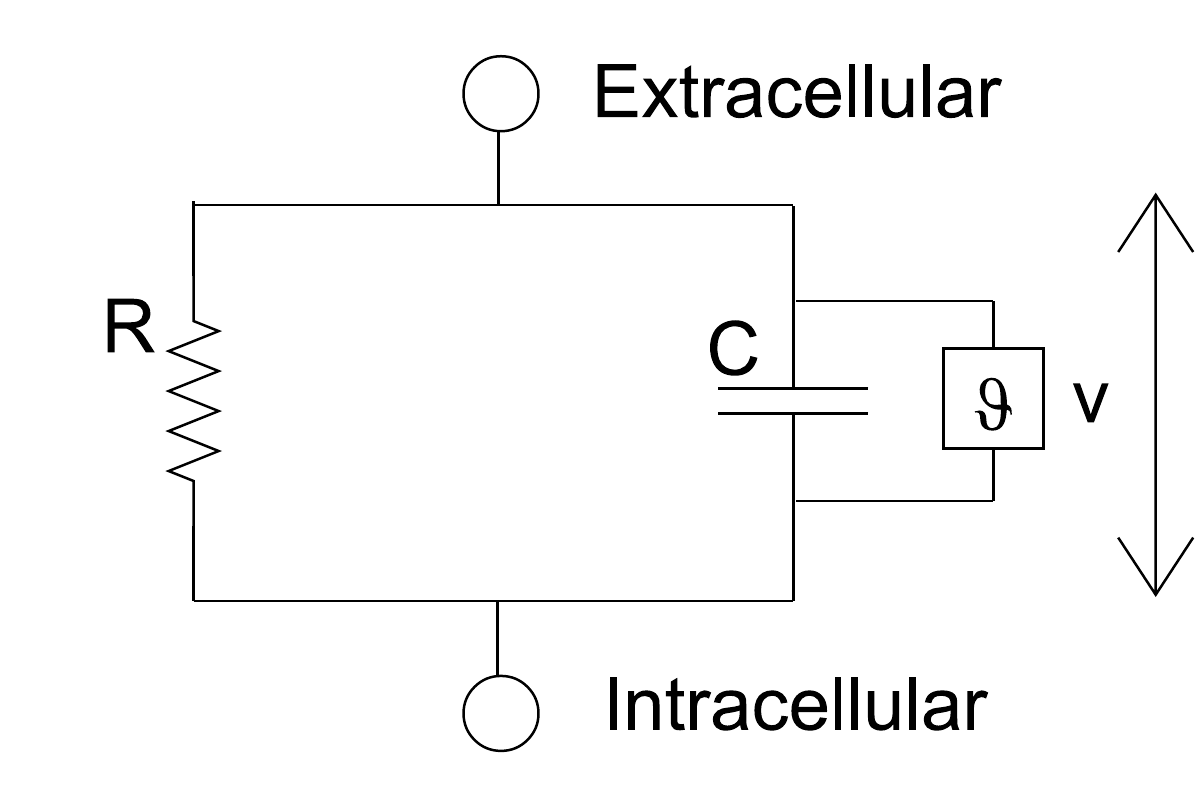}
	\caption{\label{fig:EquivalentCircuit}The equivalent circuit for the membrane potential of the leaky integrate-and-fire model for $v_r=0$.}
\end{figure}

\noindent This equation can be expressed in terms of the equivalent electronic circuit shown in Figure \ref{fig:EquivalentCircuit} \cite{abbott1999lapicque}. 

When the membrane potential reaches the spiking threshold $\vartheta$, the neuron generates an action potential. The neuron dynamics that generates the action potential is not treated by the integrate-and-fire model, as it simply defines the instant of time $t^{(f)}$ at which an action potential is generated:

\begin{equation}
	t^{(f)}:\quad v(t^{(f)})=\vartheta.
\end{equation}

\noindent After a postsynaptic spike is generated, the membrane potential is immediately changed to $v_{rs}$, the reset potential:

\begin{equation}
	\lim_{t \to {t^{(f)}}^+} v(t)=v_{rs}.
\end{equation}

\noindent We can establish that, after a spike occurs, the membrane potential does not change for a period of time corresponding to the absolute refractory period $\tau_{abs}$:

\begin{equation}
	v(t+t^{(f)})=v_{rs},\quad  0 < t \le \tau_{abs}.
\end{equation}


\subsection{Considerations}

We will henceforth adopt a single integrate-and-fire neuron with null absolute refractory period $\tau_{abs}=0ms$, subject to a single excitatory synaptic input that generates a delta function postsynaptic current, which does not depends of the membrane potential.    

These conditions are a good approximation for a fast spiking cortical neuron, because the integrate-and-fire model does not reproduces adaptation or bursts, stimulated by a single spaced spike train that generates postsynaptic currents through channels controlled by 'AMPA' receptors. However, the model can also be used, with less precision, to represent other types of neuronal dynamics.

It has been assumed that the neuron had only one synaptic input to construct the transfer function with respect to one input and output. This situation can be achieved in cases where one input of the neuron is much more active and intense than the others, or when in a small time window the neuron receives many inputs which the sum of its intensities is nearly constant.

Under these circumstances, the spike train that stimulate the neuron generates a postsynaptic current in the form:

\begin{equation}\label{eq:postsynaptic_current}
    I(t)=C\sum_{s} \alpha_E \delta(t-t^{(s)}),
\end{equation}

\noindent where $t^{(s)}$ are the arrive times of the presynaptic neurotransmitters to the postsynaptic membrane, and $\alpha_E$ is the amplitude of the excitatory postsynaptic current.

\subsection{Initial Development of the Leaky Integrate-and-Fire Model}

The postsynaptic current of a single presynaptic spike $I(t)=C\alpha_E\delta(t-t^{t-t^{(s)}})$, stimulating the neuron at rest $v(0)=v_r$, generates an alteration in the membrane potential. Solving the differential Equation \ref{eq:lif} for these circumstances, we have:

\begin{equation}\label{eq:postsynaptic_potential}
	v(t)=v_r + \alpha_E\exp \left(-\frac{t-t^{(s)}}{\tau_m}\right)\Theta(t-t^{(s)}),
\end{equation}

\noindent where $\Theta(t)$ is the Heaviside step function \cite{bracewell2000heaviside}. Therefore, a single synaptic input generates a postsynaptic potential $\epsilon(t,t^{(s)})=\alpha_E\exp(-(t-t^{(s)})/\tau_m)\Theta(t-t^{(s)})$.

Since Equation \ref{eq:lif} is linear and the synaptic stimuli does not depend on the membrane potential, it is possible to use the superposition principle to solve it for the spike train of Equation \ref{eq:postsynaptic_current}. Thus, the membrane potential at a time $t$ is given by the sum of the resting potential with the postsynaptic potentials generated by each presynaptic spike of the spike train:

\begin{equation}\label{eq:sum_postsynaptic_potential}
	v(t)=v_r + \alpha_E \sum_{s} \exp \left(-\frac{t-t^{(s)}}{\tau_m}\right)\Theta(t-t^{(s)}).
\end{equation}

Without loss of generality, the null rest potential scale $v_r = 0mV$ will be adopted. In addition, the restoration potential will be defined as the resting potential $v_{rs}=v_r=0mV$.

\subsection{Dynamic Frequency Transfer Function}

A frequency transfer function, in the framework of neuronal dynamics, is a function that associates the frequency of the input on the neuron, the presynaptic spike train, with the frequency of the output, the generated action potentials. In case the input frequency is constant, the output frequency will also be constant. In this sense, it is called the stationary frequency transfer function, associating the input frequency value with the frequency of the generated output (Appendix A).

However, in the general case, the input instantaneous frequency varies along time. In this case, it is called dynamic frequency transfer function, the function that associates the input instantaneous frequency at a given instant of time, with the output instantaneous frequency at the same instant of time.

There is many ways to define instantaneous frequency of a spike train \cite{rieke1999spikes}. Here, we adopt the following mean window method:

\begin{equation}\label{instantfreq}
	\nu(t) = \frac{1}{\Delta t}n(t;t+\Delta t),
\end{equation} 

\noindent where $n(t;t+\Delta t)$ is the number of spikes between $t$ and $t+\Delta t$. 

\begin{figure*}[!htb]
	\centering
	\begin{subfigure}[t]{\linewidth}
		\centering
		\caption{}
		\label{sfig:corrente_entrada_13-1spikes_eisd}
		\begin{subfigure}[t]{0.48\linewidth}
			\includegraphics[width=\linewidth]{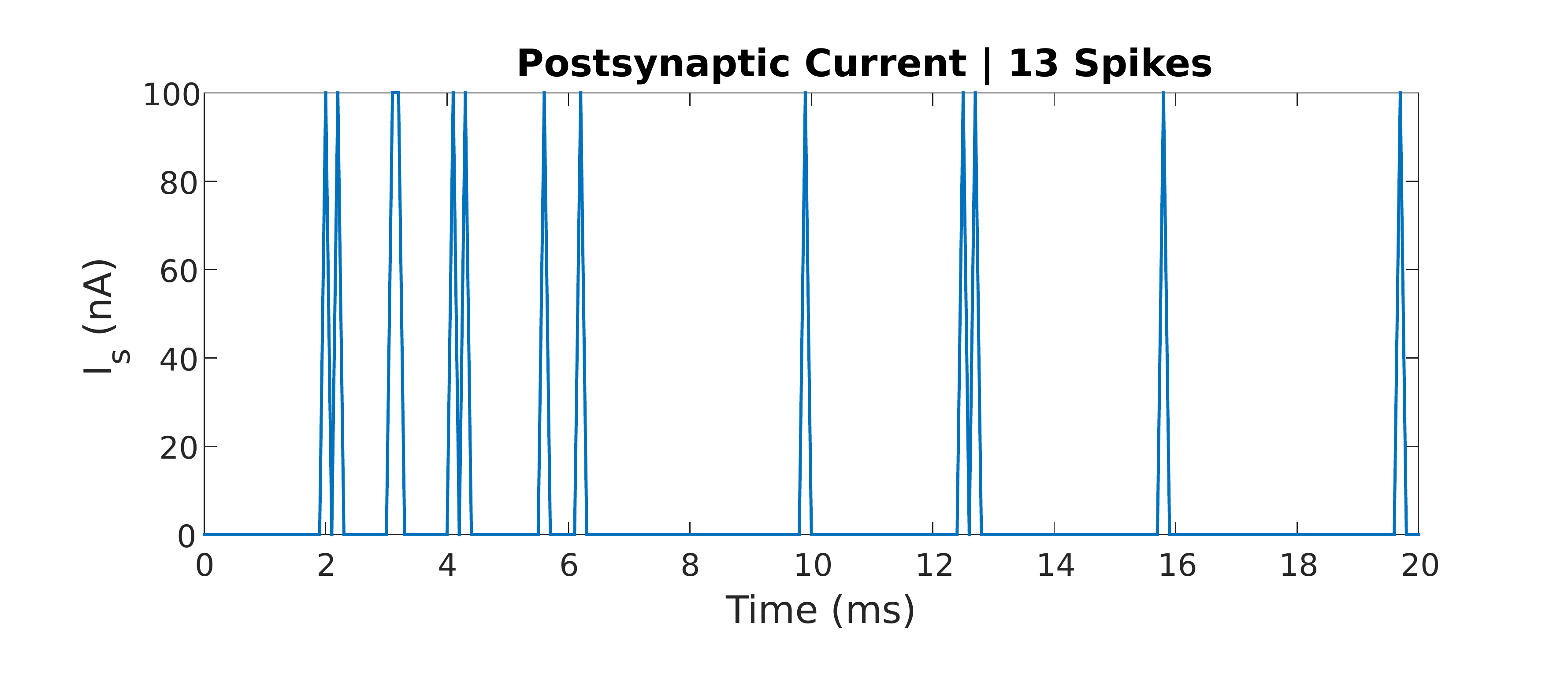}
		\end{subfigure}
		\quad
		\begin{subfigure}[t]{0.48\linewidth}	
			\includegraphics[width=\linewidth]{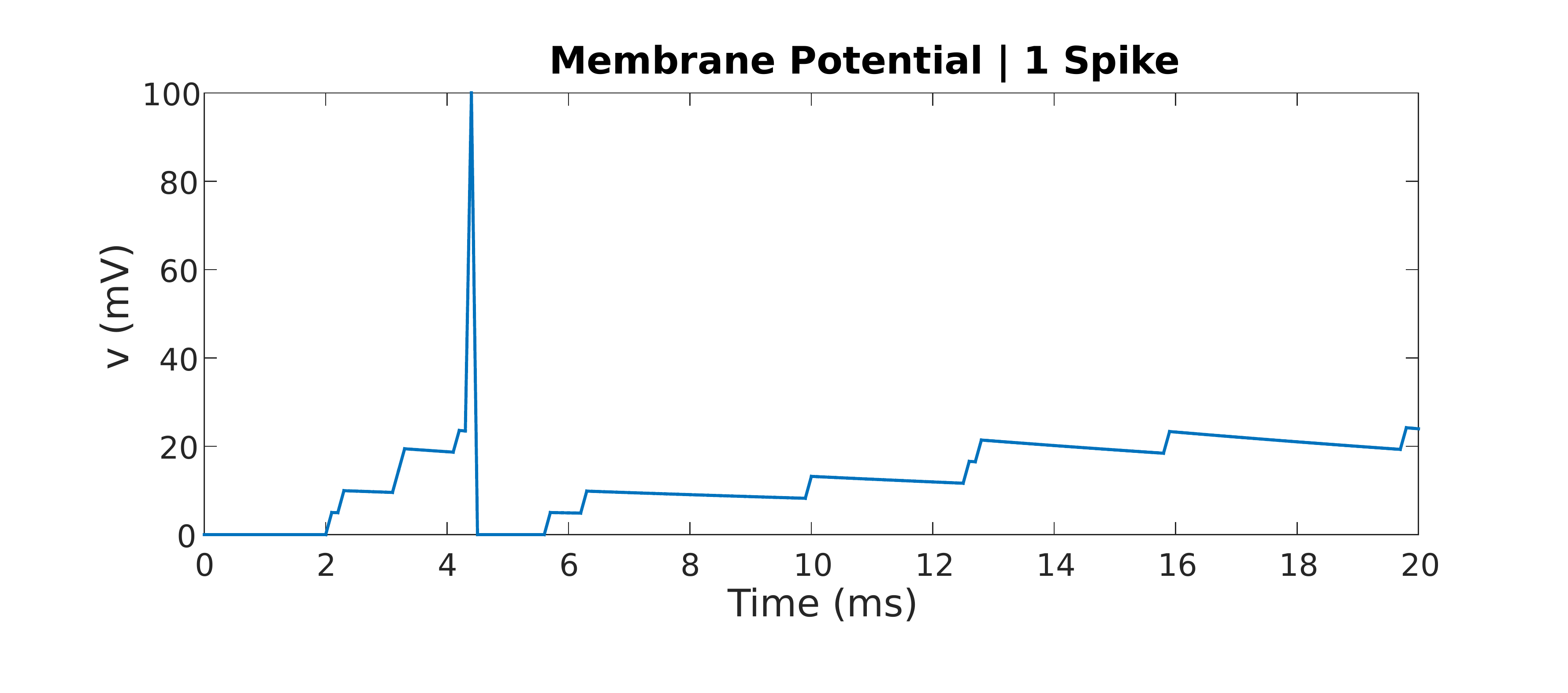}
		\end{subfigure}
	\end{subfigure}
	\quad
	\begin{subfigure}[t]{\linewidth}
		\centering
		\caption{}
		\label{sfig:corrente_entrada_13-2spikes_eisd}
		\begin{subfigure}[t]{0.48\linewidth}
			\includegraphics[width=\linewidth]{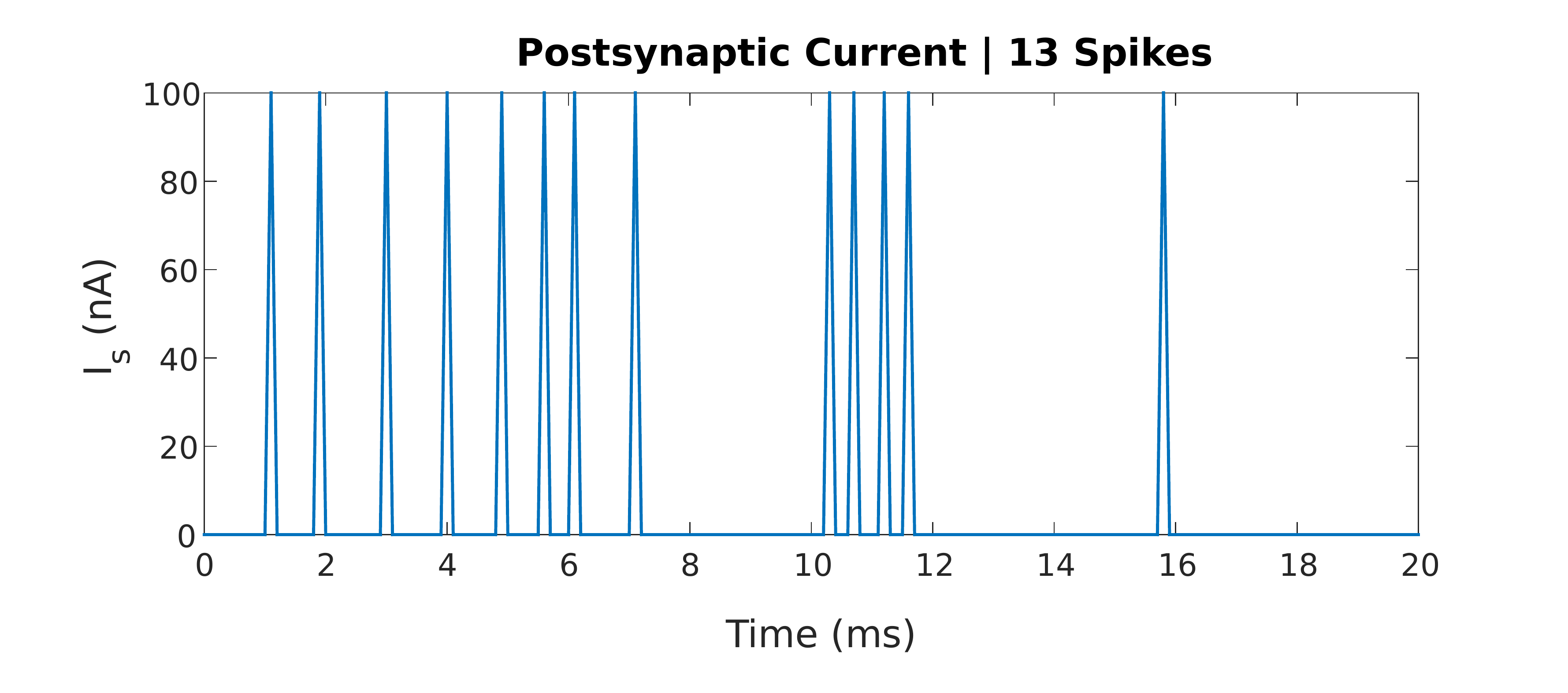}
		\end{subfigure}
		\quad
		\begin{subfigure}[t]{0.48\linewidth}
			\includegraphics[width=\linewidth]{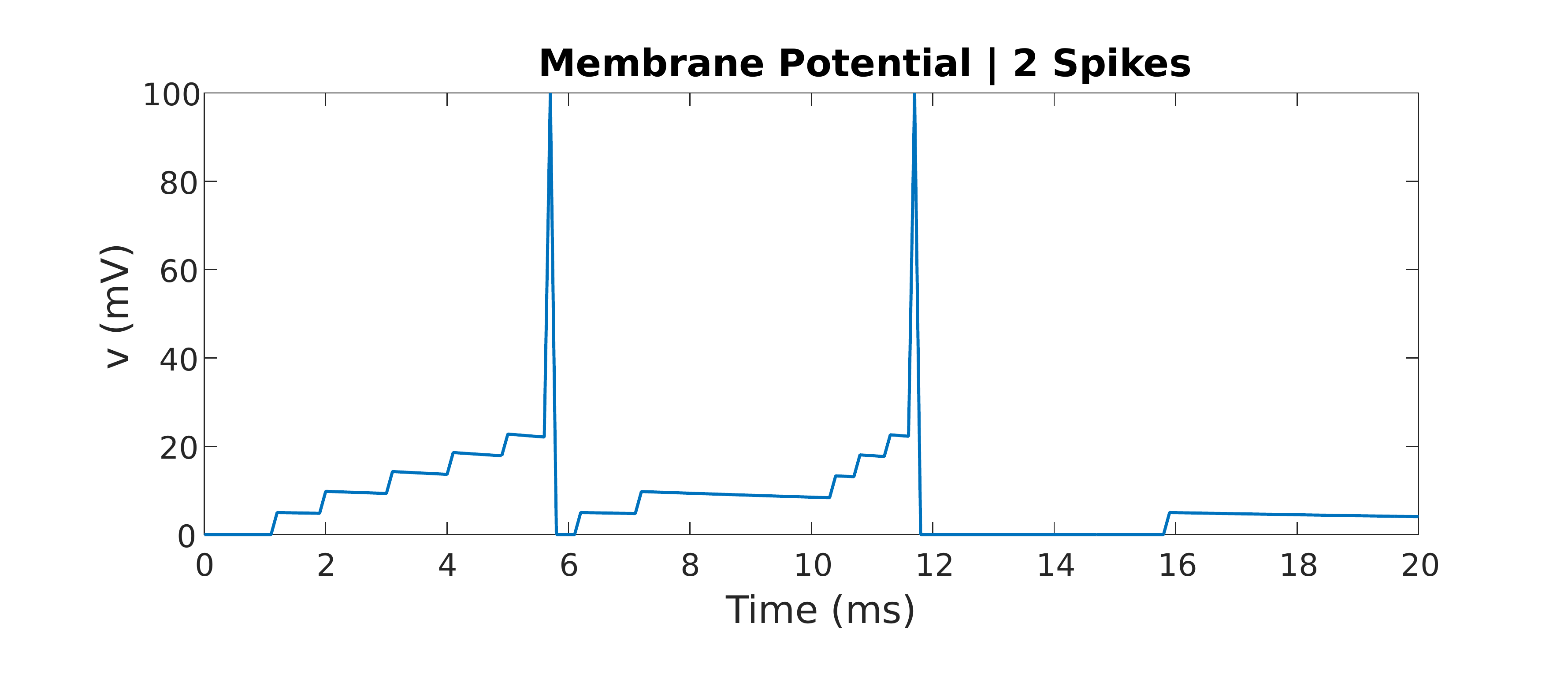}
		\end{subfigure}	
	\end{subfigure}
	\caption{\label{fig:SameInpDiffOut}Left: the presynaptic spike train. Right: the respective membrane potential variation. Considering a frequency window $\Delta t = 20ms$ both presynaptic spike trains have the same instantaneous frequencies at $t=0$, but generate outputs with different instantaneous frequencies.}
\end{figure*}

\begin{figure*}[!htb]
	\centering
	\begin{subfigure}[t]{\linewidth}
		\centering
		\caption{}
		\label{sfig:corrente_entrada_7-1spikes_edsi}
		\begin{subfigure}[t]{0.48\linewidth}
			\includegraphics[width=\linewidth]{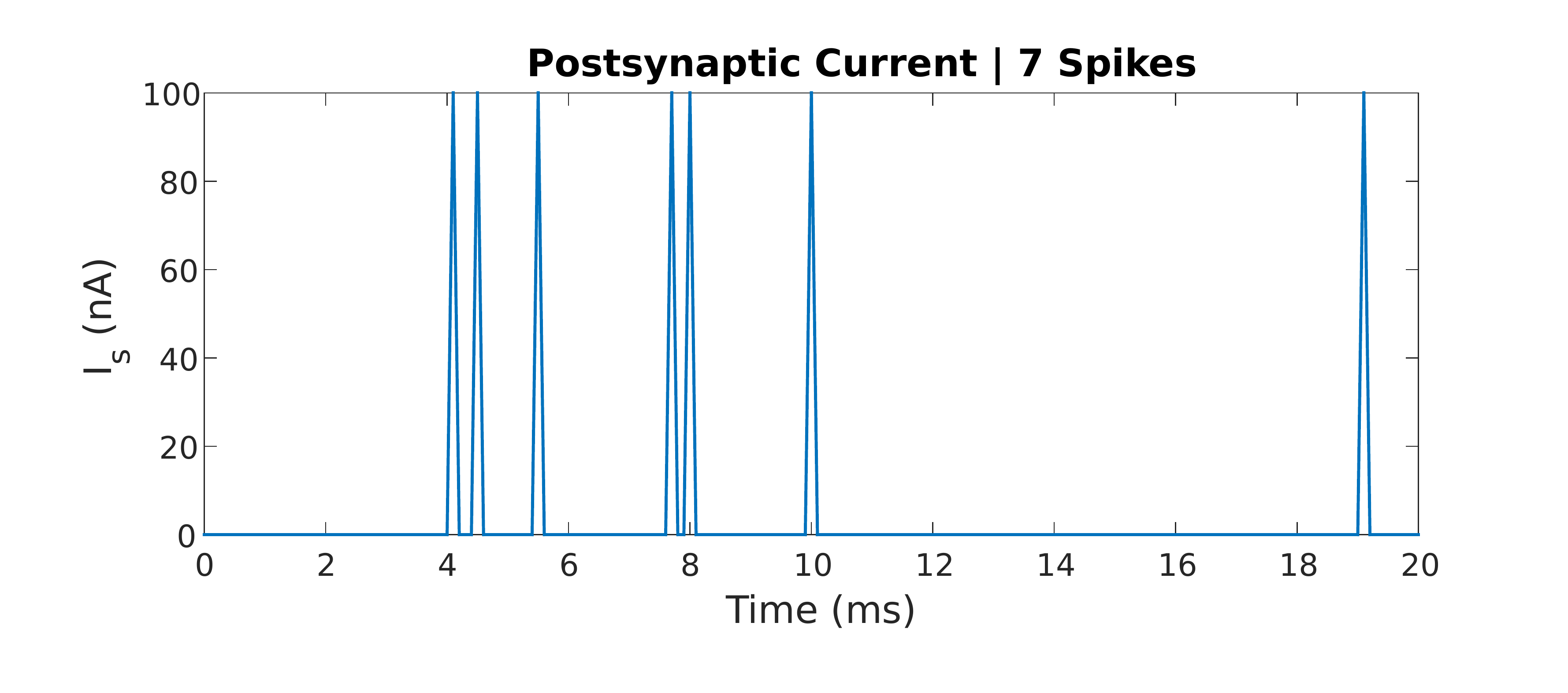}
		\end{subfigure}
		\quad
		\begin{subfigure}[t]{0.48\linewidth}
			\includegraphics[width=\linewidth]{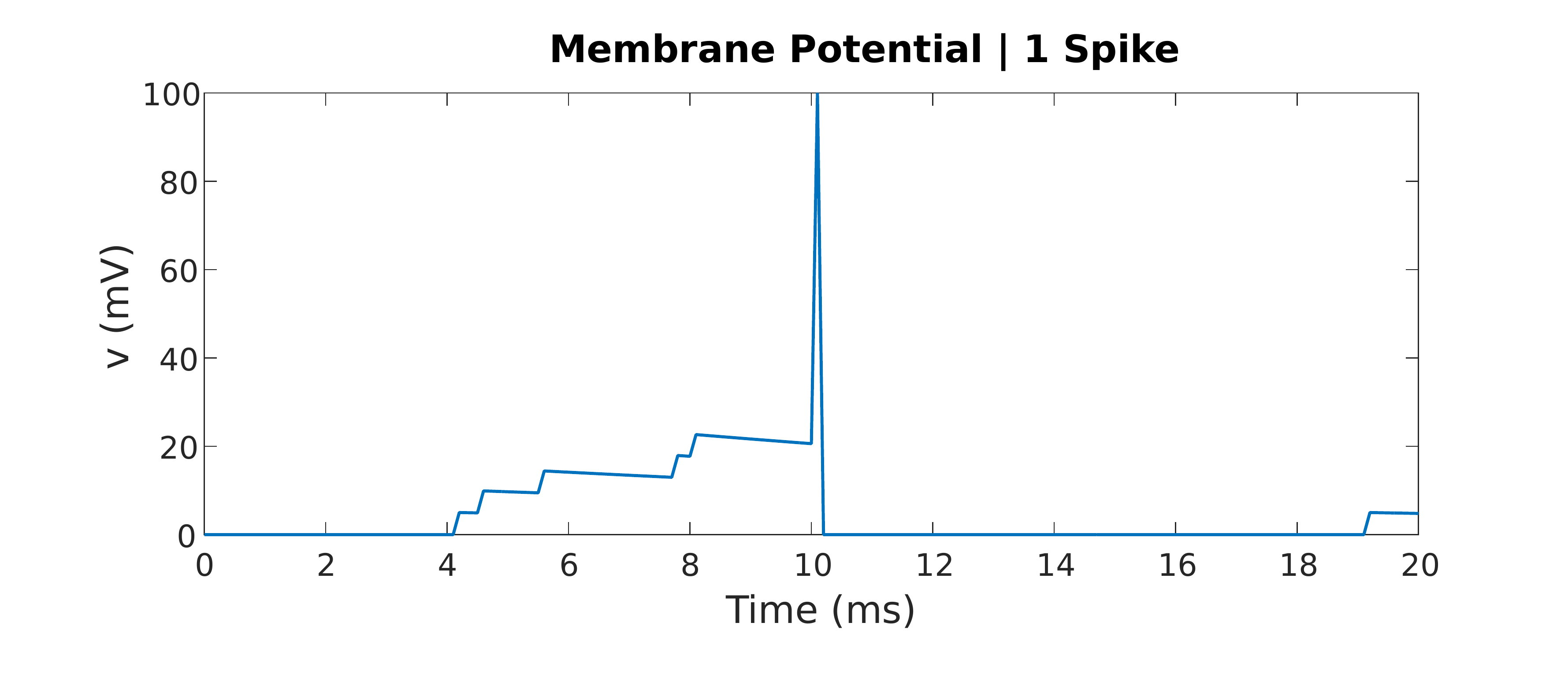}
		\end{subfigure}
	\end{subfigure}
	\quad
	\begin{subfigure}[t]{\linewidth}
		\centering
		\caption{}
		\label{sfig:corrente_entrada_12-1spikes_edsi}
		\begin{subfigure}[t]{0.48\linewidth}
			\includegraphics[width=\linewidth]{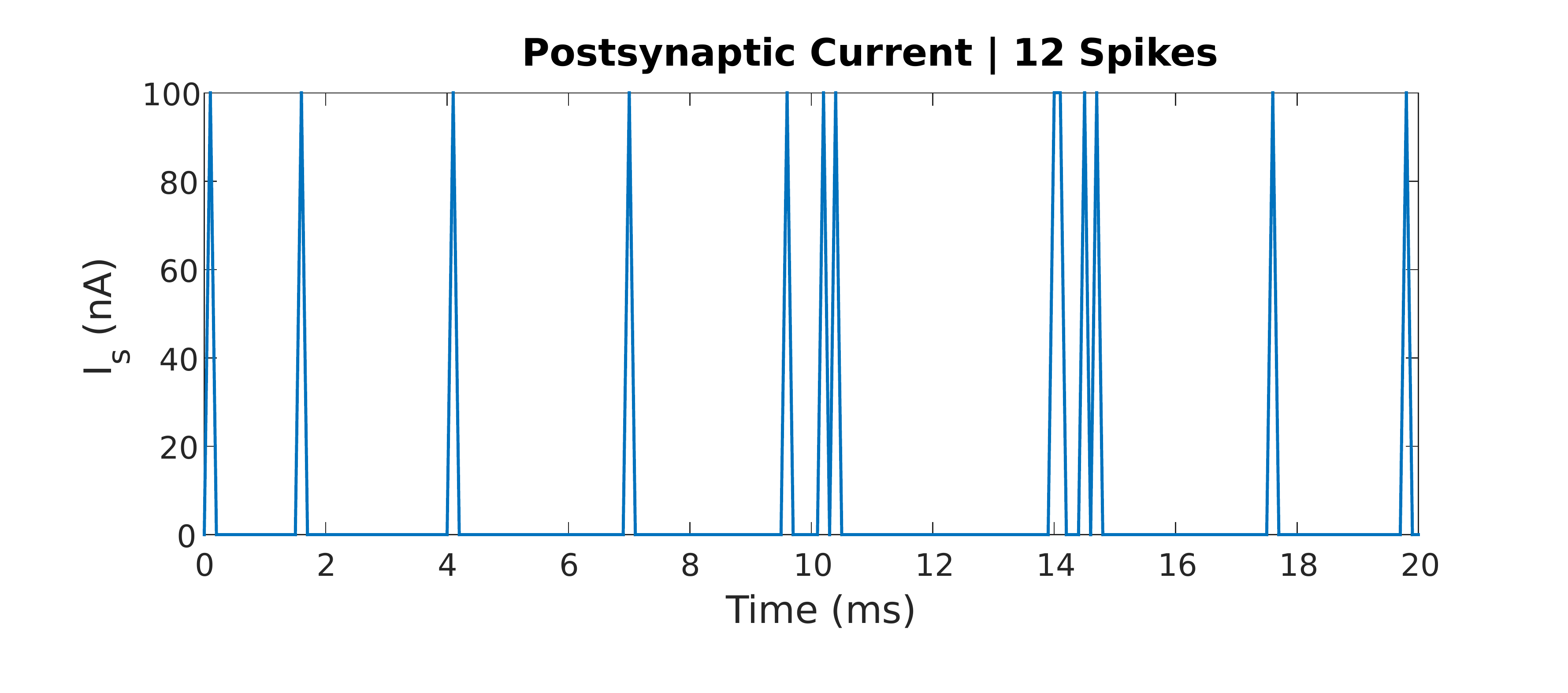}
		\end{subfigure}
		\quad
		\begin{subfigure}[t]{0.48\linewidth}
			\includegraphics[width=\linewidth]{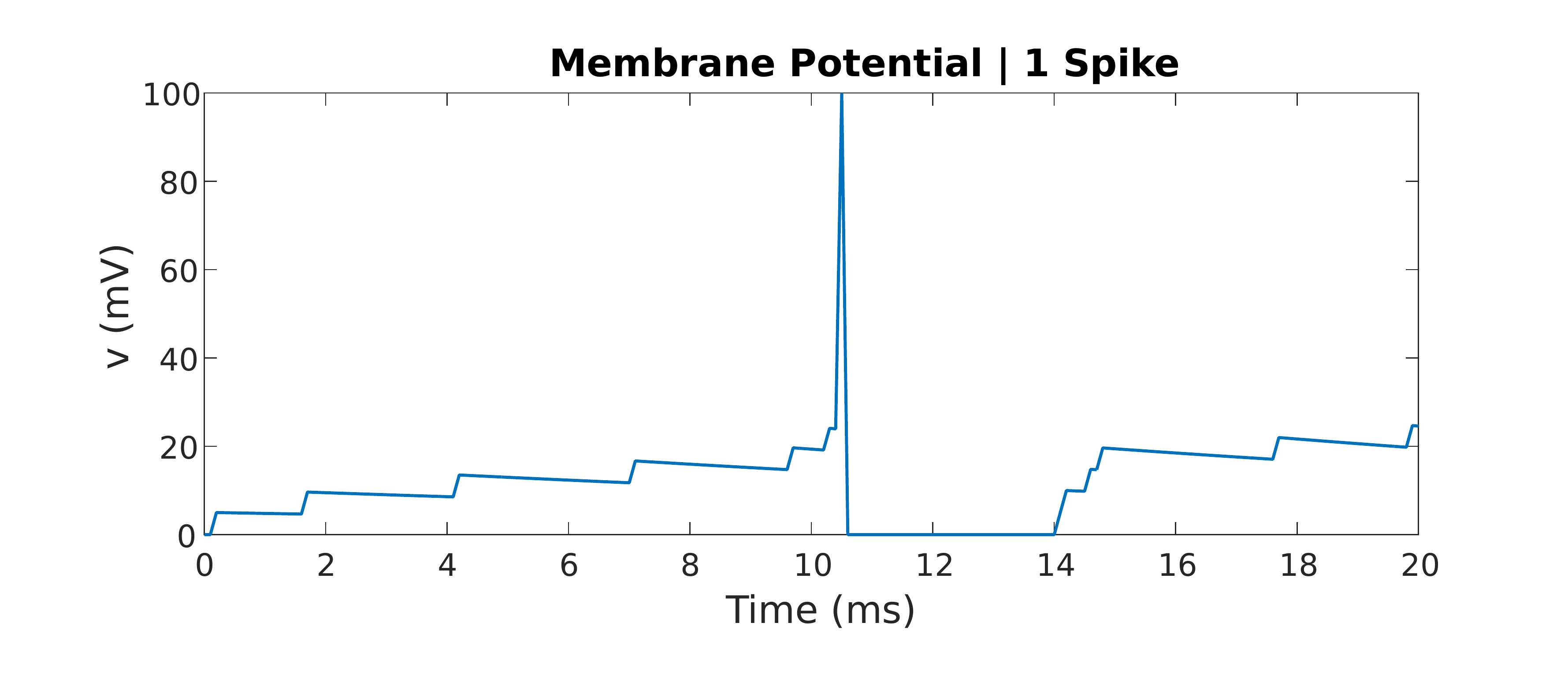}	
		\end{subfigure}	
	\end{subfigure}
	\caption{\label{fig:DifInpSameOut}Left: the presynaptic spike train. Right: the respective membrane potential variation. Considering a frequency window $\Delta t = 20ms$ both presynaptic spike train have different instantaneous frequencies at $t=0$, but generate an output with the same instantaneous frequencies.}
\end{figure*}

Observe that, in this context, it is not possible to construct a function between the input and output instantaneous frequencies. The examples in Figure \ref{fig:SameInpDiffOut} and \ref{fig:DifInpSameOut} illustrate this restriction. Considering a frequency window $\Delta t = 20ms$, the Figure \ref{fig:SameInpDiffOut} shows that two presynaptic spike trains with the same instantaneous frequencies (both have $13$ spikes in the $20ms$ interval) can generate different output instantaneous frequencies in the same interval. Contrariwise, Figure \ref{fig:DifInpSameOut} shows that two presynaptic spikes train with different instantaneous frequencies can generate equal output instantaneous frequencies. 

Despite the impossibility of constructing a function between the frequencies, it is still possible to explore characteristics of this relationship, especially its linearity, as will be developed in the following.

\section{Development}

At an instant $t$ the output instantaneous frequency of a leaky integrate-and-fire neuron depends of the input current not only at that moment but also at the moments prior of $t$. In the investigation of the instantaneous relations between the frequencies of the spikes, we do not have the complete information about the previous values of input current and, consequently, we have a non-deterministic relation between the instantaneous frequencies.

To analyze the dynamic frequency, transfer numerical simulations were used. The dynamic of the integrated-and-fire model was simulated using equation \ref{eq:sum_postsynaptic_potential}, at each simulation step the membrane potential value was calculated and compared to the spiking threshold, which was set at $25mV$, in order to define if a postsynaptic spike was generated or not. The simulations were performed with total execution time of $t_{tot}=10,000ms$ and a simulation time interval of $dt=0.1ms$, that is, every $0.1ms$ all variables were calculated and the conditions checked.


The presynaptic spike train that stimulates the neuron was generated through a homogeneous Poisson process \cite{cox2017theory,perkel1967neuronal,johnson1996point}, in which at each simulation step there is $p_S=0.05$ of chance that a presynaptic spike will be generated. Thus, the distribution of the interspike interval is an exponential one. This process was chosen because it generates a spike train with a broad spectrum of interspike intervals, while focusing on short intervals.

Under the previous assumptions and settings, the amplitude of the excitatory postsynaptic potential $\alpha_E$ and the membrane time constant $\tau_m$ are the principal parameters of the model, that can be altered for the study of its dynamic. 

\begin{figure*}[!htb]
	\centering
	\begin{subfigure}[t]{\linewidth}
		\centering
		\caption{}
		\label{sfig:demonstracao_linearidade_freqdin_alpha3}
		\begin{subfigure}[t]{0.48\linewidth}
			\includegraphics[width=\linewidth]{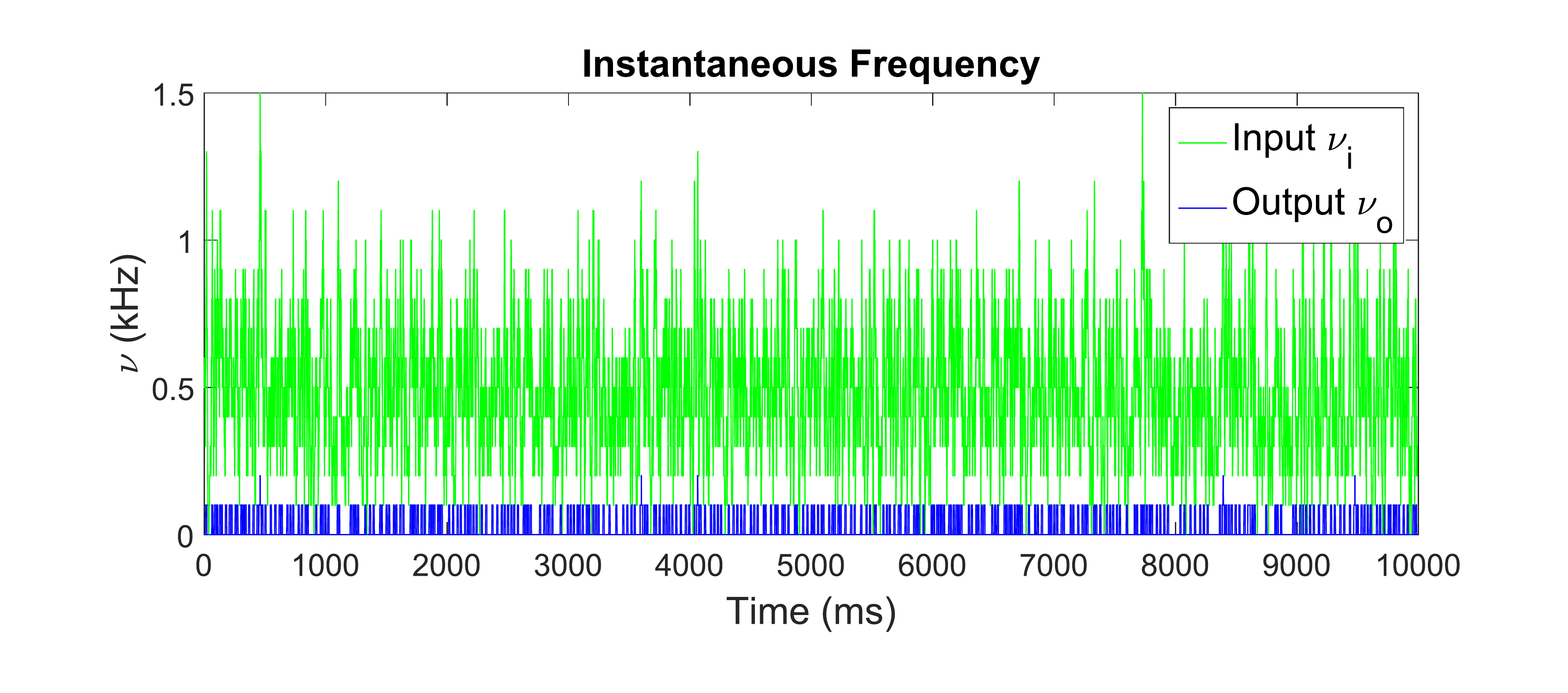}
		\end{subfigure}
		\quad
		\begin{subfigure}[t]{0.48\linewidth}
			\includegraphics[width=\linewidth]{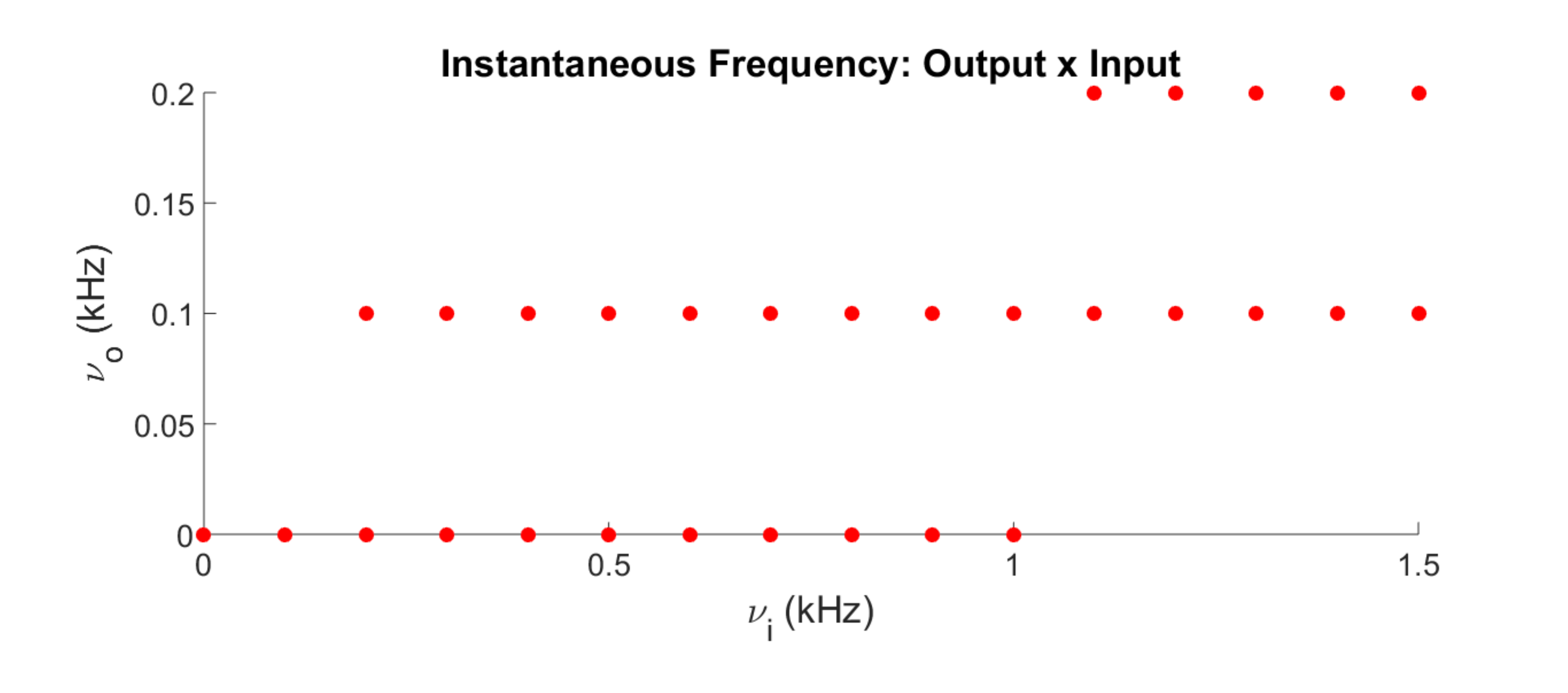}
		\end{subfigure}	
	\end{subfigure}
	\quad
	\begin{subfigure}[t]{\linewidth}
		\centering
		\caption{}
		\label{sfig:demonstracao_linearidade_freqdin_alpha5}
		\begin{subfigure}[t]{0.48\linewidth}
			\includegraphics[width=\linewidth]{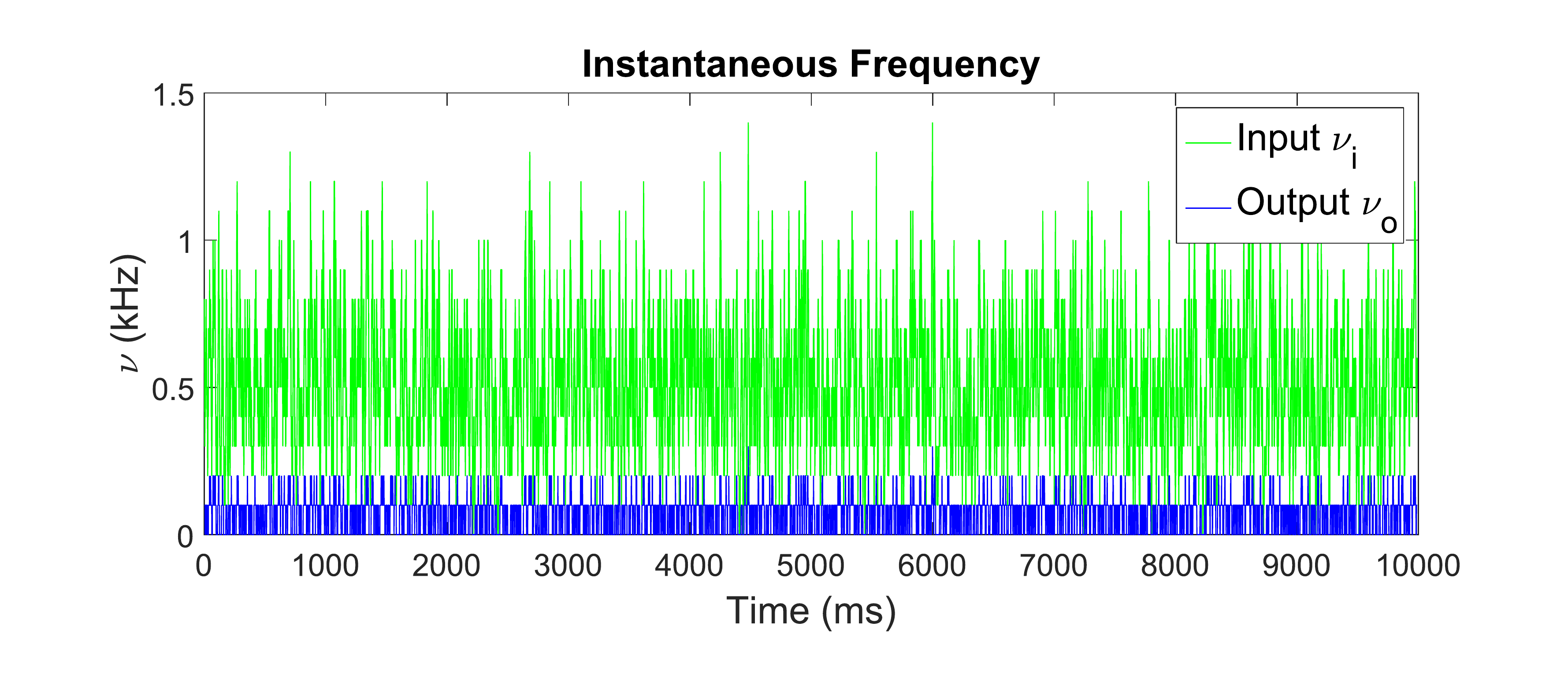}
		\end{subfigure}
		\quad
		\begin{subfigure}[t]{0.48\linewidth}
			\includegraphics[width=\linewidth]{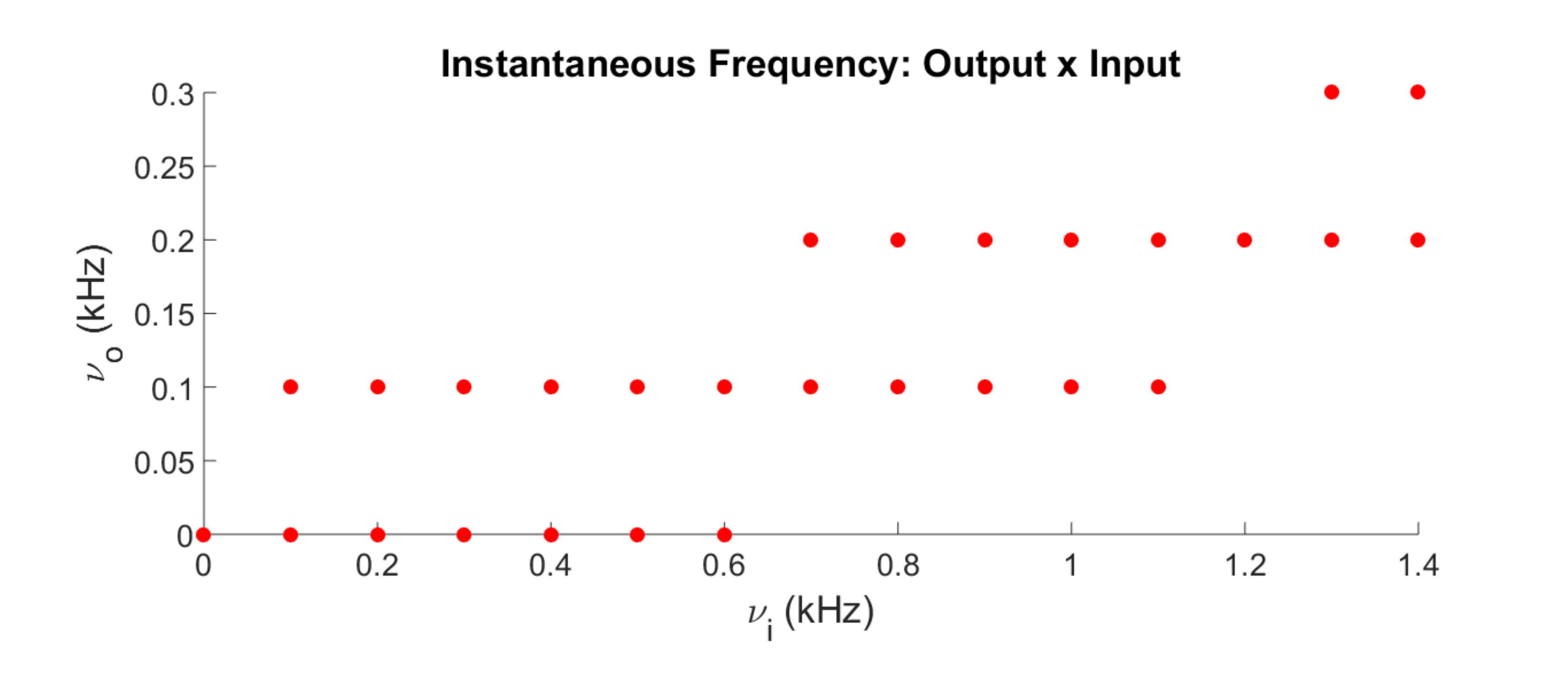}
		\end{subfigure}	
	\end{subfigure}
	\quad
	\begin{subfigure}[t]{\linewidth}
		\centering
		\caption{}
		\label{sfig:demonstracao_linearidade_freqdin_alpha10}
		\begin{subfigure}[t]{0.48\linewidth}
			\includegraphics[width=\linewidth]{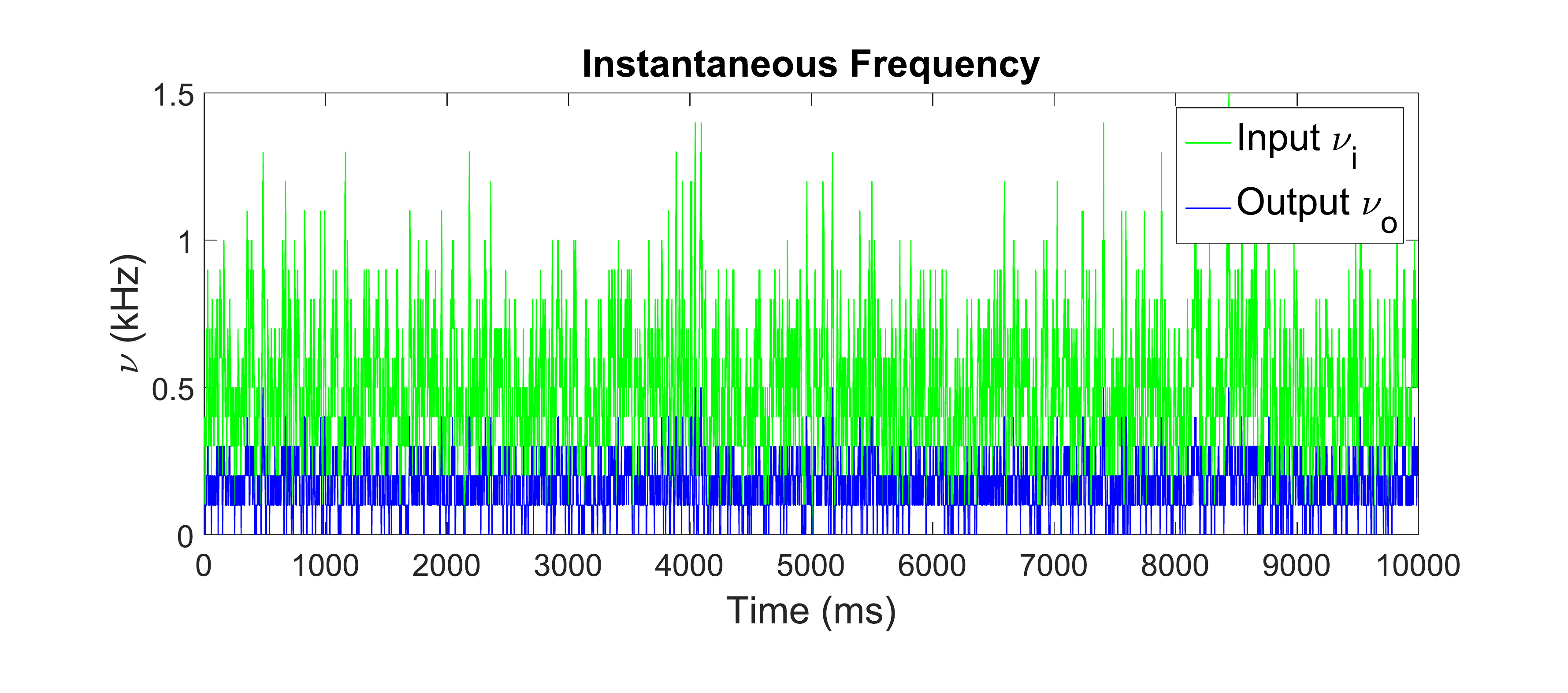}
		\end{subfigure}
		\quad
		\begin{subfigure}[t]{0.48\linewidth}
			\includegraphics[width=\linewidth]{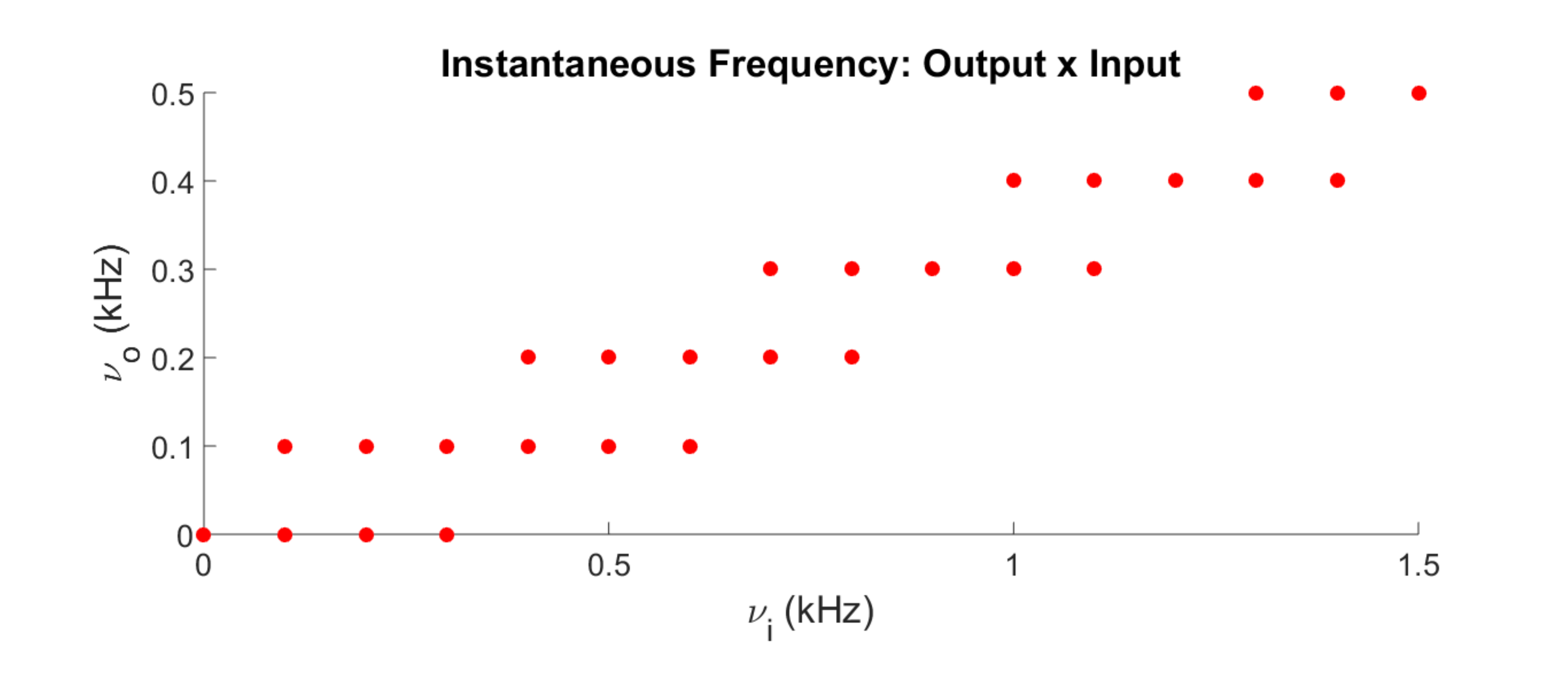}
		\end{subfigure}
	\end{subfigure}
	\quad
	\begin{subfigure}[t]{\linewidth}
		\centering
		\caption{}
		\label{sfig:demonstracao_linearidade_freqdin_alpha15}
		\begin{subfigure}[t]{0.48\linewidth}
			\includegraphics[width=\linewidth]{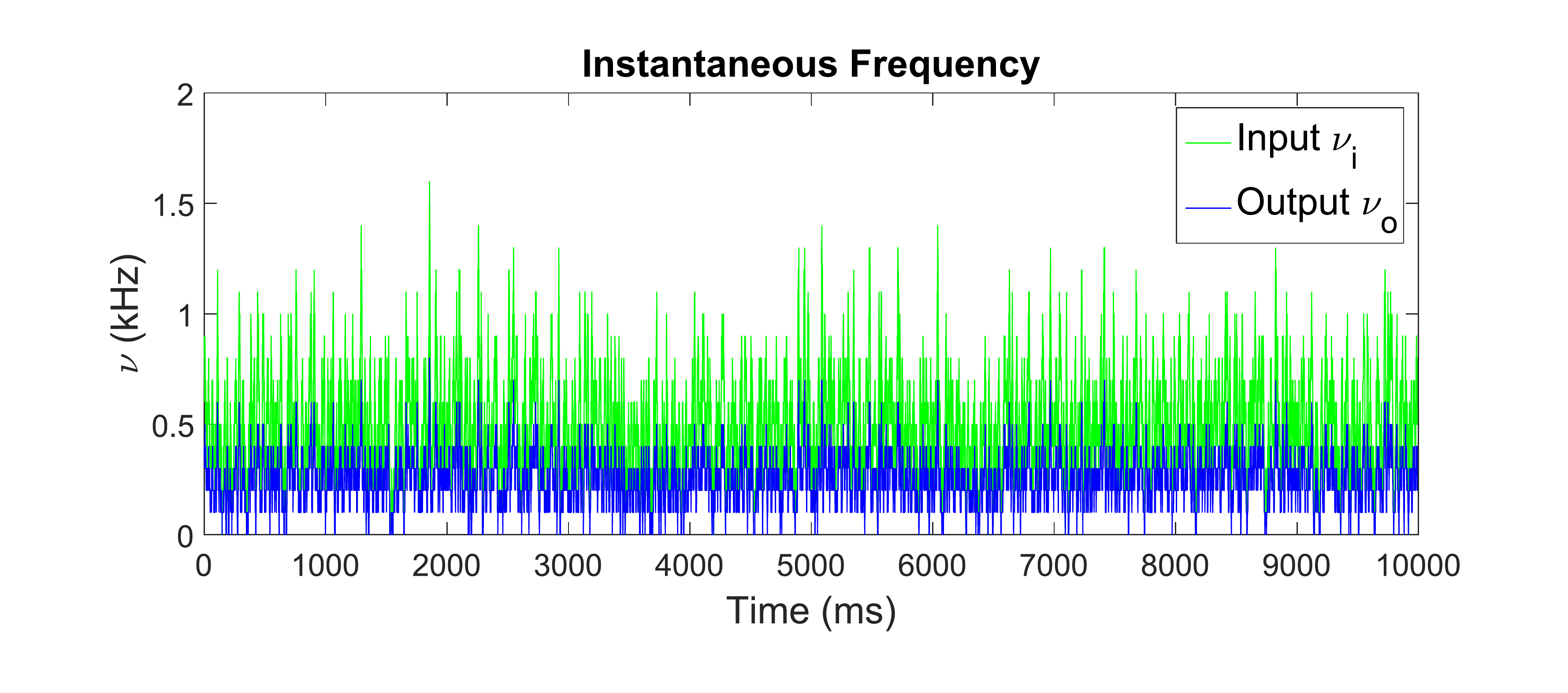}
		\end{subfigure}
		\quad
		\begin{subfigure}[t]{0.48\linewidth}
			\includegraphics[width=\linewidth]{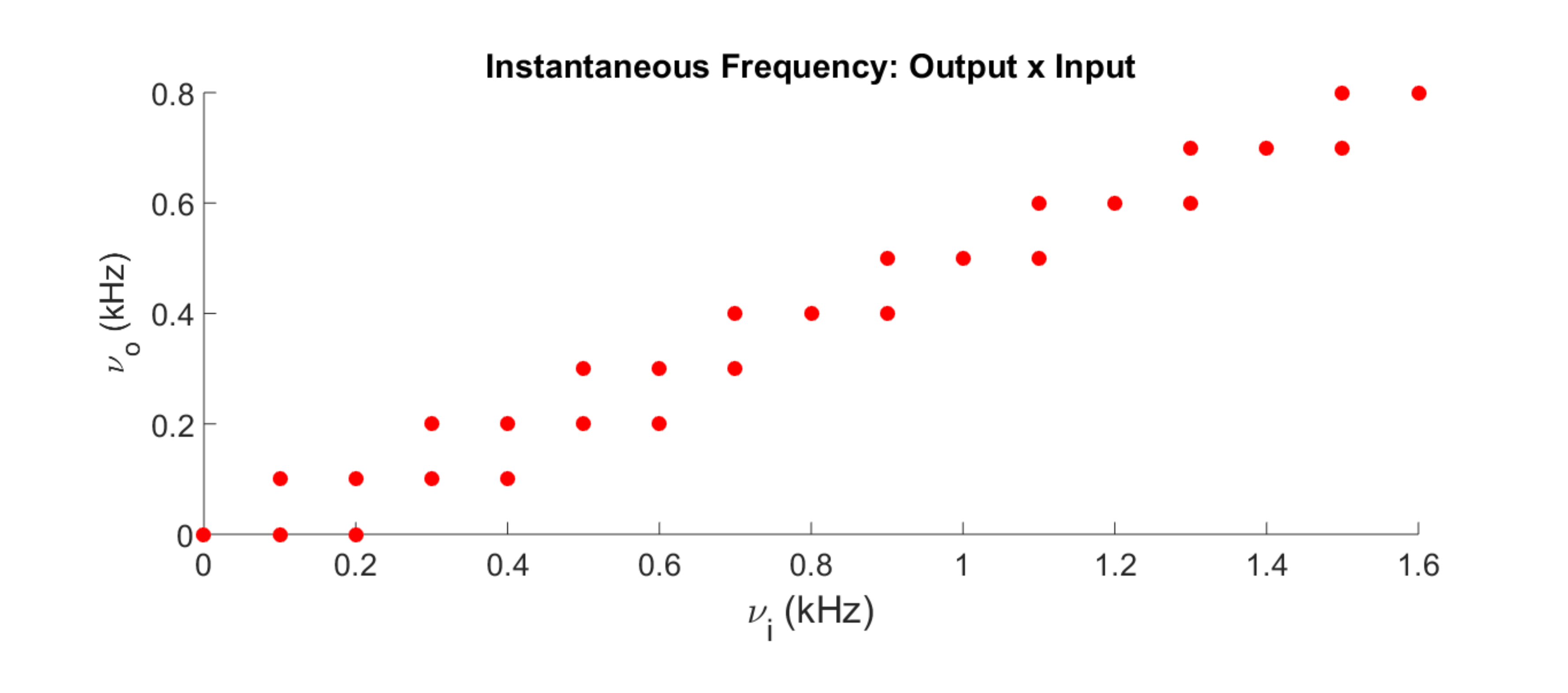}
		\end{subfigure}
	\end{subfigure}
	\caption{\label{fig:FreqTransfer}Left: plots of both instantaneous frequency input and output along time. Right: plots of the respective instantaneous frequency output in terms of the input at the same instant of time. The simulations adopted $\tau_m=20ms$, the window size of the frequency $\Delta t=10ms$, and the amplitude $\alpha_E$ varies in each subplot: a) $3mV$; b) $5mV$; c) $10mV$; d) $15mV$.}
\end{figure*}

Figure \ref{fig:FreqTransfer} presents the dynamic frequency transfer, the output instantaneous frequency in terms of the input, at the same instant of time, for several values of $\alpha_E$, that was obtained simulating the model in the circumstances described above. The possible values of the instantaneous frequency are obtained at discrete intervals, as a consequence of the instantaneous frequency mean window method (Equation \ref{instantfreq}), that takes values at $1/\Delta t$ intervals. 

In Figure \ref{fig:FreqTransfer}, it is possible to identify a linear relationship between the frequencies, which becomes more defined with the increase of the $\alpha_E$. This behavior makes sense if we analyze what is expected when $\alpha_E \ge \vartheta = 25$, in this case the system is completely determined as each presynaptic spike generates an postsynaptic spike, so the instantaneous frequency plot is a perfect line with slope $a=1$. 


Because the dynamic frequency transfer cannot be expressed in terms of a function, and there is no simple method for capturing all the information of the instantaneous frequency relationships, we resource to the strategy of characterizing the transfer according to its linearity. Among the various ways that can be used to characterize a multivalued relationship, linearity was chosen because the system has a natural tendency to linearize with the increment of its parameters. This characterization will occur by expressing the linearity of dynamic frequency transfer over a wide range of $\alpha_E $ and $\tau_m$ values, by delimiting the parameter space in relation to their linearity, and by expressing the parameters of the linear functions, which best approximates the transfer.

\begin{figure}[!htb]
	\centering
	\begin{subfigure}[t]{0.98\columnwidth}
		\centering
		\caption{}
		\label{sfig:study_window01}
		\includegraphics[width=\linewidth]{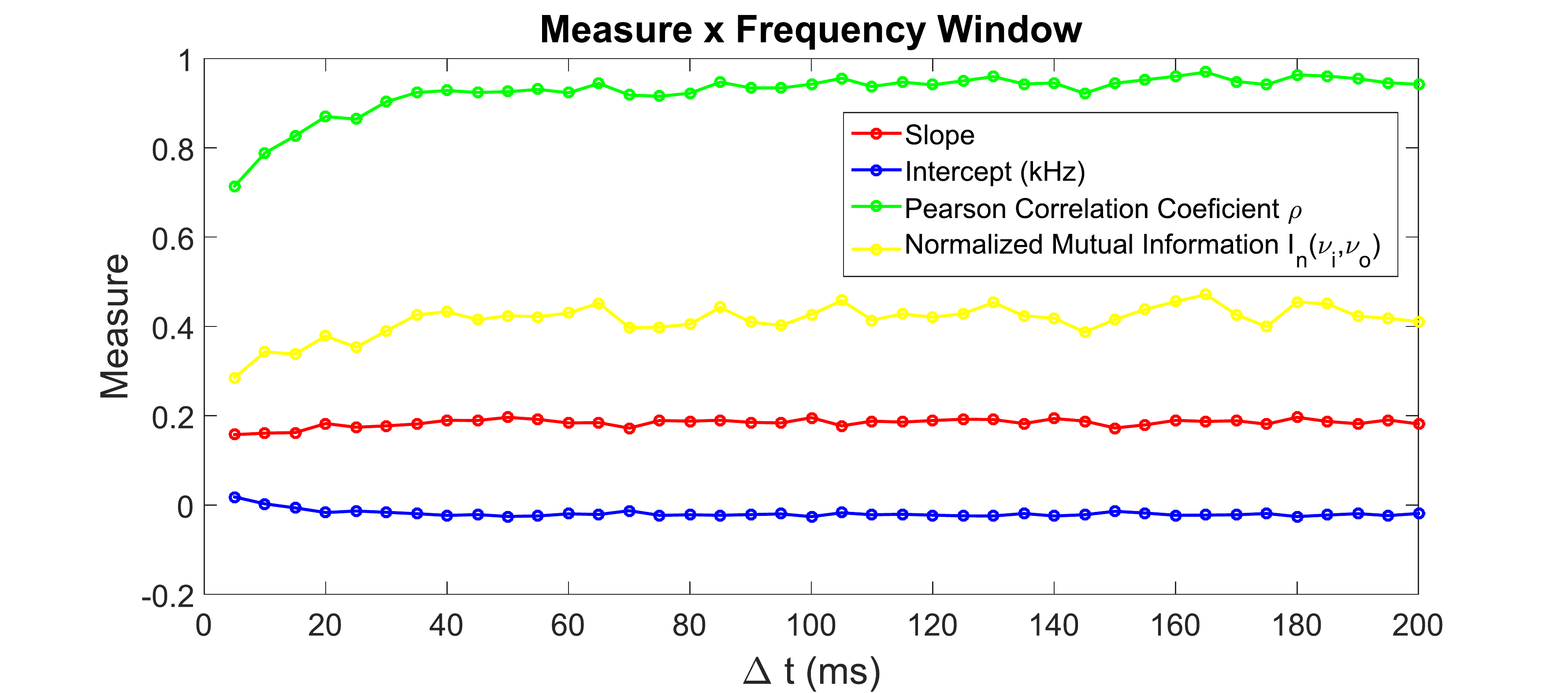}
	\end{subfigure}
	\quad
	\begin{subfigure}[t]{0.98\columnwidth}
		\centering
		\caption{}
		\label{sfig:study_window02}
		\includegraphics[width=\linewidth]{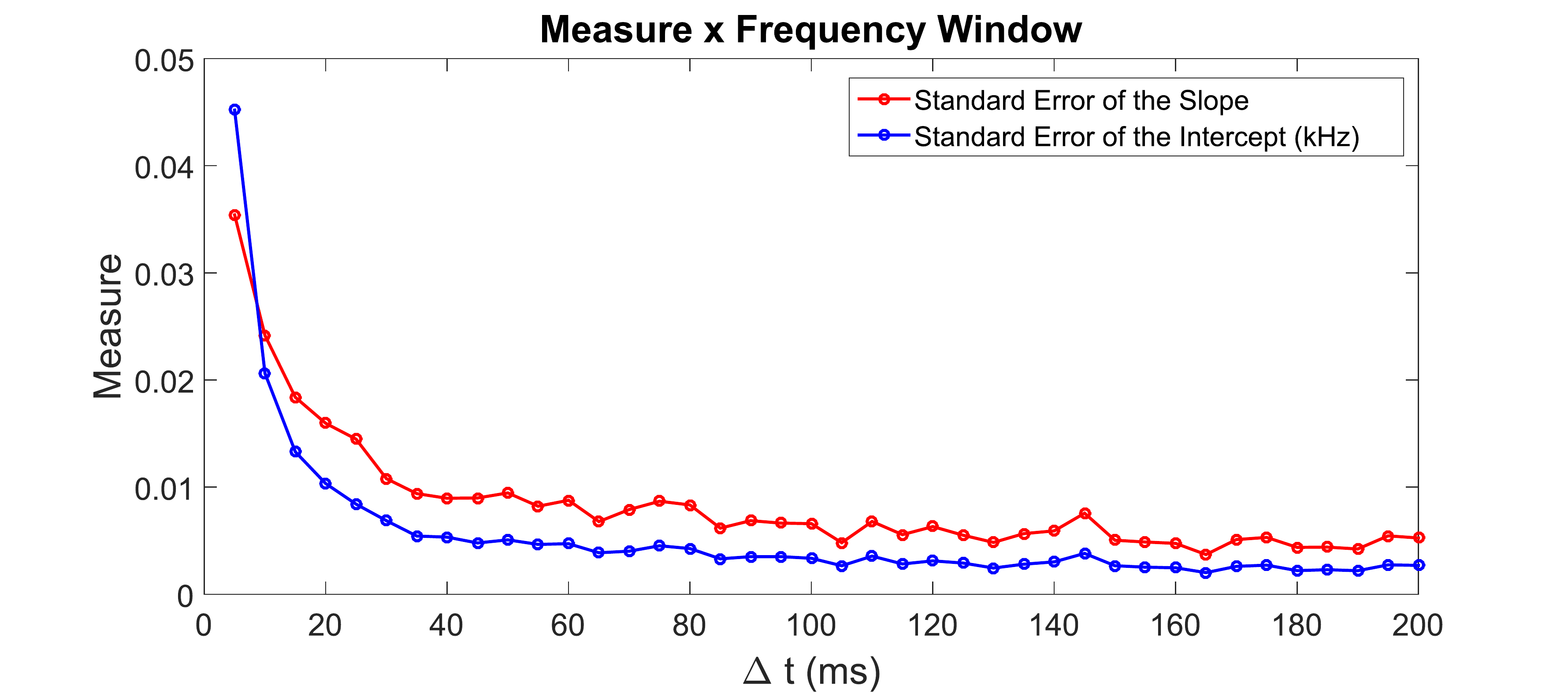}
	\end{subfigure}
	\begin{flushleft}
		\caption{\label{fig:study_window}Plots of various measurements obtained by dynamic frequency transfer in terms of the frequency window $\Delta t$. The dynamic frequency transfer used was generated by a neuron with the parameters $\alpha_E=5mV$ and $\tau_m=20ms$.}
	\end{flushleft}
\end{figure}

It is necessary to choose a frequency window $\Delta t$ in a stable region, that is, so that reasonable changes in the value of $\Delta t$ do not change the obtained results. The Figure \ref{fig:study_window} illustrates the value of different measurements obtained from the dynamic transfer function, which will receive more detail later in this text, in terms of the frequency window  $\Delta t$. Figure \ref{sfig:study_window01} shows that the values of $\Delta t$, between $5ms$ and $200ms$, do not significantly change the values obtained for the measurements.  However, Figure \ref{sfig:study_window02} shows that the dispersion of some measures decreases with the increase of $\Delta t$. As a result, $\Delta t = 80ms$ will be adopted as it is in a stable region with low dispersion.  

The way in which the instantaneous frequency has been defined implies it to assume only discrete values. After a simulation, the quantity of points found with a particular combination of input and output instantaneous frequency is based on the simulation total execution time and the probability density of the presynaptic interspike intervals, since it determines the amount of times an input pattern is repeated in an interval $\Delta t$. However, the possibility that of one of these points, with a determined combination of input and output instantaneous frequency, exist is determined only by the model dynamic and parameters.

It will be assumed that the execution time of the simulations, the probability density of the presynaptic spikes, the frequency window and the model parameters are sufficient to generate at least one result respective to most of the possible combination of input and output instantaneous frequency for small and medium frequencies values. Therefore, for a study of frequency transfer that does not depend on the probability density of the presynaptic interspike intervals, only one value of each combination of input and output instantaneous frequency needed to be considered.

\subsection{Correlation}

As discussed, the relationship between the input and output instantaneous frequencies at the same instant of time has a linear shape that depends on the parameters of the neuron (Figure \ref{fig:FreqTransfer}). A neuron for which the relationship of the instantaneous frequencies is a straight line acts as a linear frequency attenuator.

In order to determine when an integrate-and-fire neuron can be approximated by a linear frequency attenuator, and how well this approximation can be, it is necessary to measure the linearity of the instantaneous frequencies relationship of the neuron. For this, we used the Pearson's correlation as a measurement of how linear is the relationship, high values of Pearson's correlation mean that the system can be well represented by a linear function \cite{benesty2009pearson}. 

\begin{figure}[!htb]
	\centering
	\begin{subfigure}[t]{\columnwidth}
		\includegraphics[width=\linewidth]{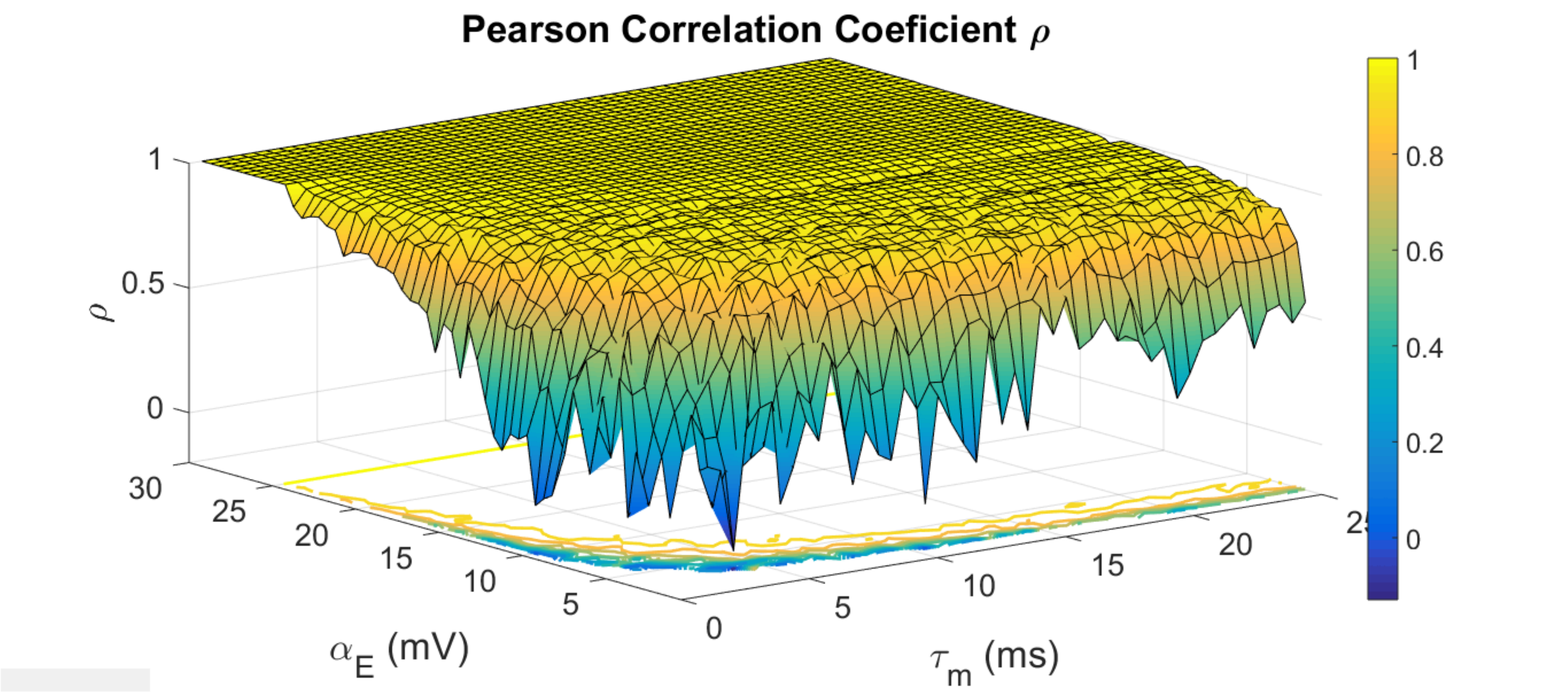}
	\end{subfigure}
	\quad
	\begin{subfigure}[t]{\columnwidth}
		\includegraphics[width=\linewidth]{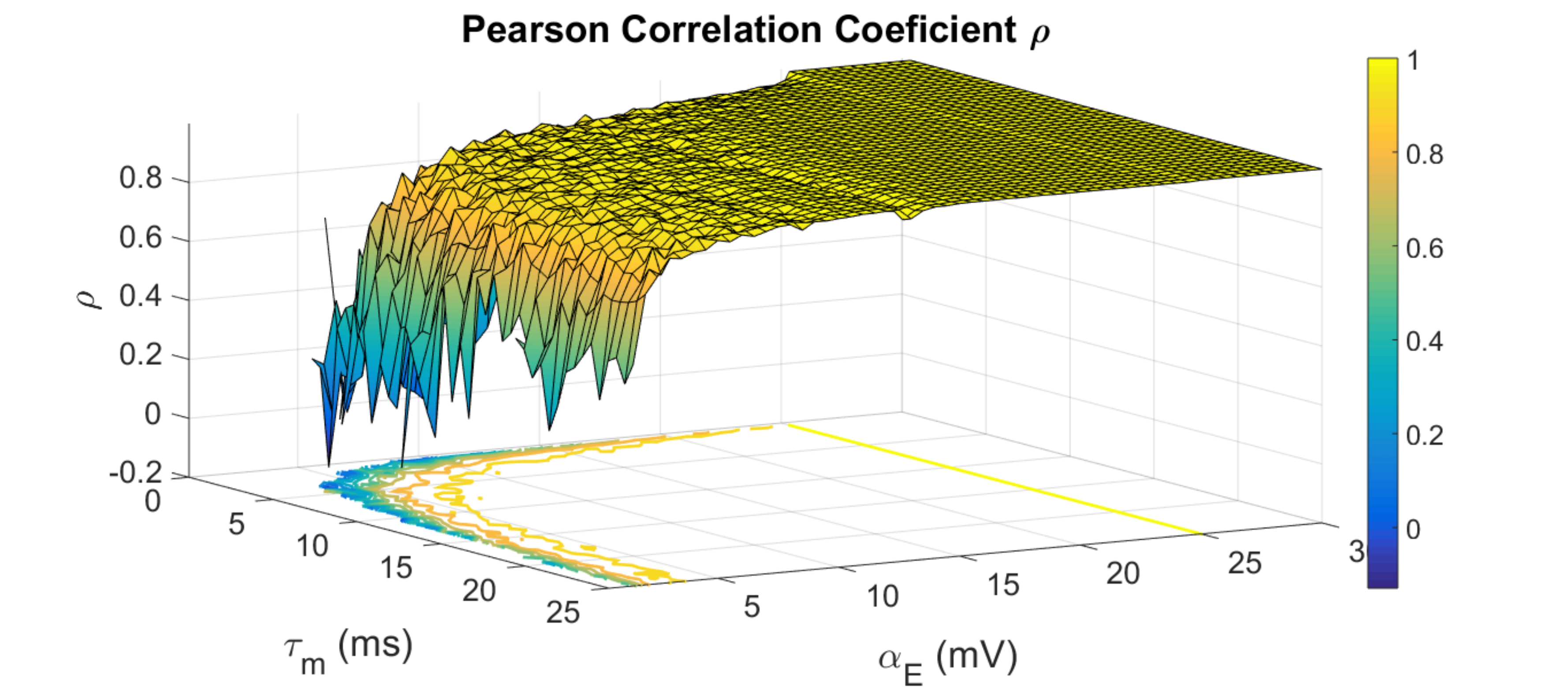}
	\end{subfigure}
	\caption{\label{fig:PearsonCorrelation}Plot of the Pearson's correlation by $\alpha_E$ and $\tau_m$ in two different angles. It was used the window size of the frequency $\Delta t=80ms$.}
\end{figure}

The plot of the Pearson's correlation coefficient in terms of $\alpha_E$ and $\tau_m$, can be seen in Figure \ref{fig:PearsonCorrelation}. The plot shows an increasing behavior of the correlation with $\alpha_E$ and $\tau_m$. This is expected because the higher the amplitude $\alpha_E$, the closer the neuron is to the saturation, which has a linear characteristic, and the higher the membrane time constant $\tau_m$, the closer the neuron is to a perfect integrator, without leak.

An important characteristic of the Person correlation is that it is a measurement of the quality of a least squares linear regression. In this context the Person correlation is called coefficient of correlation $R$ and its square is named coefficient of determination $R^2$ and both values are often used as measures of quality, each with its advantages, of a least squares linear regression \cite{nagelkerke1991note}.

It is useful to relate regions of the space parameters with the linearity of the dynamic frequency transfer, and this can be done by taking into account the fact that the Pearson's coefficient increases with $\alpha_E$ and $\tau_m$. Using a fixed value of the Pearson's coefficient, it is possible to fit the points with lowest correlation values that are greater than or equal to the fixed value in order to construct a curve of $\alpha_E$ in terms of $\tau_m$. Due to the increasing behavior of the Pearson's coefficient, it is possible to infer that for values of $\alpha_E$ and $\tau_m$ above this curve the Pearson's coefficient will result above the initially fixed value.

\begin{figure}[!htb]
	\centering
	\includegraphics[width=\columnwidth]{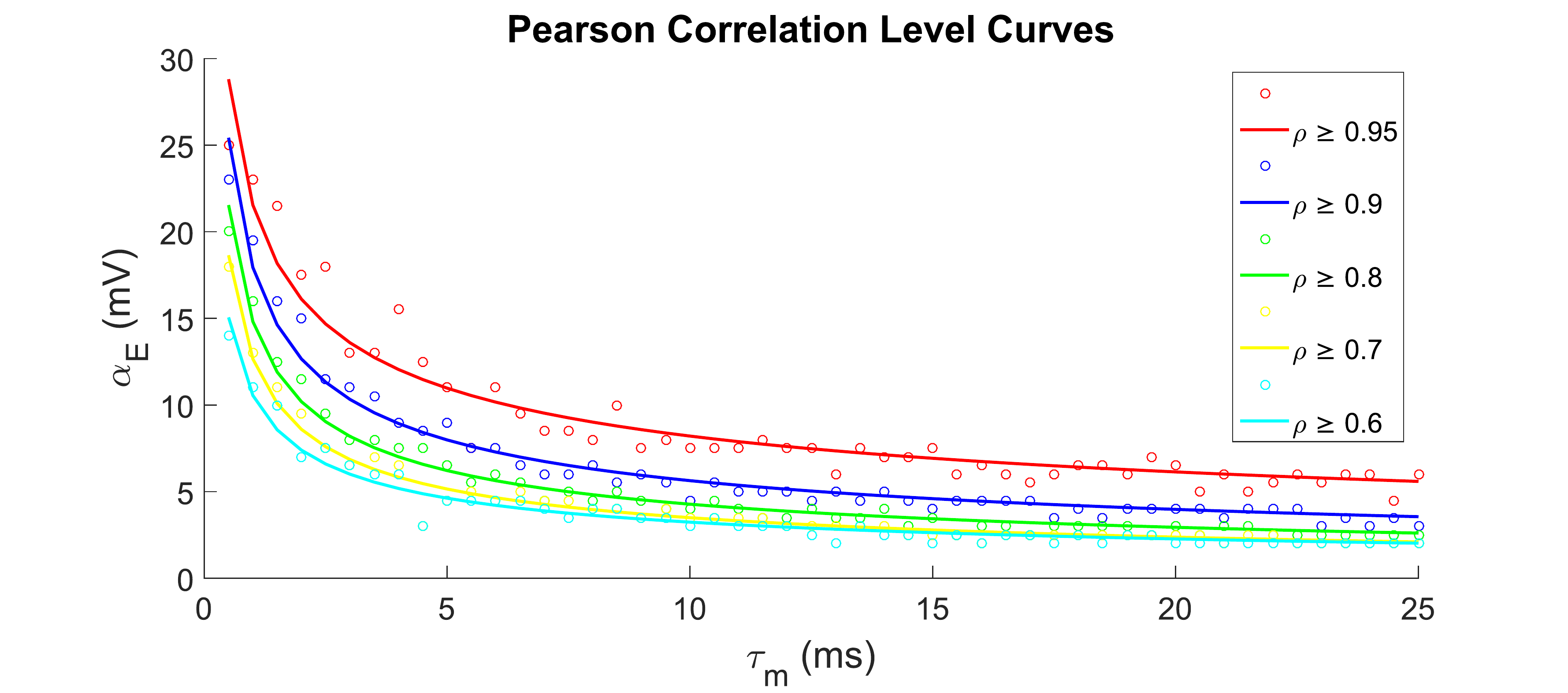}
	\caption{\label{fig:PearsonCorrelationCurves}Plot of the fitted curves of the Table \ref{tab:PearsonCorrelation}. Each curve delimits the plane ($\alpha_E$, $\tau_m$) according with the value of the Pearson's coefficient of the same color indicated in the legend.}
\end{figure}

For this curve fit, we adopted the least squares method with the model function:

\begin{equation}\label{eq:pearson_model_function}
	y=\varphi_1 x^{-\varphi_2},
\end{equation}

\noindent where $\varphi_1$ and $\varphi_2$ are the adjust parameters \cite{bjorck1990least}. The results are showed in Table \ref{tab:PearsonCorrelation} and the curves are showed in Figure \ref{fig:PearsonCorrelationCurves}.

\begin{table}[!htb]
	\centering
	\begin{tabular}{@{}cccc@{}}
		\toprule
		Pearson's Coefficient	& $\varphi_1$ & $\varphi_2$ & RMSE (mV)\\
		\midrule
		$\rho \ge 0.95$ 		& $21.5\pm1.1$	& $0.41\pm0.03$	& $1.29$\\
		$\rho \ge 0.9$ 			& $17.9\pm0.6$	& $0.50\pm0.02$	& $0.71$\\
		$\rho \ge 0.8$ 			& $14.8\pm0.4$	& $0.54\pm0.02$	& $0.50$\\
		$\rho \ge 0.7$ 			& $12.7\pm0.4$	& $0.56\pm0.02$	& $0.50$\\
		$\rho \ge 0.6$ 			& $10.5\pm0.4$	& $0.51\pm0.03$	& $0.51$\\
		\bottomrule
	\end{tabular}
    \begin{flushleft}
		\caption{\label{tab:PearsonCorrelation} Table presenting the parameters of the curves that delimit the regions in the plane of $\alpha_E$ and $\tau_m$, followed by their respective confidence intervals of $95\%$ based on Student's t-distribution, and lastly the root-mean-square deviation (RMSE) of each curve fit. For values of $\alpha_E$ and $\tau_m$ higher than one of the curves, the Pearson's correlation of the transfer is higher than the fixed value.}
	\end{flushleft}
\end{table}

Therefore, the linearity of dynamic frequency transfer can be quantified in terms of the model parameters $\alpha_E$ and $\tau_m$, yielding a method for dividing the parameter space into regions according to their linearity.

\subsection{Linear Regression}

The relationship between the frequencies is a scatter of points with a linear behavior for some model parameters. To analyze the parameters of the linear function that better approximate the dynamic frequency transfer,  the least squares method with the model function $y=\epsilon_1 x + \epsilon_2$ was employed to fit the set of points of the instantaneous frequencies relationship (Figure \ref{fig:FreqTransfer}).  The values $\epsilon_1$ and $\epsilon_2$ are the adjust coefficients,

The parameters of the fitted straight line in terms of the model parameters $\alpha_E$ and $\tau_m$ is illustrated in Figure \ref{fig:ABforEPSPTAU}.  The quality of the adjust can be evaluated by the measurements presented in Appendix B. It is possible to verify, for the slope $\epsilon_1$, that some well delimited regions are obtained, which have almost constant value.

\begin{figure*}[!htb]
	\centering
	\begin{subfigure}[t]{0.48\linewidth}
		\centering
		\caption{}
		\label{sfig:coeficiente_angular_lif}
		\begin{subfigure}[t]{\linewidth}
			\includegraphics[width=\linewidth]{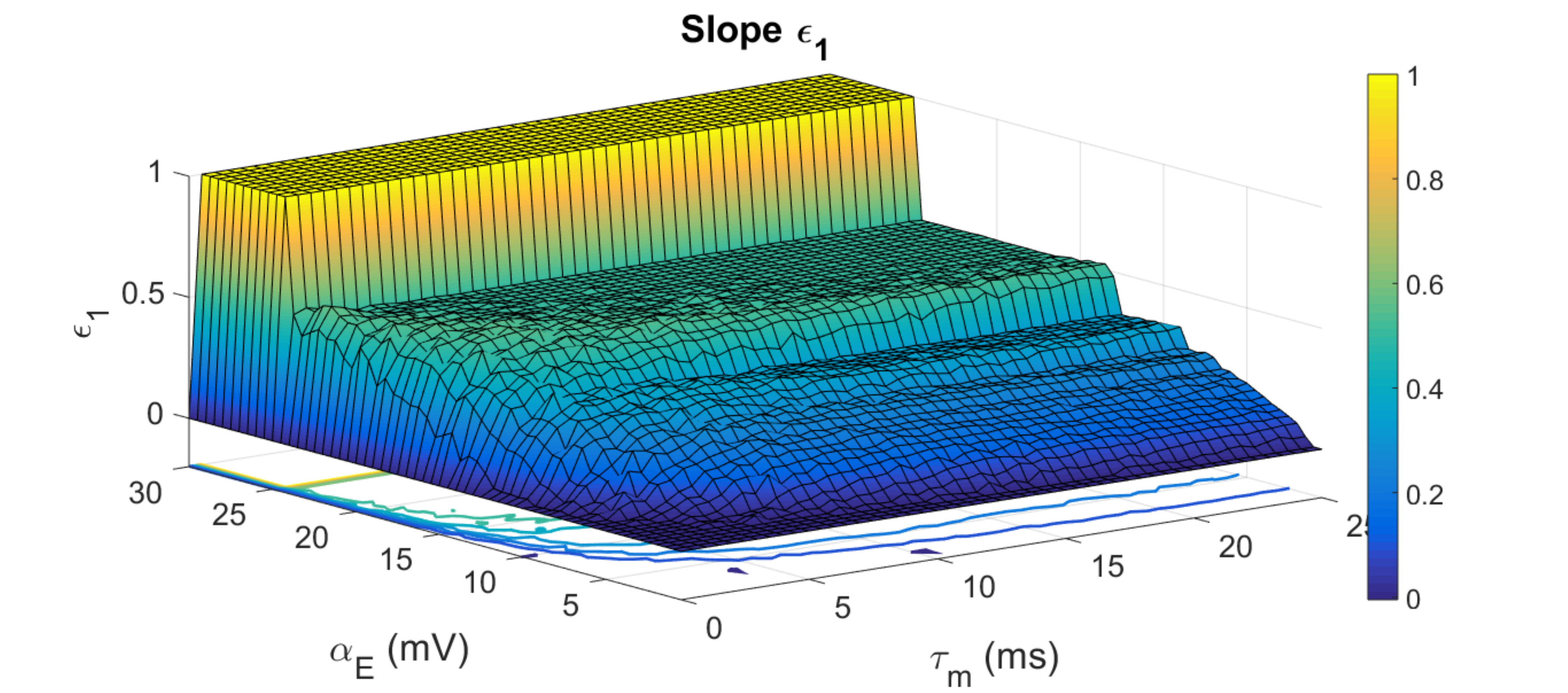}
		\end{subfigure}
		\quad
		\begin{subfigure}[t]{\linewidth}
			\includegraphics[width=\linewidth]{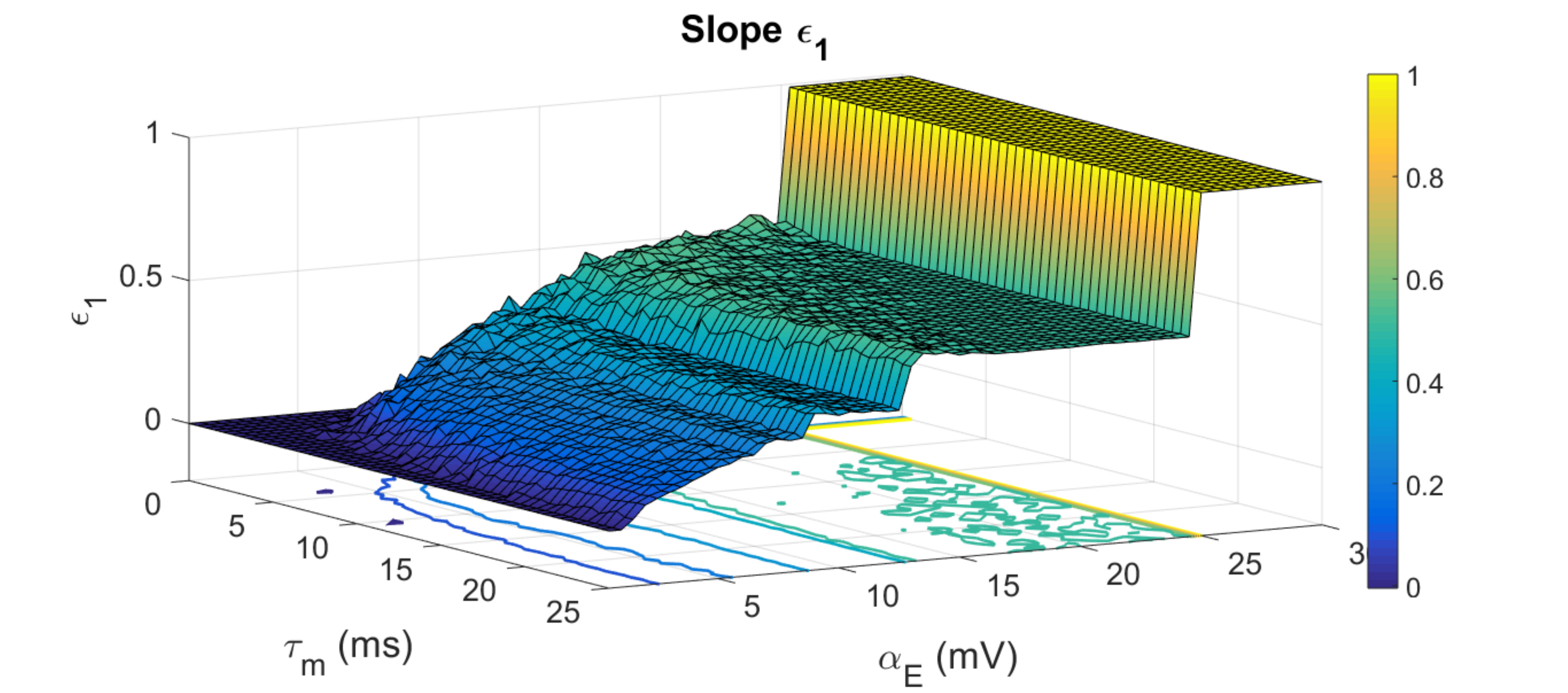}
		\end{subfigure}
		\quad
		\begin{subfigure}[t]{\linewidth}
			\includegraphics[width=\linewidth]{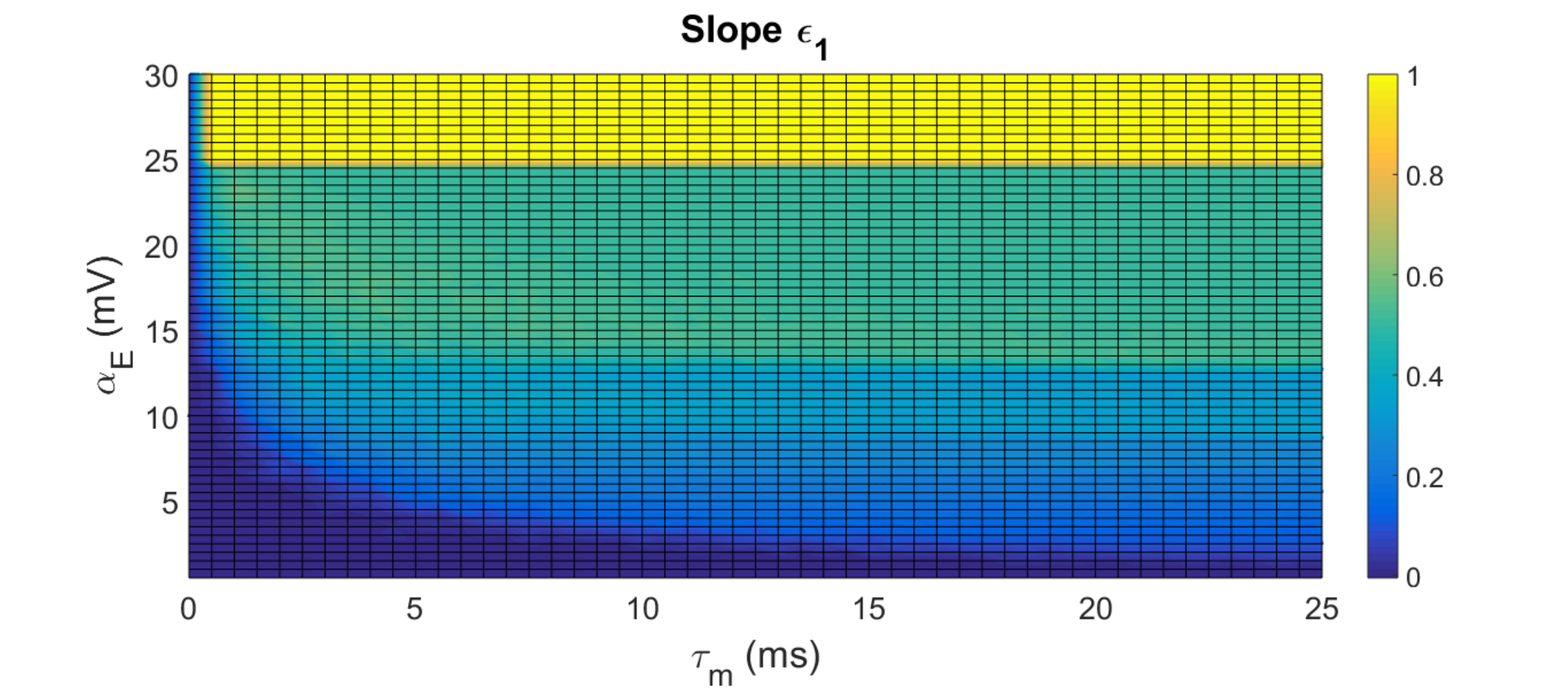}
		\end{subfigure}
	\end{subfigure}
	\quad
	\begin{subfigure}[t]{0.48\linewidth}
		\centering
		\caption{}
		\label{sfig:coeficiente_linear_lif}
		\begin{subfigure}[t]{\linewidth}
			\includegraphics[width=\linewidth]{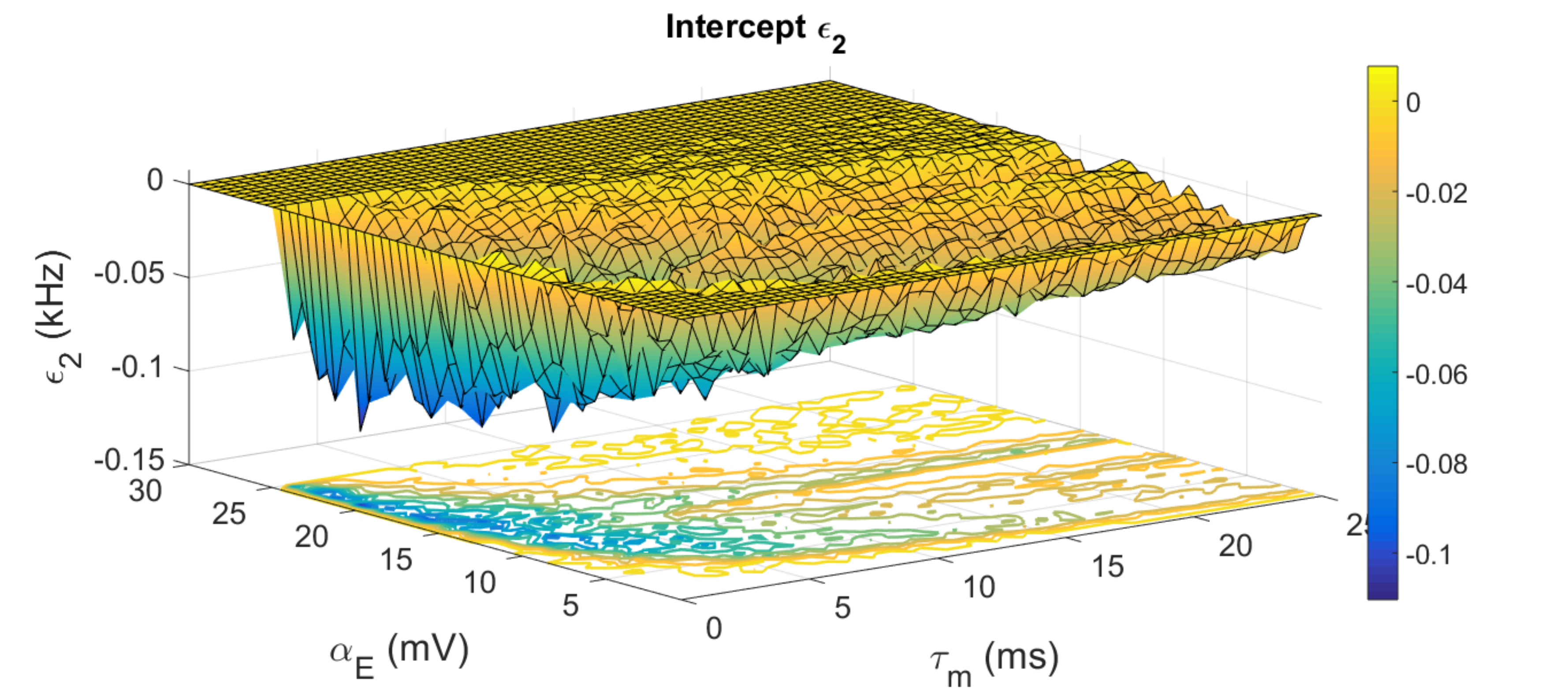}
		\end{subfigure}
		\quad
		\begin{subfigure}[t]{\linewidth}
			\includegraphics[width=\linewidth]{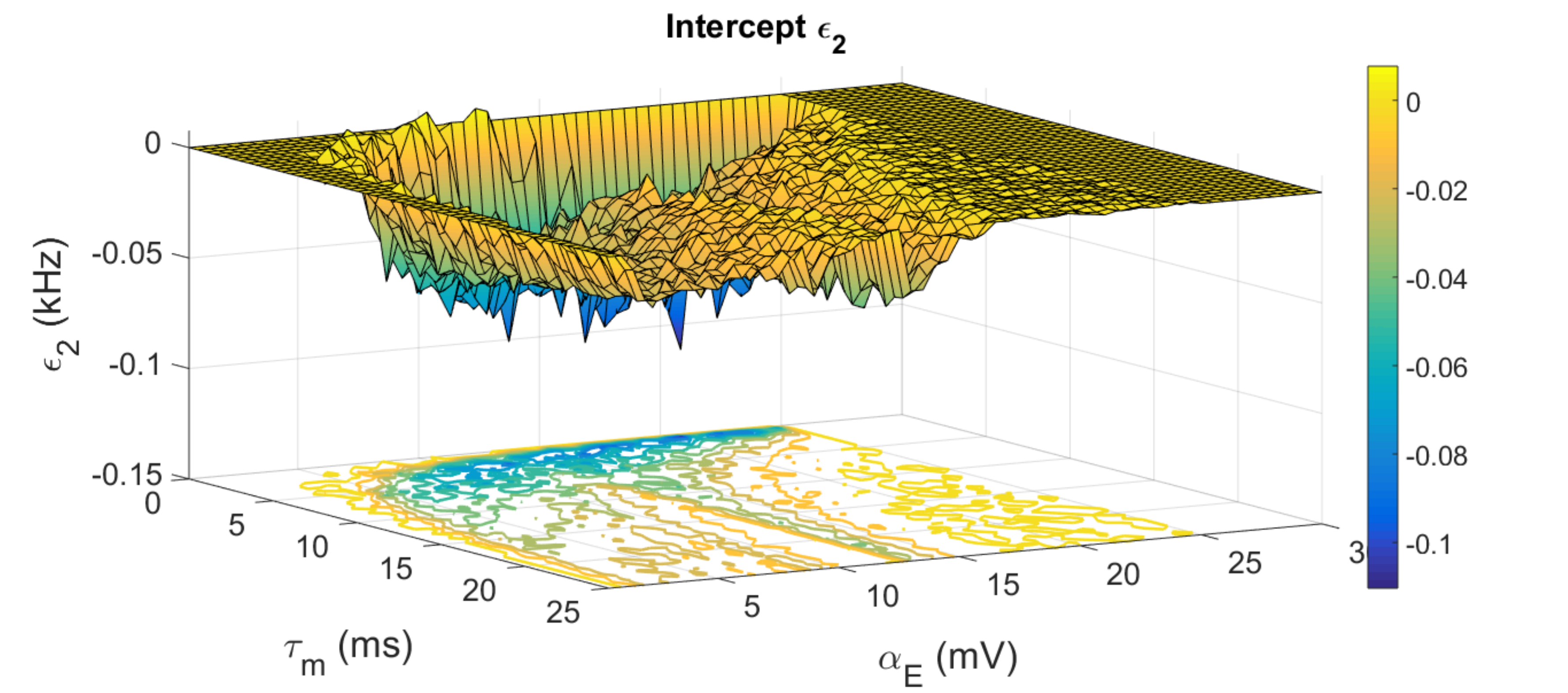}
		\end{subfigure}
		\quad
		\begin{subfigure}[t]{\linewidth}
			\includegraphics[width=\linewidth]{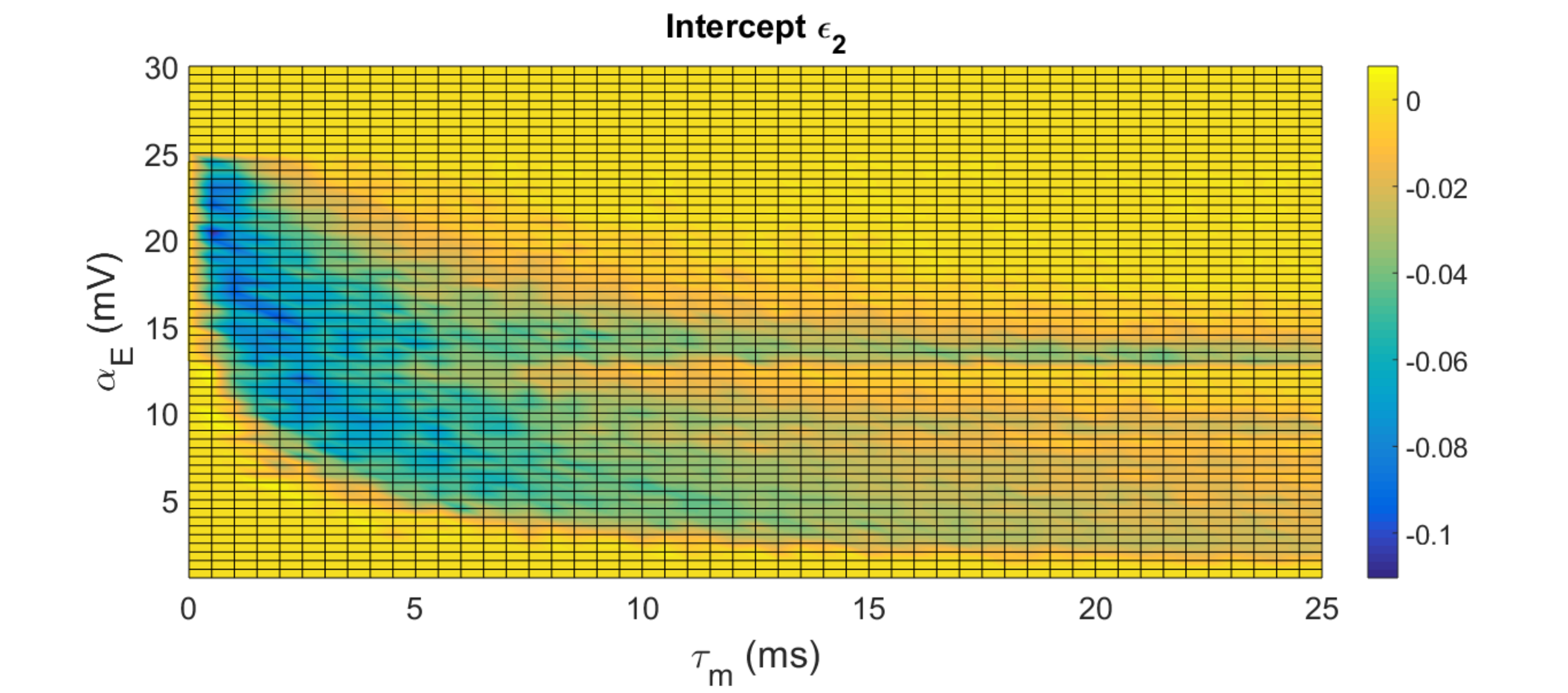}
		\end{subfigure}
	\end{subfigure}
	\caption{\label{fig:ABforEPSPTAU}a) Plot of the fitted slope $\epsilon_1$ in terms of $\alpha_E$ and $\tau_m$ in three different angles. b) Plot of the fitted intercept $\epsilon_2$ in terms of $\alpha_E$ and $\tau_m$ in three different angles. A frequency window $\Delta t=80ms$ has been adopted.}
\end{figure*}

One characteristic that can be extracted from the Figures \ref{fig:PearsonCorrelation} and \ref{fig:ABforEPSPTAU} is that the measurements vary with $\tau_m$ only for small values of $\tau_m$ and $\alpha_E$, remaining practically constant for greater values. That occurs because, given a fixed value for the amplitude $\alpha_E$ and the interspike intervals of presynaptic spikes, values of $\tau_m$ greater than a certain range do not provide the necessary leakage to change the pattern of action potentials generated by the neuron.

\begin{figure}[!htb]
	\centering
	\begin{subfigure}[t]{0.98\columnwidth}
		\centering
		\caption{}
		\label{sfig:Slope_t20}
		\includegraphics[width=\linewidth]{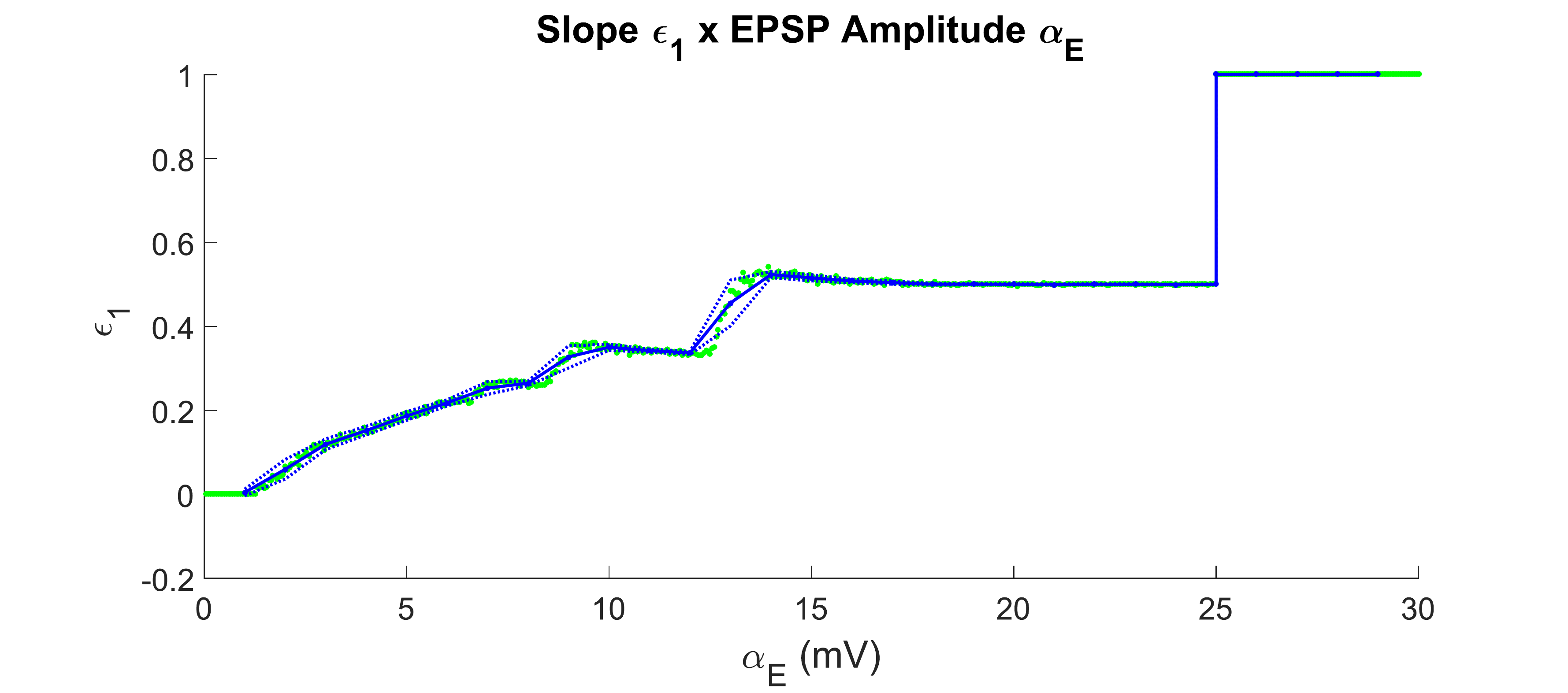}
	\end{subfigure}
	\quad
	\begin{subfigure}[t]{0.98\columnwidth}
		\centering
		\caption{}
		\label{sfig:Intercept_t20}
		\includegraphics[width=\linewidth]{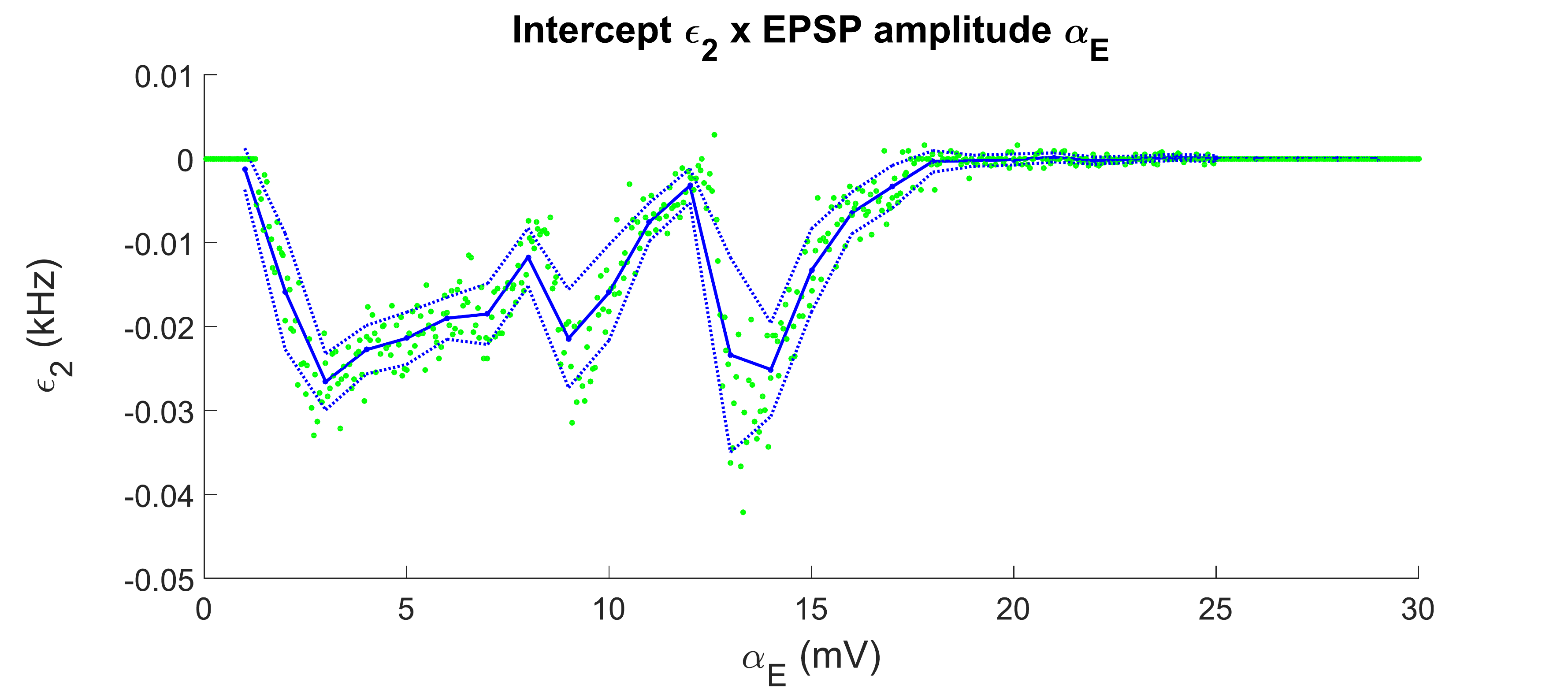}
	\end{subfigure}
	\begin{flushleft}
		\caption{\label{fig:SlopeIntercept_t20}a) Fitted slope $\epsilon_1$ of the dynamic frequency transfer in terms of $\alpha_E$.  b) Fitted intercept $\epsilon_2$ of the dynamic frequency transfer in terms of $\alpha_E$. In both plots, $\tau=20ms$. The $\alpha_E$ values varies from $0.5mV$ to $30mV$ in intervals of $0.05mV$ (green squares). The blue points define the line and are given by the mean of the squares in a $0.5mV$ interval of $\alpha_E$. The dot line is the standard error.}
	\end{flushleft}
\end{figure}

Based on the Pearson's correlation, the model does not have a frequency transfer with a linear behavior for small values of $\tau_m$ and $\alpha_E$, and therefore this region is not suitable for the linear regression. In some experiments that try to quantify the leakage of a biological neuron, founded some values of $\tau_m$ greater than $10ms$ \cite{rauch2003neocortical,lansky2006parameters}. All this motivated us to analyze the model only for $\tau_m=20ms$ (Figure \ref{fig:SlopeIntercept_t20}), i.e.~in the region where the frequency transfer exhibits some remarkable linearity and the behavior in terms of $\tau_m$ is almost constant.

Analyzing Figure \ref{sfig:Slope_t20}, it is possible to see that the slope $\epsilon_1$ has a small dispersion, and for $\alpha_E$ values above about $9mV$ it has a strong plateaus-based behavior. This occurs because the quantity of presynaptic spikes that generate an action potential is a particularly important factor to the frequency transfer function. In this way, the plateaus tend to emerge in the values of $\alpha_E$ next to $\vartheta/N$, where $N$ is the number of presynaptic spikes necessary to generate an action potential.  In the considered case the threshold is fixed at $25mV$, so the plateaus emerge principally at $25mV$, $12.5mV$, $8.3mV$ and $6.25mV$. 

As the number of spikes $N$ increases, transitions between plateaus become smaller and smoother. For small values of $\alpha_E$m, many spikes are required to generate an action potential, and in this case, because the leakage have more time to act on the membrane potential, the interspike interval becomes a more important factor in the occurrence of a postsynaptic spike, which smooth the plateaus behavior.

Based on Figure \ref{sfig:Intercept_t20}, the intercept $\epsilon_2$ has a large dispersion and varies greatly with $\alpha_E$. Their values are usually equal or smaller than zero.  This is because it is possible that presynaptic spikes do not generate an output, but postsynaptic spikes cannot occur without inputs. In addition, the values of $\epsilon_2$ are extremely small. Since $\Delta t=80ms$ was used, possible instantaneous frequency values are given at intervals of $1/  \Delta t = 0.0125ms$, in the same range as the values assumed for $\epsilon_2$. Therefore, this indicates that the intercept can be disregarded from the linear model function that fits the dynamic frequency transfer, and it can be simply described by $y=\epsilon_1 x$.

\section{Conclusions}

In the last years, there has been a great increase in the interest and application of mathematical models that describe biological neurons.  These models are often used to investigate certain characteristics of the nervous system through analytical methods, and more often numerical simulations. This is generally a consequence of the success of this approach under certain circumstances, and the large increase of the computational processing capacity \cite{gerstner2002spiking,gerstner2014neuronal,pospischil2008minimal,paun2010oxford}.

In the current work, the transfer function formalism has been applied considering the instantaneous frequencies of a single neuron. Widely used in areas such as electronic engineering, the transfer function allows the understanding of a system in terms of a graphic-functional relationship between its input and output. The leaky integrate-and-fire neuron model was adopted because of its simplicity and the fact that it have been used systematically in several previous works, often obtaining descriptions compatible with those observed in biological systems \cite{burkitt2006review,izhikevich2003simple}.

The dynamic frequency transfer, which is the relation between the instantaneous input and output frequencies at the same instant of time, was obtained through systematic numerical simulations, for several parameters of the system. In order to obtain this relationship, it was necessary to choose the window value for the calculation of the instantaneous frequency. It was found that $\Delta t = 80ms$ is a stable value in the adopted circumstances. Because the integrate-and-fire system has memory, this relationship is not deterministic and, therefore, it is possible to have one input instantaneous frequency associated to more than one output instantaneous frequency.  However, the possible values of this transfer relationship still exhibit substantial linearity that varies with the model parameters.

To quantify this linearity, we calculated the Pearson's coefficient of the relationship for different values of the model parameters: the amplitude of the excitatory postsynaptic potential $\alpha_E$ and the membrane time constant $\tau_m$ (Figure \ref{fig:PearsonCorrelation}). Based on this information, we then applied the least squares method to relate regions of the parameter space with the linearity of the transfer relationship (Table \ref{tab:PearsonCorrelation} and Figure \ref{fig:PearsonCorrelationCurves}). In this way, it became possible to determine the quality of the instantaneous frequencies transfer approximation by a straight line, which corresponds to the situation that the neuron acts as a linear frequency attenuator.

We also determined the parameters of the aforementioned straight line, which better approximate the relationship, in terms of the model parameters (Figure \ref{fig:ABforEPSPTAU}). From these data, it was possible to show that the intercept of the fitted line is usually less than or equal to zero, and has very small values with large dispersion, thus not being very relevant to the model. It was also shown that the slope has a nearly constant value in different well-delimited regions. Moreover, we showed that in the cases in which the neuron can be well approximated as a linear frequency attenuator, ie.~the instantaneous frequency transfer relationship can be well approximately by a straight line, the membrane time constant ($\tau_m$) becomes not too relevant to the frequency transfer.

This methodology can be used in simulations of neuronal experiments in the context of instantaneous frequencies under similar circumstances as those adopted.  Then, by knowing the parameters of the model, one can infer the linearity of the transfer relationship and the error associated with it. Thus, depending on the parameter values and the error one is willing to accept, the instantaneous output frequency can be more efficiently obtained directly from the input frequency without the need to simulate the entire neuronal dynamics. In addition, the frequency transfer can be used for model validation, serving as a comparison measure between the mathematical model and the biological neuron.

As future developments, it would be interesting to analyze how the behavior of the frequencies transfer in the cases in which the neuron has more than one presynaptic input with different amplitudes $\alpha_E$ and instantaneous frequencies. It would also be interesting to study the transfer relationship for other neuron models, or between neurons in a network.

\section*{Data Availability}
The data sets generated during and or analyzed during the current work are available at \url{https://figshare.com/articles/Data_Availability_-_Numerical_Frequency_Transfer_Function_Analysis_of_a_Leaky_Integrate-and-Fire_Neuron/9955535}.


\section*{Acknowledgments}

Felipe Lucas Gewers acknowledges CAPES for sponsorship. Luciano da F. Costa thanks CNPq (grant no. 307333/2013-2). This work has been supported also by FAPESP grants 11/50761-2 and 5/22308-2. 

Research carried out using the computational resources of the Center for Mathematical Sciences Applied to Industry (CeMEAI) funded by FAPESP (grant 2013/07375-0).

\section*{Appendix A - Stationary Frequency Transfer Function}
\label{s:stationary_frequency_transfer_function}

The stationary frequency transfer function relates the frequency of the presynaptic spike train with the frequency of the generated action potentials, after a sufficient adaptation period, in the case of the presynaptic spike train has constant interspike interval.


Consider the neuron initially at rest and a sequence of presynaptic spikes with interspike interval ($t_{isi}=\gamma$), a number $N$ of presynaptic spikes will be necessary to generate one postsynaptic spike.  Based on Equation \ref{eq:sum_postsynaptic_potential}, the number $N$ is defined as the smallest value that obeys the relationship:

\begin{equation}\label{eq:sum_stat_transfer}
	\sum_{i=0}^{N-1} a_E\exp{-\frac{i\gamma}{\tau_m}}\ge\vartheta,
\end{equation}

\noindent eliminating the sum we obtain:

\begin{equation}\label{eq:no_sum_stat_transfer}
	\frac{\alpha_E \exp \left(\frac{\gamma-\gamma N}{\tau_m}\right) \left[ \exp \left(\frac{\gamma N}{\tau_m}\right) - 1 \right]}{\exp \left(\frac{\gamma}{\tau_m}\right) - 1} \ge \vartheta.
\end{equation}

In the limit of $N \to \infty$ the left side of Equation \ref{eq:no_sum_stat_transfer} becomes:

\begin{equation}\label{eq:left_side_stat_transfer}
	\lim_{N \to \infty} \sum_{i=0}^{N-1} \alpha_E \exp \left(-\frac{i\gamma}{\tau_m}\right) = \frac{\alpha_E \exp \left(-\frac{\gamma}{\tau_m}\right)}{\exp \left(-\frac{\gamma}{\tau_m} \right) -1}.
\end{equation}

\noindent In case the value of Equation \ref{eq:left_side_stat_transfer} is less than or equal to the spiking threshold $\vartheta$ the neuron does not fire, since it would require at lest a infinite number of synaptic inputs to generate a single action potential.

Considering a presynaptic spike train capable of stimulating an action potential, it is possible to isolate $N$ in Equation \ref{eq:left_side_stat_transfer}:

\begin{equation}\label{eq:isolated_n_stat_transfer}
	N = ceil \left\{ \frac{\gamma-\tau_m \ln \left[- \frac{\left( 1 - \exp \frac{\gamma}{\tau_m}  \right) \left( -\frac{\alpha_E \exp \frac{\gamma}{\tau_m} }{1 - \exp \frac{\gamma}{\tau_m}} - \vartheta \right)   }{\alpha_E} \right]}{\gamma} \right\},
\end{equation}

\noindent where the function $ceil(x)$ returns the smallest integer that is greater than $x$. 

After a generation of an action potential, the membrane potential resets to the rest potential and the integration starts again. If the same regular spike train still stimulate the neuron, the same number of $N$ presynaptic spikes will be necessary to generate another action potential. Therefore, in this case, the postsynaptic interspike interval will also be constant and given by $\mu = N\gamma$, and the output frequency $\nu_o=1/\mu$ in terms of the input frequency $\nu_i=1/\gamma$, that is, the stationary frequency transfer function, is given by:

\begin{equation}
    \nu_o(\nu_i)=\frac{\nu_i}{N(\nu_i)},
\end{equation}

\noindent where $N(\nu_i)$ can be obtained from Equation \ref{eq:isolated_n_stat_transfer}.

\begin{figure}[!htb]
	\centering
	\includegraphics[width=\columnwidth]{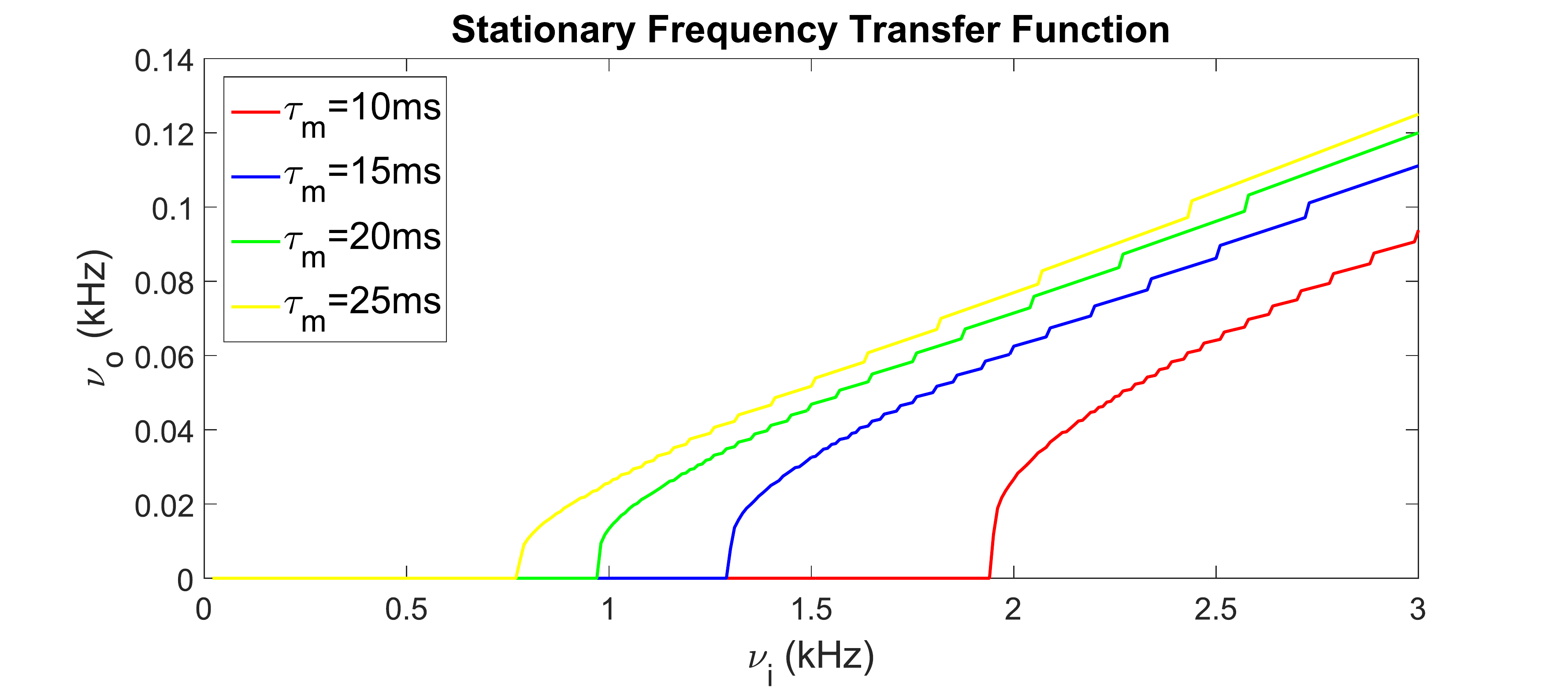}
	\caption{\label{fig:StationaryFrequencyTransferFunction} Stationary frequency transfer function of an integrate-and-fire neuron, where $v_r=v_{rs}=0mV$, $\vartheta=20mV$, $\tau_{abs}=0ms$, $\alpha_E=1mV$, and $\tau_m$ assumes several values (legend).}
\end{figure}

Figure \ref{fig:StationaryFrequencyTransferFunction} illustrates the stationary frequency transfer for different values of $\tau_m$. 

\section*{Appendix B - Linear Fit Quality Measurements}
\label{s:linear_adjust_quality_measurements}

\begin{figure*}[!htb]
	\centering
	\begin{subfigure}[t]{0.98\textwidth}
		\centering
		\caption{}
		\label{sfig:standard_error_slope}
		\begin{subfigure}[t]{0.48\textwidth}
			\includegraphics[width=\linewidth]{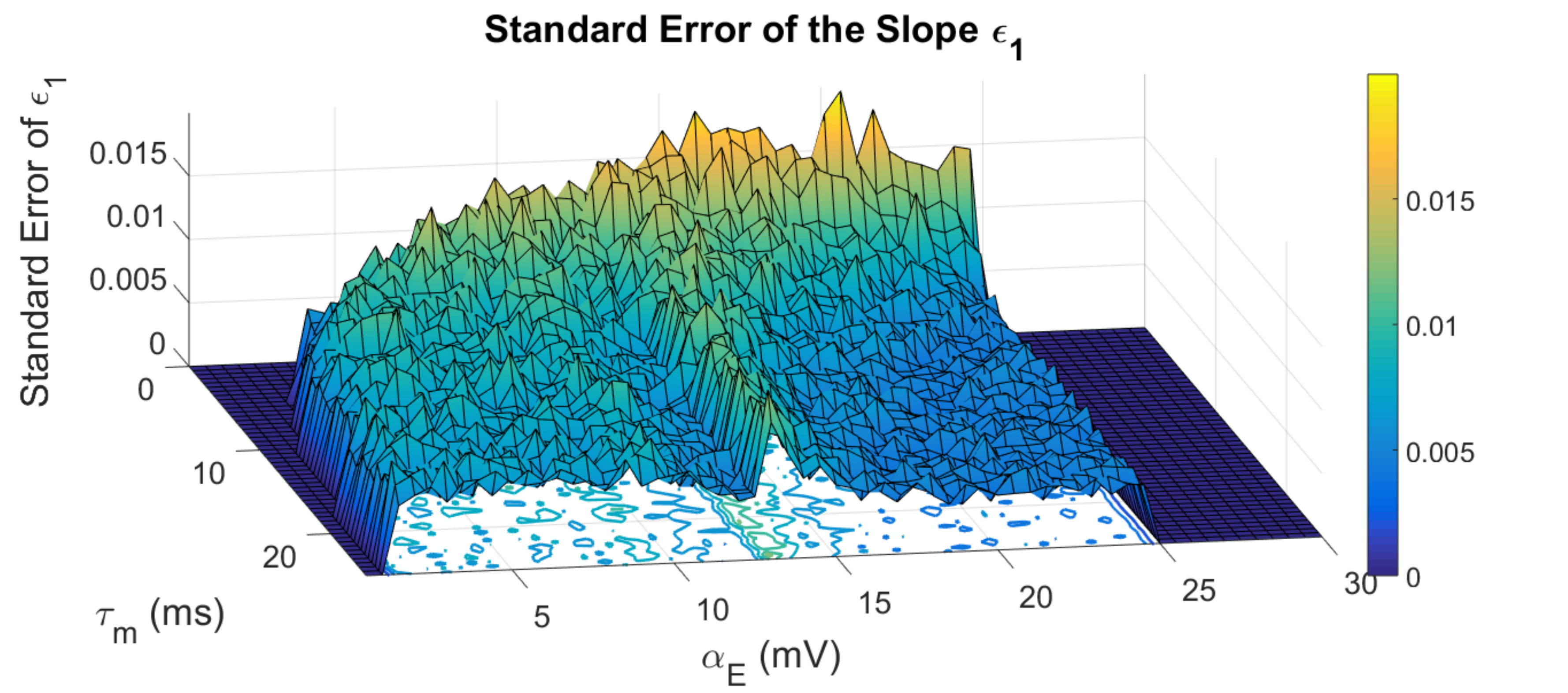}
		\end{subfigure}
		\quad
		\begin{subfigure}[t]{0.48\textwidth}
			\includegraphics[width=\linewidth]{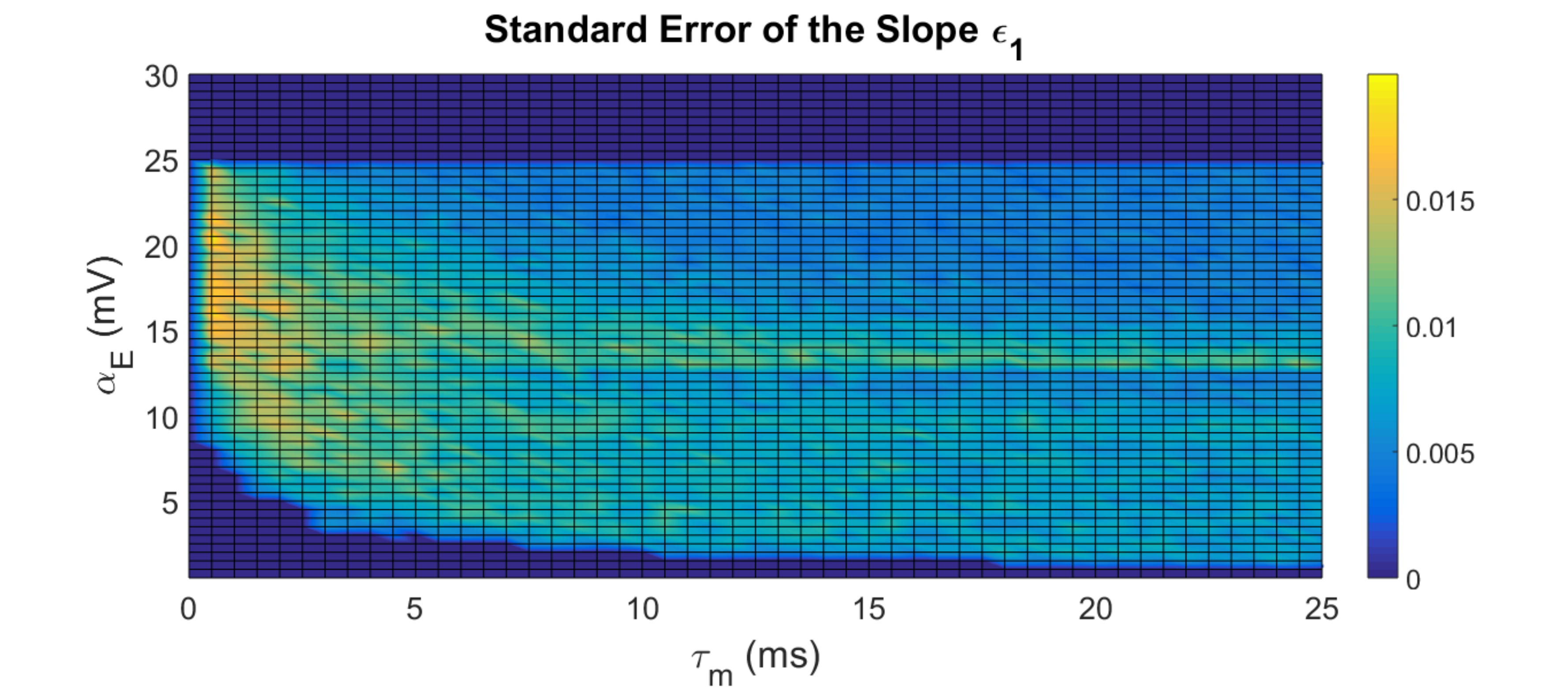}
		\end{subfigure}
	\end{subfigure}
	\quad
	\begin{subfigure}[t]{0.98\textwidth}
		\centering
		\caption{}
		\label{sfig:standard_error_intercept}
		\begin{subfigure}[t]{0.48\textwidth}
			\includegraphics[width=\linewidth]{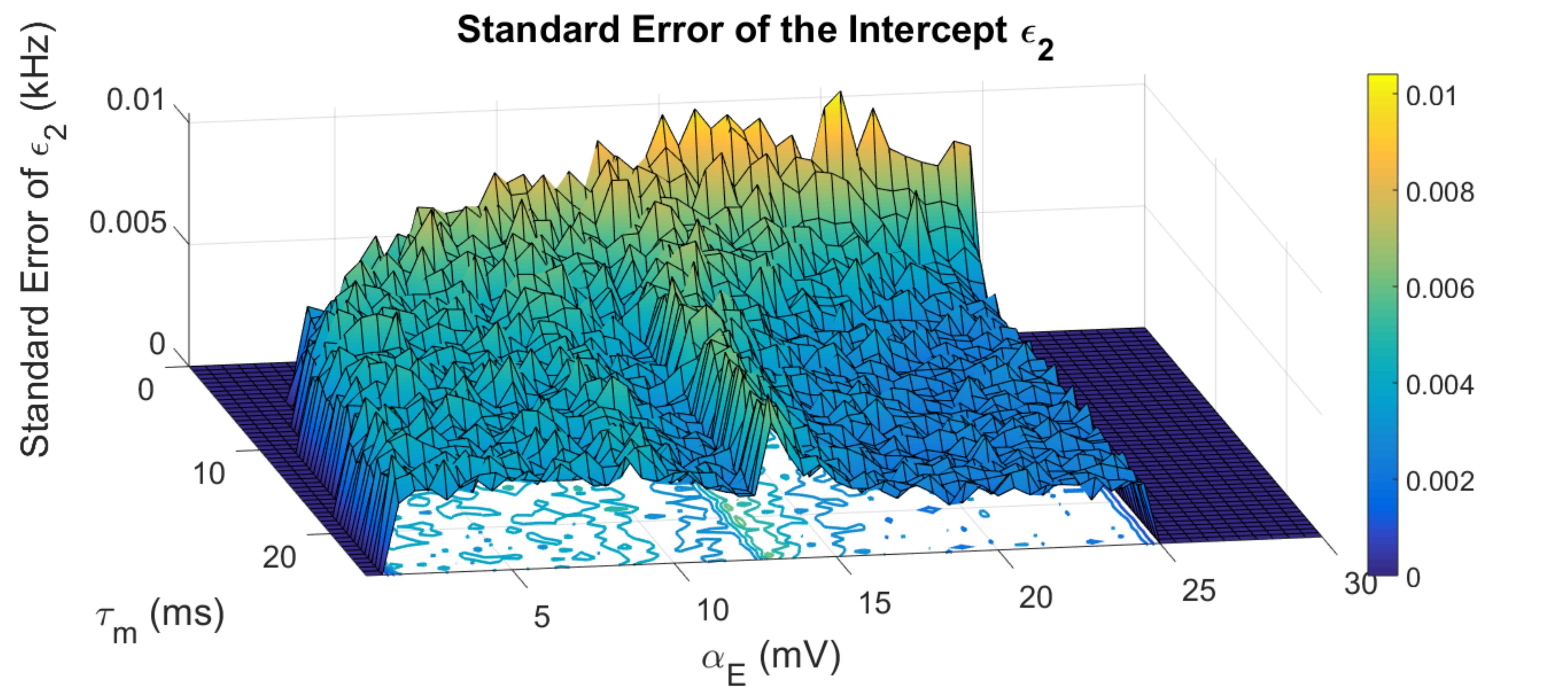}
		\end{subfigure}
		\quad
		\begin{subfigure}[t]{0.48\textwidth}
			\includegraphics[width=\linewidth]{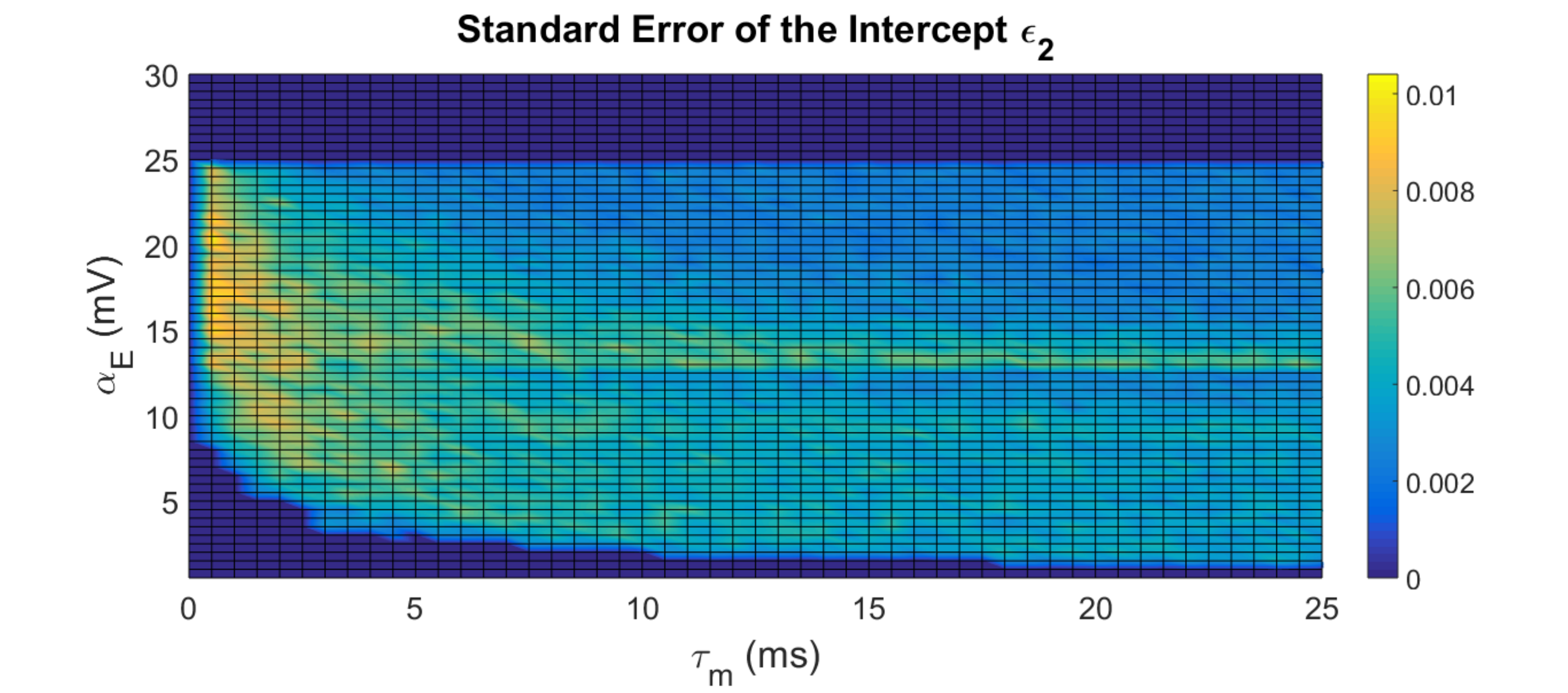}
		\end{subfigure}
	\end{subfigure}
	\quad
	\begin{subfigure}[t]{0.98\textwidth}
		\centering
		\caption{}
		\label{sfig:rmse_lif}
		\begin{subfigure}[t]{0.48\textwidth}
			\includegraphics[width=\linewidth]{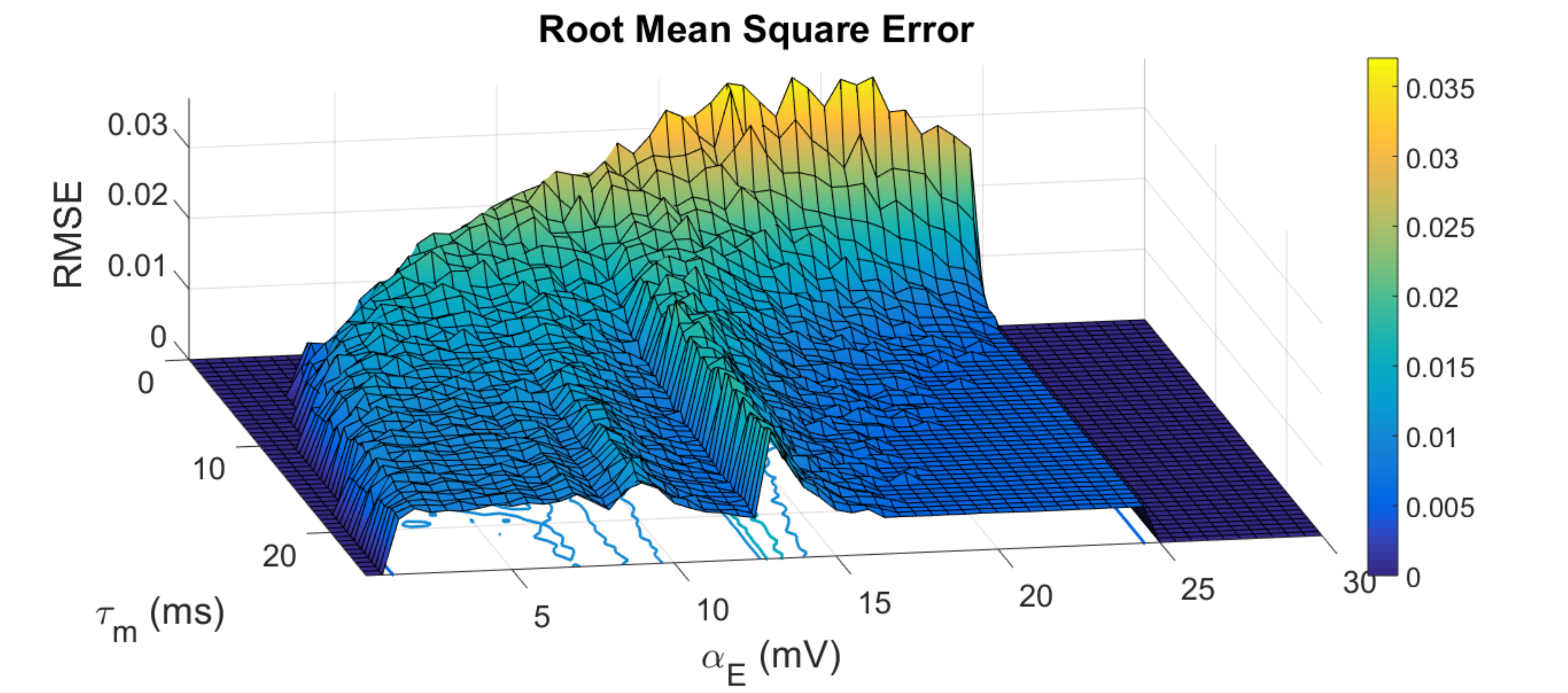}
		\end{subfigure}
		\quad
		\begin{subfigure}[t]{0.48\textwidth}
			\includegraphics[width=\linewidth]{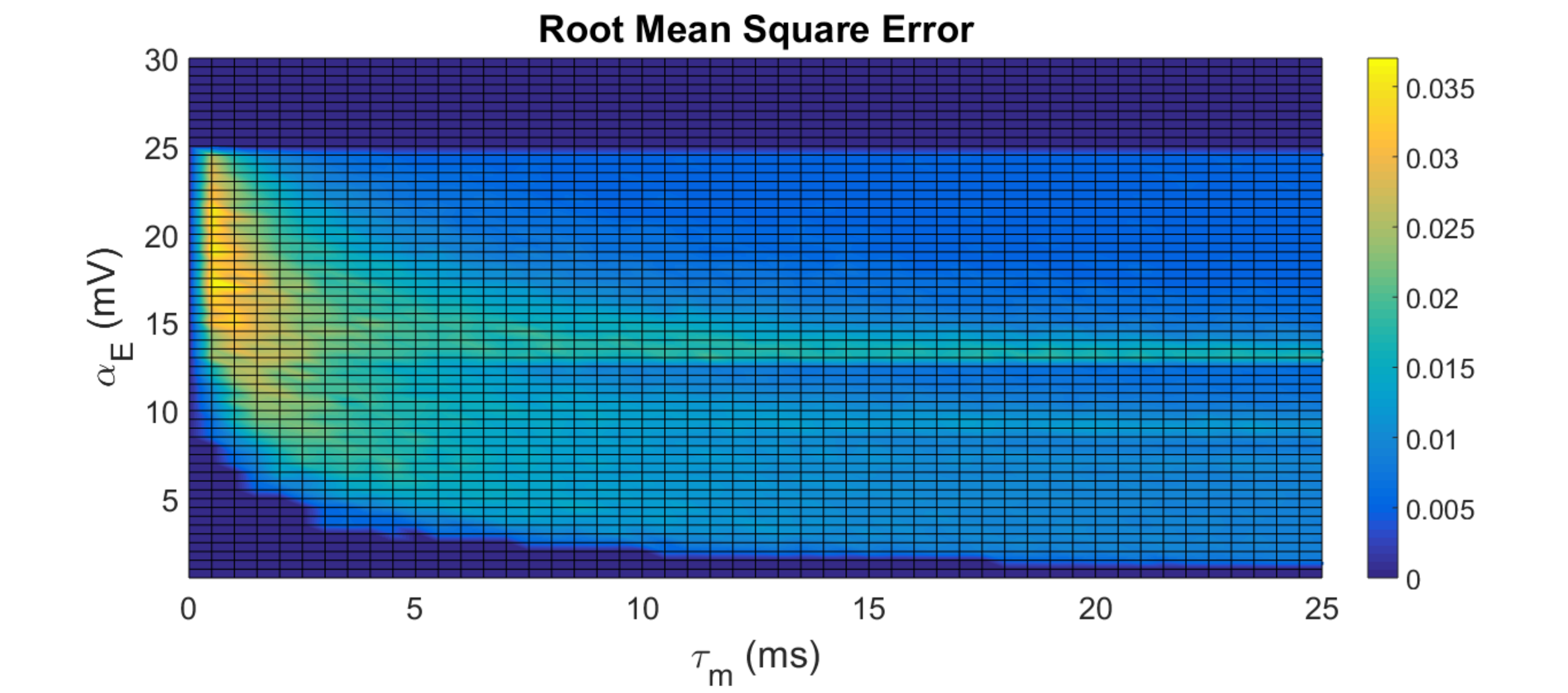}
		\end{subfigure}
	\end{subfigure}
	\quad
	\begin{subfigure}[t]{0.98\textwidth}
		\centering
		\caption{}
		\label{sfig:r2_adaptado_lif}
		\begin{subfigure}[t]{0.48\textwidth}
			\includegraphics[width=\linewidth]{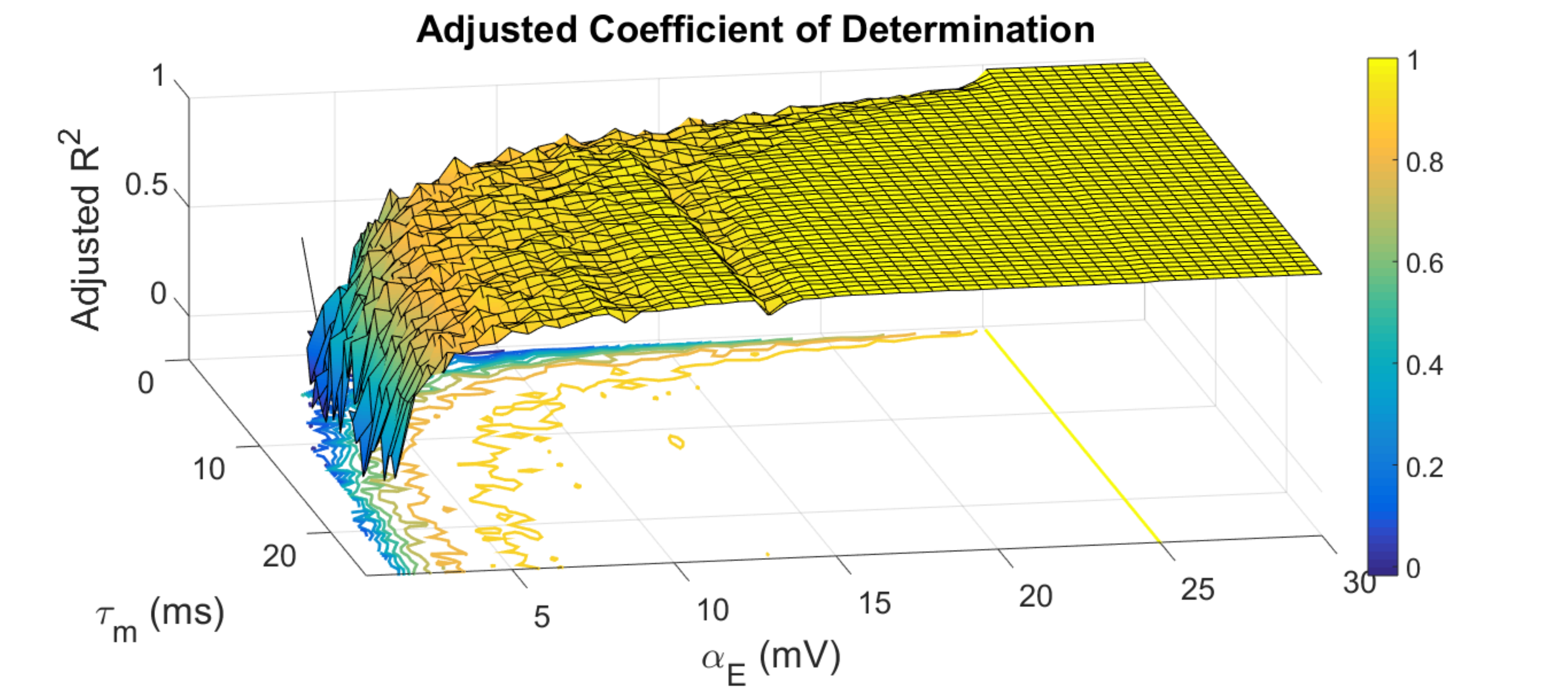}
		\end{subfigure}
		\quad
		\begin{subfigure}[t]{0.48\textwidth}
			\includegraphics[width=\linewidth]{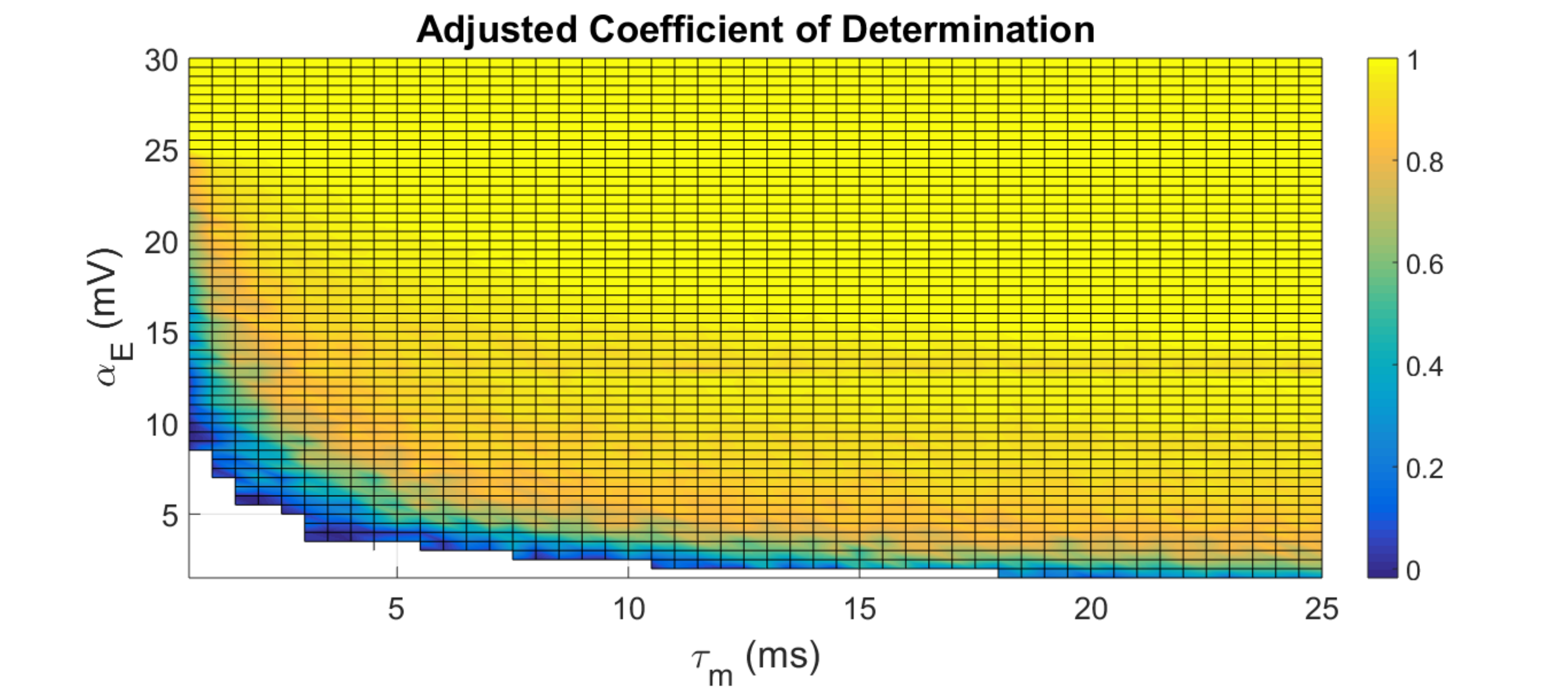}
		\end{subfigure}
	\end{subfigure}
	\caption{\label{fig:error_linear_adjust} Plots of quality measurements of the fit made using the linear least squares method. The plots are all in terms of $\alpha_E$ and $\tau_m$, and are presented at two angles. a) Standard error of the fitted slope $\epsilon_1$. b) Standard error of the fitted intercept $\epsilon_1$. c) Root mean square error. d) Adjusted coefficient of determination.}
\end{figure*}

The Figure \ref{fig:error_linear_adjust} presents the plots of different measurements associated with the quality of the linear least square fit over the dynamic frequency transfer relationship. 

\section*{Appendix C - Mutual Information Analyses}
\label{s:mutual_information_analyses}

Another, more general, way to exhibit the dependence, on dynamic frequency transfer, between the input and output frequencies is through mutual information \cite{kullback1997information}. The mutual information quantifies how much of uncertainty (entropy) the knowledge of one of the input or output frequencies removes from the other, in other words, it is a measure of how much information the input and output frequencies share.

\begin{figure}[!htb]
	\centering
	\begin{subfigure}[t]{\columnwidth}
		\includegraphics[width=\linewidth]{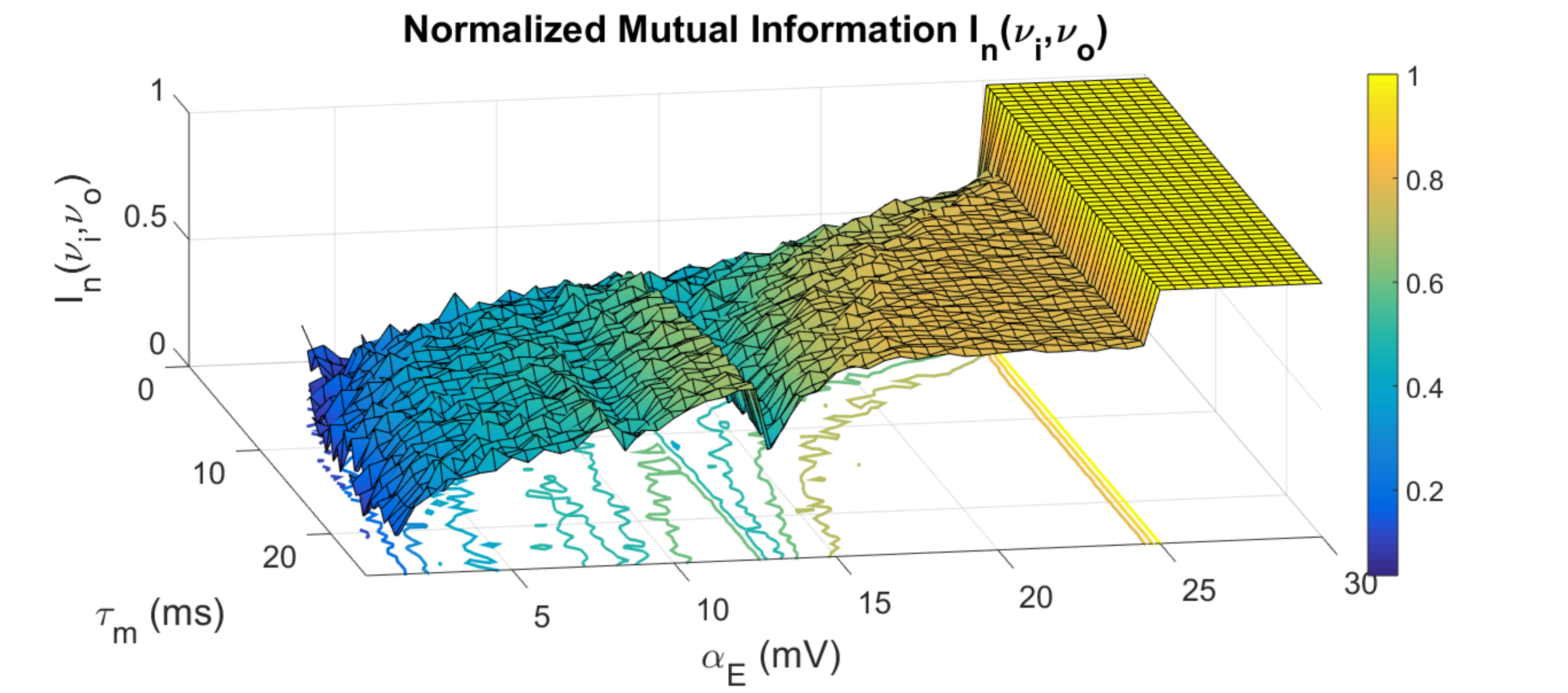}
	\end{subfigure}
	\quad
	\begin{subfigure}[t]{\columnwidth}
		\includegraphics[width=\linewidth]{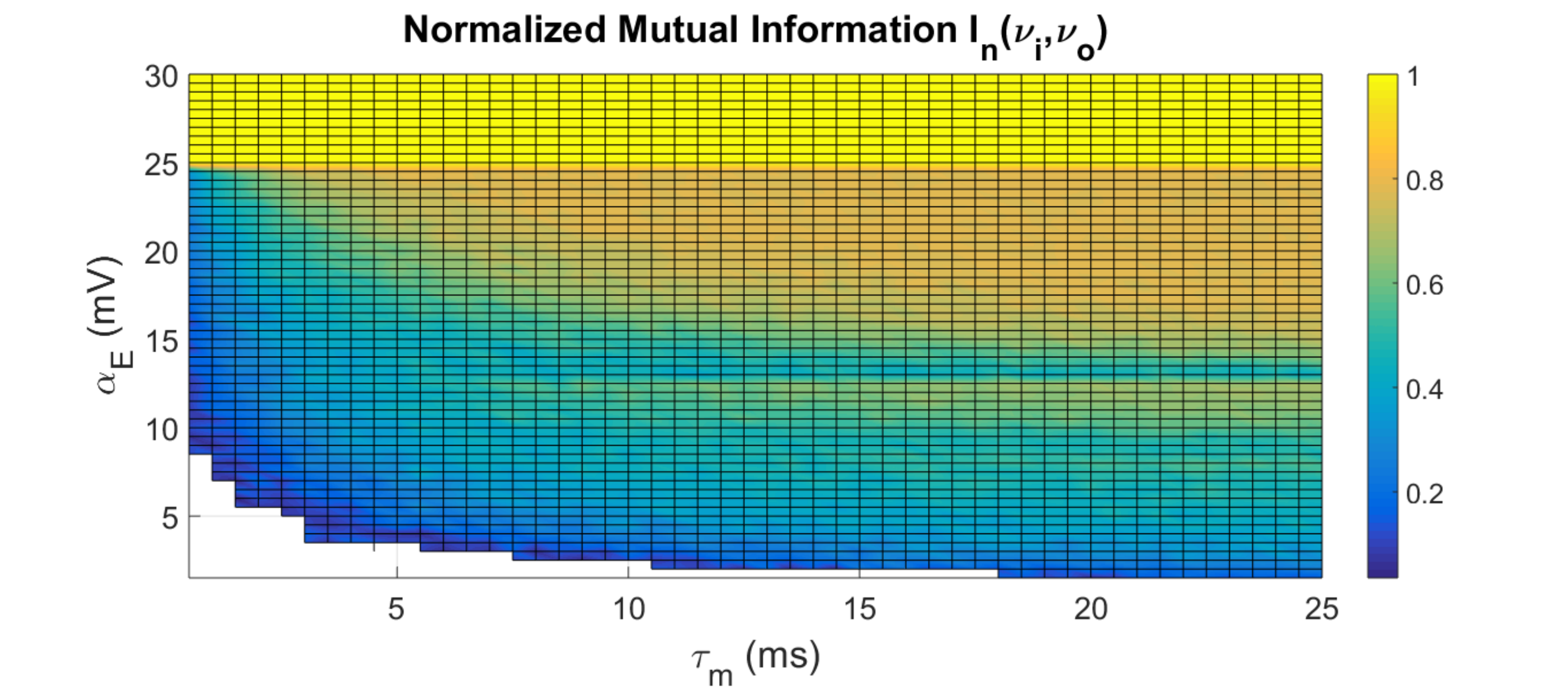}
	\end{subfigure}
	\begin{flushleft}
		\caption{\label{fig:mutual_information} Plot of the normalized mutual information, in terms of $\alpha_E$ and $\tau_m$.  A frequency window $\Delta t=80ms$ has been adopted.}
	\end{flushleft}
\end{figure}

The Figure \ref{fig:mutual_information} presents the normalized mutual information of the dynamic frequency transfer with respect to the input and output frequencies. Because the mutual information is normalized, $I(\nu_i,\nu_o)=1$ indicates that knowledge of one of the frequencies completely determines the value of the other, and $I(\nu_i,\nu_o)=0$ indicates that knowledge of one of the frequencies do not provide any information about the other, they do not share information.

It is noticeable that, for very small values of $\alpha_E$ and $\tau_m$, the input and output frequencies, under the taken circumstances, practically do not share information.  For values of $\alpha_E \ge 25mV$, because the saturation, the system is completely deterministic with respect to both input and output.  Moreover, by setting any value of $\alpha_E$, it can be observed that the mutual information increases with $\tau_m$, indicating that increments of $\tau_m$ usually increase the shared information.


\begin{thebibliography}{53}%
\makeatletter
\providecommand \@ifxundefined [1]{%
 \@ifx{#1\undefined}
}%
\providecommand \@ifnum [1]{%
 \ifnum #1\expandafter \@firstoftwo
 \else \expandafter \@secondoftwo
 \fi
}%
\providecommand \@ifx [1]{%
 \ifx #1\expandafter \@firstoftwo
 \else \expandafter \@secondoftwo
 \fi
}%
\providecommand \natexlab [1]{#1}%
\providecommand \enquote  [1]{``#1''}%
\providecommand \bibnamefont  [1]{#1}%
\providecommand \bibfnamefont [1]{#1}%
\providecommand \citenamefont [1]{#1}%
\providecommand \href@noop [0]{\@secondoftwo}%
\providecommand \href [0]{\begingroup \@sanitize@url \@href}%
\providecommand \@href[1]{\@@startlink{#1}\@@href}%
\providecommand \@@href[1]{\endgroup#1\@@endlink}%
\providecommand \@sanitize@url [0]{\catcode `\\12\catcode `\$12\catcode
  `\&12\catcode `\#12\catcode `\^12\catcode `\_12\catcode `\%12\relax}%
\providecommand \@@startlink[1]{}%
\providecommand \@@endlink[0]{}%
\providecommand \url  [0]{\begingroup\@sanitize@url \@url }%
\providecommand \@url [1]{\endgroup\@href {#1}{\urlprefix }}%
\providecommand \urlprefix  [0]{URL }%
\providecommand \Eprint [0]{\href }%
\providecommand \doibase [0]{http://dx.doi.org/}%
\providecommand \selectlanguage [0]{\@gobble}%
\providecommand \bibinfo  [0]{\@secondoftwo}%
\providecommand \bibfield  [0]{\@secondoftwo}%
\providecommand \translation [1]{[#1]}%
\providecommand \BibitemOpen [0]{}%
\providecommand \bibitemStop [0]{}%
\providecommand \bibitemNoStop [0]{.\EOS\space}%
\providecommand \EOS [0]{\spacefactor3000\relax}%
\providecommand \BibitemShut  [1]{\csname bibitem#1\endcsname}%
\let\auto@bib@innerbib\@empty
\bibitem [{\citenamefont {Dayan}\ \emph {et~al.}(2003)\citenamefont {Dayan},
  \citenamefont {Abbott} \emph {et~al.}}]{dayan2003theoretical}%
  \BibitemOpen
  \bibfield  {author} {\bibinfo {author} {\bibfnamefont {P.}~\bibnamefont
  {Dayan}}, \bibinfo {author} {\bibfnamefont {L.}~\bibnamefont {Abbott}},
  \emph {et~al.},\ }\href@noop {} {\bibfield  {journal} {\bibinfo  {journal}
  {Journal of Cognitive Neuroscience}\ }\textbf {\bibinfo {volume} {15}},\
  \bibinfo {pages} {154} (\bibinfo {year} {2003})}\BibitemShut {NoStop}%
\bibitem [{\citenamefont {Kandel}\ \emph {et~al.}(2000)\citenamefont {Kandel},
  \citenamefont {Schwartz}, \citenamefont {Jessell}, \citenamefont
  {of~Biochemistry}, \citenamefont {Jessell}, \citenamefont {Siegelbaum},\ and\
  \citenamefont {Hudspeth}}]{kandel2000principles}%
  \BibitemOpen
  \bibfield  {author} {\bibinfo {author} {\bibfnamefont {E.~R.}\ \bibnamefont
  {Kandel}}, \bibinfo {author} {\bibfnamefont {J.~H.}\ \bibnamefont
  {Schwartz}}, \bibinfo {author} {\bibfnamefont {T.~M.}\ \bibnamefont
  {Jessell}}, \bibinfo {author} {\bibfnamefont {D.}~\bibnamefont
  {of~Biochemistry}}, \bibinfo {author} {\bibfnamefont {M.~B.~T.}\ \bibnamefont
  {Jessell}}, \bibinfo {author} {\bibfnamefont {S.}~\bibnamefont {Siegelbaum}},
  \ and\ \bibinfo {author} {\bibfnamefont {A.}~\bibnamefont {Hudspeth}},\
  }\href@noop {} {\emph {\bibinfo {title} {Principles of neural science}}},\
  Vol.~\bibinfo {volume} {4}\ (\bibinfo  {publisher} {McGraw-hill New York},\
  \bibinfo {year} {2000})\BibitemShut {NoStop}%
\bibitem [{\citenamefont {Paun}\ \emph {et~al.}(2010)\citenamefont {Paun},
  \citenamefont {Rozenberg},\ and\ \citenamefont {Salomaa}}]{paun2010oxford}%
  \BibitemOpen
  \bibfield  {author} {\bibinfo {author} {\bibfnamefont {G.}~\bibnamefont
  {Paun}}, \bibinfo {author} {\bibfnamefont {G.}~\bibnamefont {Rozenberg}}, \
  and\ \bibinfo {author} {\bibfnamefont {A.}~\bibnamefont {Salomaa}},\
  }\href@noop {} {\emph {\bibinfo {title} {The Oxford handbook of membrane
  computing}}}\ (\bibinfo  {publisher} {Oxford University Press, Inc.},\
  \bibinfo {year} {2010})\BibitemShut {NoStop}%
\bibitem [{\citenamefont {Ganong}\ and\ \citenamefont
  {Ganong}(1995)}]{ganong1995review}%
  \BibitemOpen
  \bibfield  {author} {\bibinfo {author} {\bibfnamefont {W.~F.}\ \bibnamefont
  {Ganong}}\ and\ \bibinfo {author} {\bibfnamefont {W.}~\bibnamefont
  {Ganong}},\ }\href@noop {} {\emph {\bibinfo {title} {Review of medical
  physiology}}}\ (\bibinfo  {publisher} {Appleton \& Lange Norwalk, CT},\
  \bibinfo {year} {1995})\BibitemShut {NoStop}%
\bibitem [{\citenamefont {Hobson}\ and\ \citenamefont
  {Pace-Schott}(2002)}]{hobson2002cognitive}%
  \BibitemOpen
  \bibfield  {author} {\bibinfo {author} {\bibfnamefont {J.~A.}\ \bibnamefont
  {Hobson}}\ and\ \bibinfo {author} {\bibfnamefont {E.~F.}\ \bibnamefont
  {Pace-Schott}},\ }\href@noop {} {\bibfield  {journal} {\bibinfo  {journal}
  {Nature Reviews Neuroscience}\ }\textbf {\bibinfo {volume} {3}},\ \bibinfo
  {pages} {679} (\bibinfo {year} {2002})}\BibitemShut {NoStop}%
\bibitem [{\citenamefont {Hodgkin}\ and\ \citenamefont
  {Huxley}(1952)}]{hodgkin1952quantitative}%
  \BibitemOpen
  \bibfield  {author} {\bibinfo {author} {\bibfnamefont {A.~L.}\ \bibnamefont
  {Hodgkin}}\ and\ \bibinfo {author} {\bibfnamefont {A.~F.}\ \bibnamefont
  {Huxley}},\ }\href@noop {} {\bibfield  {journal} {\bibinfo  {journal} {The
  Journal of physiology}\ }\textbf {\bibinfo {volume} {117}},\ \bibinfo {pages}
  {500} (\bibinfo {year} {1952})}\BibitemShut {NoStop}%
\bibitem [{\citenamefont {Pospischil}\ \emph {et~al.}(2008)\citenamefont
  {Pospischil}, \citenamefont {Toledo-Rodriguez}, \citenamefont {Monier},
  \citenamefont {Piwkowska}, \citenamefont {Bal}, \citenamefont {Fr{\'e}gnac},
  \citenamefont {Markram},\ and\ \citenamefont
  {Destexhe}}]{pospischil2008minimal}%
  \BibitemOpen
  \bibfield  {author} {\bibinfo {author} {\bibfnamefont {M.}~\bibnamefont
  {Pospischil}}, \bibinfo {author} {\bibfnamefont {M.}~\bibnamefont
  {Toledo-Rodriguez}}, \bibinfo {author} {\bibfnamefont {C.}~\bibnamefont
  {Monier}}, \bibinfo {author} {\bibfnamefont {Z.}~\bibnamefont {Piwkowska}},
  \bibinfo {author} {\bibfnamefont {T.}~\bibnamefont {Bal}}, \bibinfo {author}
  {\bibfnamefont {Y.}~\bibnamefont {Fr{\'e}gnac}}, \bibinfo {author}
  {\bibfnamefont {H.}~\bibnamefont {Markram}}, \ and\ \bibinfo {author}
  {\bibfnamefont {A.}~\bibnamefont {Destexhe}},\ }\href@noop {} {\bibfield
  {journal} {\bibinfo  {journal} {Biological cybernetics}\ }\textbf {\bibinfo
  {volume} {99}},\ \bibinfo {pages} {427} (\bibinfo {year} {2008})}\BibitemShut
  {NoStop}%
\bibitem [{\citenamefont {Izhikevich}(2003)}]{izhikevich2003simple}%
  \BibitemOpen
  \bibfield  {author} {\bibinfo {author} {\bibfnamefont {E.~M.}\ \bibnamefont
  {Izhikevich}},\ }\href@noop {} {\bibfield  {journal} {\bibinfo  {journal}
  {IEEE Transactions on Neural Networks}\ }\textbf {\bibinfo {volume} {14}},\
  \bibinfo {pages} {1569} (\bibinfo {year} {2003})}\BibitemShut {NoStop}%
\bibitem [{\citenamefont {Burkitt}(2006)}]{burkitt2006review}%
  \BibitemOpen
  \bibfield  {author} {\bibinfo {author} {\bibfnamefont {A.~N.}\ \bibnamefont
  {Burkitt}},\ }\href@noop {} {\bibfield  {journal} {\bibinfo  {journal}
  {Biological cybernetics}\ }\textbf {\bibinfo {volume} {95}},\ \bibinfo
  {pages} {1} (\bibinfo {year} {2006})}\BibitemShut {NoStop}%
\bibitem [{\citenamefont {Izhikevich}(2004)}]{izhikevich2004model}%
  \BibitemOpen
  \bibfield  {author} {\bibinfo {author} {\bibfnamefont {E.~M.}\ \bibnamefont
  {Izhikevich}},\ }\href@noop {} {\bibfield  {journal} {\bibinfo  {journal}
  {IEEE transactions on neural networks}\ }\textbf {\bibinfo {volume} {15}},\
  \bibinfo {pages} {1063} (\bibinfo {year} {2004})}\BibitemShut {NoStop}%
\bibitem [{\citenamefont {Brette}\ \emph {et~al.}(2007)\citenamefont {Brette},
  \citenamefont {Rudolph}, \citenamefont {Carnevale}, \citenamefont {Hines},
  \citenamefont {Beeman}, \citenamefont {Bower}, \citenamefont {Diesmann},
  \citenamefont {Morrison}, \citenamefont {Goodman}, \citenamefont {Harris}
  \emph {et~al.}}]{brette2007simulation}%
  \BibitemOpen
  \bibfield  {author} {\bibinfo {author} {\bibfnamefont {R.}~\bibnamefont
  {Brette}}, \bibinfo {author} {\bibfnamefont {M.}~\bibnamefont {Rudolph}},
  \bibinfo {author} {\bibfnamefont {T.}~\bibnamefont {Carnevale}}, \bibinfo
  {author} {\bibfnamefont {M.}~\bibnamefont {Hines}}, \bibinfo {author}
  {\bibfnamefont {D.}~\bibnamefont {Beeman}}, \bibinfo {author} {\bibfnamefont
  {J.~M.}\ \bibnamefont {Bower}}, \bibinfo {author} {\bibfnamefont
  {M.}~\bibnamefont {Diesmann}}, \bibinfo {author} {\bibfnamefont
  {A.}~\bibnamefont {Morrison}}, \bibinfo {author} {\bibfnamefont {P.~H.}\
  \bibnamefont {Goodman}}, \bibinfo {author} {\bibfnamefont {F.~C.}\
  \bibnamefont {Harris}},  \emph {et~al.},\ }\href@noop {} {\bibfield
  {journal} {\bibinfo  {journal} {Journal of computational neuroscience}\
  }\textbf {\bibinfo {volume} {23}},\ \bibinfo {pages} {349} (\bibinfo {year}
  {2007})}\BibitemShut {NoStop}%
\bibitem [{\citenamefont {Baddeley}\ \emph {et~al.}(1997)\citenamefont
  {Baddeley}, \citenamefont {Abbott}, \citenamefont {Booth}, \citenamefont
  {Sengpiel}, \citenamefont {Freeman}, \citenamefont {Wakeman},\ and\
  \citenamefont {Rolls}}]{baddeley1997responses}%
  \BibitemOpen
  \bibfield  {author} {\bibinfo {author} {\bibfnamefont {R.}~\bibnamefont
  {Baddeley}}, \bibinfo {author} {\bibfnamefont {L.~F.}\ \bibnamefont
  {Abbott}}, \bibinfo {author} {\bibfnamefont {M.~C.}\ \bibnamefont {Booth}},
  \bibinfo {author} {\bibfnamefont {F.}~\bibnamefont {Sengpiel}}, \bibinfo
  {author} {\bibfnamefont {T.}~\bibnamefont {Freeman}}, \bibinfo {author}
  {\bibfnamefont {E.~A.}\ \bibnamefont {Wakeman}}, \ and\ \bibinfo {author}
  {\bibfnamefont {E.~T.}\ \bibnamefont {Rolls}},\ }\href@noop {} {\bibfield
  {journal} {\bibinfo  {journal} {Proceedings of the Royal Society of London B:
  Biological Sciences}\ }\textbf {\bibinfo {volume} {264}},\ \bibinfo {pages}
  {1775} (\bibinfo {year} {1997})}\BibitemShut {NoStop}%
\bibitem [{\citenamefont {Salinas}\ \emph {et~al.}(2000)\citenamefont
  {Salinas}, \citenamefont {Hernandez}, \citenamefont {Zainos},\ and\
  \citenamefont {Romo}}]{salinas2000periodicity}%
  \BibitemOpen
  \bibfield  {author} {\bibinfo {author} {\bibfnamefont {E.}~\bibnamefont
  {Salinas}}, \bibinfo {author} {\bibfnamefont {A.}~\bibnamefont {Hernandez}},
  \bibinfo {author} {\bibfnamefont {A.}~\bibnamefont {Zainos}}, \ and\ \bibinfo
  {author} {\bibfnamefont {R.}~\bibnamefont {Romo}},\ }\href@noop {} {\bibfield
   {journal} {\bibinfo  {journal} {Journal of neuroscience}\ }\textbf {\bibinfo
  {volume} {20}},\ \bibinfo {pages} {5503} (\bibinfo {year}
  {2000})}\BibitemShut {NoStop}%
\bibitem [{\citenamefont {Ferster}\ and\ \citenamefont
  {Spruston}(1995)}]{ferster1995cracking}%
  \BibitemOpen
  \bibfield  {author} {\bibinfo {author} {\bibfnamefont {D.}~\bibnamefont
  {Ferster}}\ and\ \bibinfo {author} {\bibfnamefont {N.}~\bibnamefont
  {Spruston}},\ }\href@noop {} {\bibfield  {journal} {\bibinfo  {journal}
  {Science}\ }\textbf {\bibinfo {volume} {270}},\ \bibinfo {pages} {756}
  (\bibinfo {year} {1995})}\BibitemShut {NoStop}%
\bibitem [{\citenamefont {Lu}\ \emph {et~al.}(2001)\citenamefont {Lu},
  \citenamefont {Liang},\ and\ \citenamefont {Wang}}]{lu2001temporal}%
  \BibitemOpen
  \bibfield  {author} {\bibinfo {author} {\bibfnamefont {T.}~\bibnamefont
  {Lu}}, \bibinfo {author} {\bibfnamefont {L.}~\bibnamefont {Liang}}, \ and\
  \bibinfo {author} {\bibfnamefont {X.}~\bibnamefont {Wang}},\ }\href@noop {}
  {\bibfield  {journal} {\bibinfo  {journal} {Nature neuroscience}\ }\textbf
  {\bibinfo {volume} {4}},\ \bibinfo {pages} {1131} (\bibinfo {year}
  {2001})}\BibitemShut {NoStop}%
\bibitem [{\citenamefont {Colgin}\ \emph {et~al.}(2009)\citenamefont {Colgin},
  \citenamefont {Denninger}, \citenamefont {Fyhn}, \citenamefont {Hafting},
  \citenamefont {Bonnevie}, \citenamefont {Jensen}, \citenamefont {Moser},\
  and\ \citenamefont {Moser}}]{colgin2009frequency}%
  \BibitemOpen
  \bibfield  {author} {\bibinfo {author} {\bibfnamefont {L.~L.}\ \bibnamefont
  {Colgin}}, \bibinfo {author} {\bibfnamefont {T.}~\bibnamefont {Denninger}},
  \bibinfo {author} {\bibfnamefont {M.}~\bibnamefont {Fyhn}}, \bibinfo {author}
  {\bibfnamefont {T.}~\bibnamefont {Hafting}}, \bibinfo {author} {\bibfnamefont
  {T.}~\bibnamefont {Bonnevie}}, \bibinfo {author} {\bibfnamefont
  {O.}~\bibnamefont {Jensen}}, \bibinfo {author} {\bibfnamefont {M.-B.}\
  \bibnamefont {Moser}}, \ and\ \bibinfo {author} {\bibfnamefont {E.~I.}\
  \bibnamefont {Moser}},\ }\href@noop {} {\bibfield  {journal} {\bibinfo
  {journal} {Nature}\ }\textbf {\bibinfo {volume} {462}},\ \bibinfo {pages}
  {353} (\bibinfo {year} {2009})}\BibitemShut {NoStop}%
\bibitem [{\citenamefont {Mitra}\ and\ \citenamefont
  {Kuo}(2006)}]{mitra2006digital}%
  \BibitemOpen
  \bibfield  {author} {\bibinfo {author} {\bibfnamefont {S.~K.}\ \bibnamefont
  {Mitra}}\ and\ \bibinfo {author} {\bibfnamefont {Y.}~\bibnamefont {Kuo}},\
  }\href@noop {} {\emph {\bibinfo {title} {Digital signal processing: a
  computer-based approach}}},\ Vol.~\bibinfo {volume} {2}\ (\bibinfo
  {publisher} {McGraw-Hill New York},\ \bibinfo {year} {2006})\BibitemShut
  {NoStop}%
\bibitem [{\citenamefont {Keesman}(2011)}]{keesman2011system}%
  \BibitemOpen
  \bibfield  {author} {\bibinfo {author} {\bibfnamefont {K.~J.}\ \bibnamefont
  {Keesman}},\ }\href@noop {} {\emph {\bibinfo {title} {System identification:
  an introduction}}}\ (\bibinfo  {publisher} {Springer Science \& Business
  Media},\ \bibinfo {year} {2011})\BibitemShut {NoStop}%
\bibitem [{\citenamefont {Brunel}\ and\ \citenamefont
  {Van~Rossum}(2007)}]{brunel2007lapicque}%
  \BibitemOpen
  \bibfield  {author} {\bibinfo {author} {\bibfnamefont {N.}~\bibnamefont
  {Brunel}}\ and\ \bibinfo {author} {\bibfnamefont {M.~C.}\ \bibnamefont
  {Van~Rossum}},\ }\href@noop {} {\bibfield  {journal} {\bibinfo  {journal}
  {Biological Cybernetics}\ }\textbf {\bibinfo {volume} {97}},\ \bibinfo
  {pages} {337} (\bibinfo {year} {2007})}\BibitemShut {NoStop}%
\bibitem [{\citenamefont {Quiroga}\ and\ \citenamefont
  {Panzeri}(2013)}]{quiroga2013principles}%
  \BibitemOpen
  \bibfield  {author} {\bibinfo {author} {\bibfnamefont {R.~Q.}\ \bibnamefont
  {Quiroga}}\ and\ \bibinfo {author} {\bibfnamefont {S.}~\bibnamefont
  {Panzeri}},\ }\href@noop {} {\emph {\bibinfo {title} {Principles of neural
  coding}}}\ (\bibinfo  {publisher} {New York: CRC Press},\ \bibinfo {year}
  {2013})\BibitemShut {NoStop}%
\bibitem [{\citenamefont {Vreeswijk}\ and\ \citenamefont
  {Sompolinsky}(1998)}]{vreeswijk1998chaotic}%
  \BibitemOpen
  \bibfield  {author} {\bibinfo {author} {\bibfnamefont {C.~v.}\ \bibnamefont
  {Vreeswijk}}\ and\ \bibinfo {author} {\bibfnamefont {H.}~\bibnamefont
  {Sompolinsky}},\ }\href@noop {} {\bibfield  {journal} {\bibinfo  {journal}
  {Neural Computation}\ }\textbf {\bibinfo {volume} {10}},\ \bibinfo {pages}
  {1321} (\bibinfo {year} {1998})}\BibitemShut {NoStop}%
\bibitem [{\citenamefont {Brunel}(2000)}]{brunel2000dynamics}%
  \BibitemOpen
  \bibfield  {author} {\bibinfo {author} {\bibfnamefont {N.}~\bibnamefont
  {Brunel}},\ }\href@noop {} {\bibfield  {journal} {\bibinfo  {journal}
  {Journal of Computational Neuroscience}\ }\textbf {\bibinfo {volume} {8}},\
  \bibinfo {pages} {183} (\bibinfo {year} {2000})}\BibitemShut {NoStop}%
\bibitem [{\citenamefont {Compte}\ \emph {et~al.}(2003)\citenamefont {Compte},
  \citenamefont {Sanchez-Vives}, \citenamefont {McCormick},\ and\ \citenamefont
  {Wang}}]{compte2003cellular}%
  \BibitemOpen
  \bibfield  {author} {\bibinfo {author} {\bibfnamefont {A.}~\bibnamefont
  {Compte}}, \bibinfo {author} {\bibfnamefont {M.~V.}\ \bibnamefont
  {Sanchez-Vives}}, \bibinfo {author} {\bibfnamefont {D.~A.}\ \bibnamefont
  {McCormick}}, \ and\ \bibinfo {author} {\bibfnamefont {X.-J.}\ \bibnamefont
  {Wang}},\ }\href@noop {} {\bibfield  {journal} {\bibinfo  {journal} {Journal
  of Neurophysiology}\ }\textbf {\bibinfo {volume} {89}},\ \bibinfo {pages}
  {2707} (\bibinfo {year} {2003})}\BibitemShut {NoStop}%
\bibitem [{\citenamefont {Parga}\ and\ \citenamefont
  {Abbott}(2007)}]{parga2007network}%
  \BibitemOpen
  \bibfield  {author} {\bibinfo {author} {\bibfnamefont {N.}~\bibnamefont
  {Parga}}\ and\ \bibinfo {author} {\bibfnamefont {L.~F.}\ \bibnamefont
  {Abbott}},\ }\href@noop {} {\bibfield  {journal} {\bibinfo  {journal}
  {Frontiers in Neuroscience}\ }\textbf {\bibinfo {volume} {1}},\ \bibinfo
  {pages} {57} (\bibinfo {year} {2007})}\BibitemShut {NoStop}%
\bibitem [{\citenamefont {Holcman}\ and\ \citenamefont
  {Tsodyks}(2006)}]{holcman2006emergence}%
  \BibitemOpen
  \bibfield  {author} {\bibinfo {author} {\bibfnamefont {D.}~\bibnamefont
  {Holcman}}\ and\ \bibinfo {author} {\bibfnamefont {M.}~\bibnamefont
  {Tsodyks}},\ }\href@noop {} {\bibfield  {journal} {\bibinfo  {journal} {PLoS
  Computational Biology}\ }\textbf {\bibinfo {volume} {2}},\ \bibinfo {pages}
  {e23} (\bibinfo {year} {2006})}\BibitemShut {NoStop}%
\bibitem [{\citenamefont {Kriener}\ \emph {et~al.}(2014)\citenamefont
  {Kriener}, \citenamefont {Enger}, \citenamefont {Tetzlaff}, \citenamefont
  {Plesser}, \citenamefont {Gewaltig},\ and\ \citenamefont
  {Einevoll}}]{kriener2014dynamics}%
  \BibitemOpen
  \bibfield  {author} {\bibinfo {author} {\bibfnamefont {B.}~\bibnamefont
  {Kriener}}, \bibinfo {author} {\bibfnamefont {H.}~\bibnamefont {Enger}},
  \bibinfo {author} {\bibfnamefont {T.}~\bibnamefont {Tetzlaff}}, \bibinfo
  {author} {\bibfnamefont {H.~E.}\ \bibnamefont {Plesser}}, \bibinfo {author}
  {\bibfnamefont {M.-O.}\ \bibnamefont {Gewaltig}}, \ and\ \bibinfo {author}
  {\bibfnamefont {G.~T.}\ \bibnamefont {Einevoll}},\ }\href@noop {} {\bibfield
  {journal} {\bibinfo  {journal} {Frontiers in Computational Neuroscience}\
  }\textbf {\bibinfo {volume} {8}},\ \bibinfo {pages} {136} (\bibinfo {year}
  {2014})},\ \bibinfo {note} {$\,$doi: 10.3389/fncom.2014.00136}\BibitemShut
  {NoStop}%
\bibitem [{\citenamefont {Hopfield}\ and\ \citenamefont
  {Herz}(1995)}]{hopfield1995rapid}%
  \BibitemOpen
  \bibfield  {author} {\bibinfo {author} {\bibfnamefont {J.~J.}\ \bibnamefont
  {Hopfield}}\ and\ \bibinfo {author} {\bibfnamefont {A.~V.}\ \bibnamefont
  {Herz}},\ }\href@noop {} {\bibfield  {journal} {\bibinfo  {journal}
  {Proceedings of the National Academy of Sciences}\ }\textbf {\bibinfo
  {volume} {92}},\ \bibinfo {pages} {6655} (\bibinfo {year}
  {1995})}\BibitemShut {NoStop}%
\bibitem [{\citenamefont {Griffith}(1963)}]{griffith1963stability}%
  \BibitemOpen
  \bibfield  {author} {\bibinfo {author} {\bibfnamefont {J.}~\bibnamefont
  {Griffith}},\ }\href@noop {} {\bibfield  {journal} {\bibinfo  {journal}
  {Biophysical Journal}\ }\textbf {\bibinfo {volume} {3}},\ \bibinfo {pages}
  {299} (\bibinfo {year} {1963})}\BibitemShut {NoStop}%
\bibitem [{\citenamefont {El~Boustani}\ and\ \citenamefont
  {Destexhe}(2009)}]{el2009master}%
  \BibitemOpen
  \bibfield  {author} {\bibinfo {author} {\bibfnamefont {S.}~\bibnamefont
  {El~Boustani}}\ and\ \bibinfo {author} {\bibfnamefont {A.}~\bibnamefont
  {Destexhe}},\ }\href@noop {} {\bibfield  {journal} {\bibinfo  {journal}
  {Neural Computation}\ }\textbf {\bibinfo {volume} {21}},\ \bibinfo {pages}
  {46} (\bibinfo {year} {2009})}\BibitemShut {NoStop}%
\bibitem [{\citenamefont {Compte}\ \emph {et~al.}(2000)\citenamefont {Compte},
  \citenamefont {Brunel}, \citenamefont {Goldman-Rakic},\ and\ \citenamefont
  {Wang}}]{compte2000synaptic}%
  \BibitemOpen
  \bibfield  {author} {\bibinfo {author} {\bibfnamefont {A.}~\bibnamefont
  {Compte}}, \bibinfo {author} {\bibfnamefont {N.}~\bibnamefont {Brunel}},
  \bibinfo {author} {\bibfnamefont {P.~S.}\ \bibnamefont {Goldman-Rakic}}, \
  and\ \bibinfo {author} {\bibfnamefont {X.-J.}\ \bibnamefont {Wang}},\
  }\href@noop {} {\bibfield  {journal} {\bibinfo  {journal} {Cerebral Cortex}\
  }\textbf {\bibinfo {volume} {10}},\ \bibinfo {pages} {910} (\bibinfo {year}
  {2000})}\BibitemShut {NoStop}%
\bibitem [{\citenamefont {Wang}(1999)}]{wang1999synaptic}%
  \BibitemOpen
  \bibfield  {author} {\bibinfo {author} {\bibfnamefont {X.-J.}\ \bibnamefont
  {Wang}},\ }\href@noop {} {\bibfield  {journal} {\bibinfo  {journal} {Journal
  of Neuroscience}\ }\textbf {\bibinfo {volume} {19}},\ \bibinfo {pages} {9587}
  (\bibinfo {year} {1999})}\BibitemShut {NoStop}%
\bibitem [{\citenamefont {Mongillo}\ \emph {et~al.}(2008)\citenamefont
  {Mongillo}, \citenamefont {Barak},\ and\ \citenamefont
  {Tsodyks}}]{mongillo2008synaptic}%
  \BibitemOpen
  \bibfield  {author} {\bibinfo {author} {\bibfnamefont {G.}~\bibnamefont
  {Mongillo}}, \bibinfo {author} {\bibfnamefont {O.}~\bibnamefont {Barak}}, \
  and\ \bibinfo {author} {\bibfnamefont {M.}~\bibnamefont {Tsodyks}},\
  }\href@noop {} {\bibfield  {journal} {\bibinfo  {journal} {Science}\ }\textbf
  {\bibinfo {volume} {319}},\ \bibinfo {pages} {1543} (\bibinfo {year}
  {2008})}\BibitemShut {NoStop}%
\bibitem [{\citenamefont {Buehlmann}\ and\ \citenamefont
  {Deco}(2010)}]{buehlmann2010optimal}%
  \BibitemOpen
  \bibfield  {author} {\bibinfo {author} {\bibfnamefont {A.}~\bibnamefont
  {Buehlmann}}\ and\ \bibinfo {author} {\bibfnamefont {G.}~\bibnamefont
  {Deco}},\ }\href@noop {} {\bibfield  {journal} {\bibinfo  {journal} {PLoS
  Computational Biology}\ }\textbf {\bibinfo {volume} {6}},\ \bibinfo {pages}
  {e1000934} (\bibinfo {year} {2010})}\BibitemShut {NoStop}%
\bibitem [{\citenamefont {Deco}\ and\ \citenamefont
  {Thiele}(2011)}]{deco2011cholinergic}%
  \BibitemOpen
  \bibfield  {author} {\bibinfo {author} {\bibfnamefont {G.}~\bibnamefont
  {Deco}}\ and\ \bibinfo {author} {\bibfnamefont {A.}~\bibnamefont {Thiele}},\
  }\href@noop {} {\bibfield  {journal} {\bibinfo  {journal} {European Journal
  of Neuroscience}\ }\textbf {\bibinfo {volume} {34}},\ \bibinfo {pages} {146}
  (\bibinfo {year} {2011})}\BibitemShut {NoStop}%
\bibitem [{\citenamefont {Wang}(2002)}]{wang2002probabilistic}%
  \BibitemOpen
  \bibfield  {author} {\bibinfo {author} {\bibfnamefont {X.-J.}\ \bibnamefont
  {Wang}},\ }\href@noop {} {\bibfield  {journal} {\bibinfo  {journal} {Neuron}\
  }\textbf {\bibinfo {volume} {36}},\ \bibinfo {pages} {955} (\bibinfo {year}
  {2002})}\BibitemShut {NoStop}%
\bibitem [{\citenamefont {Brunel}\ and\ \citenamefont
  {Wang}(2003)}]{brunel2003determines}%
  \BibitemOpen
  \bibfield  {author} {\bibinfo {author} {\bibfnamefont {N.}~\bibnamefont
  {Brunel}}\ and\ \bibinfo {author} {\bibfnamefont {X.-J.}\ \bibnamefont
  {Wang}},\ }\href@noop {} {\bibfield  {journal} {\bibinfo  {journal} {Journal
  of Neurophysiology}\ }\textbf {\bibinfo {volume} {90}},\ \bibinfo {pages}
  {415} (\bibinfo {year} {2003})}\BibitemShut {NoStop}%
\bibitem [{\citenamefont {Mazzoni}\ \emph {et~al.}(2008)\citenamefont
  {Mazzoni}, \citenamefont {Panzeri}, \citenamefont {Logothetis},\ and\
  \citenamefont {Brunel}}]{mazzoni2008encoding}%
  \BibitemOpen
  \bibfield  {author} {\bibinfo {author} {\bibfnamefont {A.}~\bibnamefont
  {Mazzoni}}, \bibinfo {author} {\bibfnamefont {S.}~\bibnamefont {Panzeri}},
  \bibinfo {author} {\bibfnamefont {N.~K.}\ \bibnamefont {Logothetis}}, \ and\
  \bibinfo {author} {\bibfnamefont {N.}~\bibnamefont {Brunel}},\ }\href@noop {}
  {\bibfield  {journal} {\bibinfo  {journal} {PLoS Computational Biology}\
  }\textbf {\bibinfo {volume} {4}},\ \bibinfo {pages} {e1000239} (\bibinfo
  {year} {2008})}\BibitemShut {NoStop}%
\bibitem [{\citenamefont {Mazzoni}\ \emph {et~al.}(2011)\citenamefont
  {Mazzoni}, \citenamefont {Brunel}, \citenamefont {Cavallari}, \citenamefont
  {Logothetis},\ and\ \citenamefont {Panzeri}}]{mazzoni2011cortical}%
  \BibitemOpen
  \bibfield  {author} {\bibinfo {author} {\bibfnamefont {A.}~\bibnamefont
  {Mazzoni}}, \bibinfo {author} {\bibfnamefont {N.}~\bibnamefont {Brunel}},
  \bibinfo {author} {\bibfnamefont {S.}~\bibnamefont {Cavallari}}, \bibinfo
  {author} {\bibfnamefont {N.~K.}\ \bibnamefont {Logothetis}}, \ and\ \bibinfo
  {author} {\bibfnamefont {S.}~\bibnamefont {Panzeri}},\ }\href@noop {}
  {\bibfield  {journal} {\bibinfo  {journal} {Journal of Physiology}\ }\textbf
  {\bibinfo {volume} {105}},\ \bibinfo {pages} {2} (\bibinfo {year}
  {2011})}\BibitemShut {NoStop}%
\bibitem [{\citenamefont {Hansel}\ \emph {et~al.}(1998)\citenamefont {Hansel},
  \citenamefont {Mato}, \citenamefont {Meunier},\ and\ \citenamefont
  {Neltner}}]{hansel1998numerical}%
  \BibitemOpen
  \bibfield  {author} {\bibinfo {author} {\bibfnamefont {D.}~\bibnamefont
  {Hansel}}, \bibinfo {author} {\bibfnamefont {G.}~\bibnamefont {Mato}},
  \bibinfo {author} {\bibfnamefont {C.}~\bibnamefont {Meunier}}, \ and\
  \bibinfo {author} {\bibfnamefont {L.}~\bibnamefont {Neltner}},\ }\href@noop
  {} {\bibfield  {journal} {\bibinfo  {journal} {Neural Computation}\ }\textbf
  {\bibinfo {volume} {10}},\ \bibinfo {pages} {467} (\bibinfo {year}
  {1998})}\BibitemShut {NoStop}%
\bibitem [{\citenamefont {Abbott}(1999)}]{abbott1999lapicque}%
  \BibitemOpen
  \bibfield  {author} {\bibinfo {author} {\bibfnamefont {L.~F.}\ \bibnamefont
  {Abbott}},\ }\href@noop {} {\bibfield  {journal} {\bibinfo  {journal} {Brain
  research bulletin}\ }\textbf {\bibinfo {volume} {50}},\ \bibinfo {pages}
  {303} (\bibinfo {year} {1999})}\BibitemShut {NoStop}%
\bibitem [{\citenamefont {Bracewell}(2000)}]{bracewell2000heaviside}%
  \BibitemOpen
  \bibfield  {author} {\bibinfo {author} {\bibfnamefont {R.}~\bibnamefont
  {Bracewell}},\ }\href@noop {} {\bibfield  {journal} {\bibinfo  {journal} {The
  Fourier Transform and Its Applications}\ ,\ \bibinfo {pages} {61}} (\bibinfo
  {year} {2000})}\BibitemShut {NoStop}%
\bibitem [{\citenamefont {Rieke}\ \emph {et~al.}(1999)\citenamefont {Rieke},
  \citenamefont {Warland}, \citenamefont {Van~Steveninck}, \citenamefont
  {Bialek} \emph {et~al.}}]{rieke1999spikes}%
  \BibitemOpen
  \bibfield  {author} {\bibinfo {author} {\bibfnamefont {F.}~\bibnamefont
  {Rieke}}, \bibinfo {author} {\bibfnamefont {D.}~\bibnamefont {Warland}},
  \bibinfo {author} {\bibfnamefont {R.~d.~R.}\ \bibnamefont {Van~Steveninck}},
  \bibinfo {author} {\bibfnamefont {W.~S.}\ \bibnamefont {Bialek}},  \emph
  {et~al.},\ }\href@noop {} {\emph {\bibinfo {title} {Spikes: exploring the
  neural code}}},\ Vol.~\bibinfo {volume} {7}\ (\bibinfo  {publisher} {MIT
  press Cambridge},\ \bibinfo {year} {1999})\BibitemShut {NoStop}%
\bibitem [{\citenamefont {Cox}(2017)}]{cox2017theory}%
  \BibitemOpen
  \bibfield  {author} {\bibinfo {author} {\bibfnamefont {D.~R.}\ \bibnamefont
  {Cox}},\ }\href@noop {} {\emph {\bibinfo {title} {The theory of stochastic
  processes}}}\ (\bibinfo  {publisher} {Routledge},\ \bibinfo {year}
  {2017})\BibitemShut {NoStop}%
\bibitem [{\citenamefont {Perkel}\ \emph {et~al.}(1967)\citenamefont {Perkel},
  \citenamefont {Gerstein},\ and\ \citenamefont {Moore}}]{perkel1967neuronal}%
  \BibitemOpen
  \bibfield  {author} {\bibinfo {author} {\bibfnamefont {D.~H.}\ \bibnamefont
  {Perkel}}, \bibinfo {author} {\bibfnamefont {G.~L.}\ \bibnamefont
  {Gerstein}}, \ and\ \bibinfo {author} {\bibfnamefont {G.~P.}\ \bibnamefont
  {Moore}},\ }\href@noop {} {\bibfield  {journal} {\bibinfo  {journal}
  {Biophysical journal}\ }\textbf {\bibinfo {volume} {7}},\ \bibinfo {pages}
  {391} (\bibinfo {year} {1967})}\BibitemShut {NoStop}%
\bibitem [{\citenamefont {Johnson}(1996)}]{johnson1996point}%
  \BibitemOpen
  \bibfield  {author} {\bibinfo {author} {\bibfnamefont {D.~H.}\ \bibnamefont
  {Johnson}},\ }\href@noop {} {\bibfield  {journal} {\bibinfo  {journal}
  {Journal of computational neuroscience}\ }\textbf {\bibinfo {volume} {3}},\
  \bibinfo {pages} {275} (\bibinfo {year} {1996})}\BibitemShut {NoStop}%
\bibitem [{\citenamefont {Benesty}\ \emph {et~al.}(2009)\citenamefont
  {Benesty}, \citenamefont {Chen}, \citenamefont {Huang},\ and\ \citenamefont
  {Cohen}}]{benesty2009pearson}%
  \BibitemOpen
  \bibfield  {author} {\bibinfo {author} {\bibfnamefont {J.}~\bibnamefont
  {Benesty}}, \bibinfo {author} {\bibfnamefont {J.}~\bibnamefont {Chen}},
  \bibinfo {author} {\bibfnamefont {Y.}~\bibnamefont {Huang}}, \ and\ \bibinfo
  {author} {\bibfnamefont {I.}~\bibnamefont {Cohen}},\ }in\ \href@noop {}
  {\emph {\bibinfo {booktitle} {Noise reduction in speech processing}}}\
  (\bibinfo  {publisher} {Springer},\ \bibinfo {year} {2009})\ pp.\ \bibinfo
  {pages} {1--4}\BibitemShut {NoStop}%
\bibitem [{\citenamefont {Nagelkerke}\ \emph {et~al.}(1991)\citenamefont
  {Nagelkerke} \emph {et~al.}}]{nagelkerke1991note}%
  \BibitemOpen
  \bibfield  {author} {\bibinfo {author} {\bibfnamefont {N.~J.}\ \bibnamefont
  {Nagelkerke}} \emph {et~al.},\ }\href@noop {} {\bibfield  {journal} {\bibinfo
   {journal} {Biometrika}\ }\textbf {\bibinfo {volume} {78}},\ \bibinfo {pages}
  {691} (\bibinfo {year} {1991})}\BibitemShut {NoStop}%
\bibitem [{\citenamefont {Bj{\"o}rck}(1990)}]{bjorck1990least}%
  \BibitemOpen
  \bibfield  {author} {\bibinfo {author} {\bibfnamefont {{\AA}.}~\bibnamefont
  {Bj{\"o}rck}},\ }\href@noop {} {\bibfield  {journal} {\bibinfo  {journal}
  {Handbook of numerical analysis}\ }\textbf {\bibinfo {volume} {1}},\ \bibinfo
  {pages} {465} (\bibinfo {year} {1990})}\BibitemShut {NoStop}%
\bibitem [{\citenamefont {Rauch}\ \emph {et~al.}(2003)\citenamefont {Rauch},
  \citenamefont {La~Camera}, \citenamefont {Luscher}, \citenamefont {Senn},\
  and\ \citenamefont {Fusi}}]{rauch2003neocortical}%
  \BibitemOpen
  \bibfield  {author} {\bibinfo {author} {\bibfnamefont {A.}~\bibnamefont
  {Rauch}}, \bibinfo {author} {\bibfnamefont {G.}~\bibnamefont {La~Camera}},
  \bibinfo {author} {\bibfnamefont {H.-R.}\ \bibnamefont {Luscher}}, \bibinfo
  {author} {\bibfnamefont {W.}~\bibnamefont {Senn}}, \ and\ \bibinfo {author}
  {\bibfnamefont {S.}~\bibnamefont {Fusi}},\ }\href@noop {} {\bibfield
  {journal} {\bibinfo  {journal} {Journal of neurophysiology}\ }\textbf
  {\bibinfo {volume} {90}},\ \bibinfo {pages} {1598} (\bibinfo {year}
  {2003})}\BibitemShut {NoStop}%
\bibitem [{\citenamefont {Lansky}\ \emph {et~al.}(2006)\citenamefont {Lansky},
  \citenamefont {Sanda},\ and\ \citenamefont {He}}]{lansky2006parameters}%
  \BibitemOpen
  \bibfield  {author} {\bibinfo {author} {\bibfnamefont {P.}~\bibnamefont
  {Lansky}}, \bibinfo {author} {\bibfnamefont {P.}~\bibnamefont {Sanda}}, \
  and\ \bibinfo {author} {\bibfnamefont {J.}~\bibnamefont {He}},\ }\href@noop
  {} {\bibfield  {journal} {\bibinfo  {journal} {Journal of Computational
  Neuroscience}\ }\textbf {\bibinfo {volume} {21}},\ \bibinfo {pages} {211}
  (\bibinfo {year} {2006})}\BibitemShut {NoStop}%
\bibitem [{\citenamefont {Gerstner}\ and\ \citenamefont
  {Kistler}(2002)}]{gerstner2002spiking}%
  \BibitemOpen
  \bibfield  {author} {\bibinfo {author} {\bibfnamefont {W.}~\bibnamefont
  {Gerstner}}\ and\ \bibinfo {author} {\bibfnamefont {W.~M.}\ \bibnamefont
  {Kistler}},\ }\href@noop {} {\emph {\bibinfo {title} {Spiking neuron models:
  Single neurons, populations, plasticity}}}\ (\bibinfo  {publisher} {Cambridge
  university press},\ \bibinfo {year} {2002})\BibitemShut {NoStop}%
\bibitem [{\citenamefont {Gerstner}\ \emph {et~al.}(2014)\citenamefont
  {Gerstner}, \citenamefont {Kistler}, \citenamefont {Naud},\ and\
  \citenamefont {Paninski}}]{gerstner2014neuronal}%
  \BibitemOpen
  \bibfield  {author} {\bibinfo {author} {\bibfnamefont {W.}~\bibnamefont
  {Gerstner}}, \bibinfo {author} {\bibfnamefont {W.~M.}\ \bibnamefont
  {Kistler}}, \bibinfo {author} {\bibfnamefont {R.}~\bibnamefont {Naud}}, \
  and\ \bibinfo {author} {\bibfnamefont {L.}~\bibnamefont {Paninski}},\
  }\href@noop {} {\emph {\bibinfo {title} {Neuronal dynamics}}}\ (\bibinfo
  {publisher} {Cambridge: Cambridge University Press},\ \bibinfo {year}
  {2014})\BibitemShut {NoStop}%
\bibitem [{\citenamefont {Kullback}(1997)}]{kullback1997information}%
  \BibitemOpen
  \bibfield  {author} {\bibinfo {author} {\bibfnamefont {S.}~\bibnamefont
  {Kullback}},\ }\href@noop {} {\emph {\bibinfo {title} {Information theory and
  statistics}}}\ (\bibinfo  {publisher} {Courier Corporation},\ \bibinfo {year}
  {1997})\BibitemShut {NoStop}%
\end{thebibliography}
\end{document}